\PassOptionsToPackage{unicode}{hyperref}
\PassOptionsToPackage{hyphens}{url}
\PassOptionsToPackage{dvipsnames,svgnames,x11names}{xcolor}
\documentclass[
  12pt]{article}

\newsavebox{\GraphRecoveryTableBox}

\usepackage{amsmath,amssymb, amsthm}
\usepackage{iftex}
\usepackage{rotating}
\ifPDFTeX
  \usepackage[T1]{fontenc}
  \usepackage[utf8]{inputenc}
  \usepackage{textcomp} 
\else 
  \usepackage{unicode-math}
  \defaultfontfeatures{Scale=MatchLowercase}
  \defaultfontfeatures[\rmfamily]{Ligatures=TeX,Scale=1}
\fi
\usepackage{lmodern}
\ifPDFTeX\else  
\fi
\IfFileExists{upquote.sty}{\usepackage{upquote}}{}
\IfFileExists{microtype.sty}{
  \usepackage[]{microtype}
  \UseMicrotypeSet[protrusion]{basicmath} 
}{}
\usepackage{xcolor}
\makeatletter
\ifx\paragraph\undefined\else
  \let\oldparagraph\paragraph
  \renewcommand{\paragraph}{
    \@ifstar
      \xxxParagraphStar
      \xxxParagraphNoStar
  }
  \newcommand{\xxxParagraphStar}[1]{\oldparagraph*{#1}\mbox{}}
  \newcommand{\xxxParagraphNoStar}[1]{\oldparagraph{#1}\mbox{}}
\fi
\ifx\subparagraph\undefined\else
  \let\oldsubparagraph\subparagraph
  \renewcommand{\subparagraph}{
    \@ifstar
      \xxxSubParagraphStar
      \xxxSubParagraphNoStar
  }
  \newcommand{\xxxSubParagraphStar}[1]{\oldsubparagraph*{#1}\mbox{}}
  \newcommand{\xxxSubParagraphNoStar}[1]{\oldsubparagraph{#1}\mbox{}}
\fi
\makeatother

\newcommand{\simrowlabel}[1]{%
  \raisebox{\dimexpr.095455\linewidth-.5\height\relax}{%
    \rotatebox{90}{\textbf{#1}}}}

\usepackage{longtable,booktabs,array}
\usepackage{calc} 
\usepackage{etoolbox}
\makeatletter
\patchcmd\longtable{\par}{\if@noskipsec\mbox{}\fi\par}{}{}
\makeatother
\IfFileExists{footnotehyper.sty}{\usepackage{footnotehyper}}{\usepackage{footnote}}
\makesavenoteenv{longtable}
\usepackage{graphicx}
\makeatletter
\def\maxwidth{\ifdim\Gin@nat@width>\linewidth\linewidth\else\Gin@nat@width\fi}
\def\maxheight{\ifdim\Gin@nat@height>\textheight\textheight\else\Gin@nat@height\fi}
\makeatother
\setkeys{Gin}{width=\maxwidth,height=\maxheight,keepaspectratio}
\makeatletter
\def\fps@figure{htbp}
\makeatother

\makeatletter
\@ifpackageloaded{caption}{}{\usepackage{caption}}
\AtBeginDocument{%
\ifdefined\contentsname
  \renewcommand*\contentsname{Table of contents}
\else
  \newcommand\contentsname{Table of contents}
\fi
\ifdefined\listfigurename
  \renewcommand*\listfigurename{List of Figures}
\else
  \newcommand\listfigurename{List of Figures}
\fi
\ifdefined\listtablename
  \renewcommand*\listtablename{List of Tables}
\else
  \newcommand\listtablename{List of Tables}
\fi
\ifdefined\figurename
  \renewcommand*\figurename{Figure}
\else
  \newcommand\figurename{Figure}
\fi
\ifdefined\tablename
  \renewcommand*\tablename{Table}
\else
  \newcommand\tablename{Table}
\fi
}
\@ifpackageloaded{float}{}{\usepackage{float}}
\floatstyle{ruled}
\@ifundefined{c@chapter}{\newfloat{codelisting}{h}{lop}}{\newfloat{codelisting}{h}{lop}[chapter]}
\floatname{codelisting}{Listing}

\makeatother
\makeatletter
\@ifpackageloaded{caption}{}{\usepackage{caption}}
\@ifpackageloaded{subcaption}{}{\usepackage{subcaption}}
\makeatother

\ifLuaTeX
  \usepackage{selnolig}  
\fi
\usepackage[]{natbib}
\usepackage{bookmark}

\IfFileExists{xurl.sty}{\usepackage{xurl}}{} 
\definecolor{darkblue}{rgb}{0.0,0.0,0.6}
\definecolor{red}{rgb}{0.75,0.0,0.0}
\hypersetup{
  pdftitle={Title},
  pdfauthor={Author 1; Author 2},
  pdfkeywords={3 to 6 keywords, that do not appear in the title},
  colorlinks=true,
  linkcolor={darkblue},
  filecolor={Maroon},
  citecolor={darkblue},
  urlcolor={darkblue},
  pdfcreator={LaTeX via pandoc}}

\newcommand{\anon}{1}

\usepackage{bbm}

\newcommand\independent{\protect\mathpalette{\protect\independenT}{\perp}}
\def\independenT#1#2{\mathrel{\rlap{$#1#2$}\mkern2mu{#1#2}}}

\def\I {\mathbbm i}

\DeclareMathOperator*{\argmin}{argmin} 
\DeclareMathOperator*{\argmax}{arg\,max}

\DeclareMathOperator{\sd}{\partial}

\newcommand{\mb}[1]{\boldsymbol{#1}}

\usepackage[noend, boxed, linesnumbered]{algorithm2e}
\SetKwFor{For}{for}{}{end for}%

\newtheorem{proposition}{Proposition}
\usepackage{amsmath,amssymb}
\graphicspath{{../}}
\usepackage[export]{adjustbox}
\usepackage{graphbox}

\usepackage{amsfonts}
\usepackage{geometry}
\usepackage{enumitem}
\usepackage{xltabular}
\usepackage{url}
\usepackage{cleveref}

\usepackage{caption}
\DeclareCaptionFont{capspacing}{\linespread{1}\selectfont}   
\newtheorem{theorem}{Theorem}
\newtheorem{lemma}[theorem]{Lemma}

\theoremstyle{definition}

\newtheorem{assumption}{Assumption}

\crefname{assumption}{Assumption}{Assumptions}
\theoremstyle{remark}

\newcommand{\precision}{\mb{\sigma}^{-1}}
\newcommand{\E}{\mathbb{E}}
\newcommand{\Prob}{\mathbb{P}}

\newcommand{\suffstat}{u}

\newcommand{\Lstar}{\Theta_P^\star}

\newcommand{\dist}{\operatorname{dist}}

\newcommand{\norm}[1]{\left\lVert#1\right\rVert}
\newcommand{\inner}[2]{\left\langle#1,#2\right\rangle}

\newcommand{\lgL}{\ell_{g,L}}

\newcommand{\ellg}{\ell_g}
\newcommand{\ellgP}{\ell_{g,P}}
\newcommand{\stationary}{\Theta_{g,P}^{\star}}
\newcommand{\stationaryL}{\Theta_{g,L,P}^{\star}}

\date{}
\begin{document}

\def\spacingset#1{\renewcommand{\baselinestretch}%
{#1}\small\normalsize} \spacingset{1}


\if1\anon
{
  \title{Penalized Likelihood Inference in Highly Multivariate Sparse Spectral Graphical Gaussian Processes}
  \author{Bo-yu Chen, Drew Yarger\thanks{Correspondence: anyarger@purdue.edu}, and Anindya Bhadra\hspace{.2cm}\\
    Department of Statistics, Purdue University}
  \maketitle
} \fi

\if0\anon
{
  \bigskip
  \bigskip
  \bigskip
  \begin{center}
    {\LARGE\bf Penalized Likelihood Inference in highly multivariate Gaussian Processes via Random Fourier Features and Sparse Spectral Graphical Modeling}
\end{center}
  \medskip
} \fi

\bigskip
\begin{abstract}
We develop a penalized likelihood inference approach for \emph{highly multivariate} Gaussian stochastic processes. Since zeros in the inverse cross-spectral density encode process-level conditional independence, sparse modeling of this object provides a natural route to scalable inference in the highly multivariate case. We introduce a latent variable representation which, combined with random Fourier features, yields a stochastic approximation expectation--maximization algorithm for penalized maximum likelihood estimation of a sparse inverse cross-spectral density matrix under a graphical LASSO penalty. The augmentation is crucial because sparsity is present in the inverse spectral density of the latent variable, whereas the marginal inverse covariance remains dense in the presence of a spatial nugget. 
Our frequency domain formulation enables joint penalized likelihood estimation of process-specific spatial covariance parameters and sparse process-level conditional dependence without requiring a pre-specified reference set of locations.
The method is demonstrated on both synthetic and the MERRA-2 aerosol reanalysis data sets, yielding benefit over existing alternatives, especially for prediction at unobserved locations under spatial misalignment.  
 \end{abstract}

\noindent%
\vfill

\newpage
\spacingset{1.8} 

\section{Introduction}\label{sec:intro}
Construction of \emph{highly multivariate} cross-covariances is of great contemporary interest in disparate applications, including environmental sciences, where recent technological advances have enabled measurements of multiple correlated variables over a spatial domain, and joint modeling is desired. 
One example is NASA’s MERRA-2 reanalysis, which incorporates observations from satellite instruments and provides global outputs of atmospheric variables at multiple vertical levels \citep{gelaro2017modern}. Statistical models have been used to study dependence in MERRA-2 fields \citep{porcu2018modeling,zhang2021assessing,salvana2023spatio}, and jointly predicting multiple MERRA-2 fields is also of interest \citep{gui2026advancing}.

However, flexible approaches to multivariate cross-covariances \citep{gneiting2010matern, apanasovich2012valid} unfortunately do not efficiently scale to a number of variables beyond single digits while retaining flexibility for the individual processes. 
As evident from the comprehensive review by \citet{genton15}, a key reason is that the traditional approaches construct a \emph{fully dense} cross-covariance, and there is no notion of \emph{sparsity}, in a suitable sense to be defined, that enables statistical estimation with respect to three main challenges: managing computation, reducing memory usage, and handling model parsimony and interpretability.
The estimation of a fully dense highly multivariate cross-covariance is infeasible both computationally and statistically. 
In particular, likelihood estimation requires the repeated inversion of a matrix of size $np\times np$, where $p$ is the number of variables or processes, and $n$ is the average number of data locations per process, incurring a nominal computational cost of $O(n^3p^3)$ floating point operations per likelihood evaluation, without further simplifications. 
Furthermore, memory storage of $O(n^2p^2)$ is needed. 
Finally, the number of parameters grows quadratically in $p$, complicating numerical optimization of the likelihood for large $p$, and proposing a model that is both flexible and valid can be challenging. The large $n$ problem has traditionally received attention in the Gaussian process (GP) literature \citep{vecchia1988estimation,datta2016hierarchical,Katzfuss2021Framework}, as have the problems with asymmetry \citep[see, e.g.,][]{li_approach_2011,yarger2025multivariate} or nonstationarity \citep[see, e.g.,][]{kleiber2013spatially} in cross-covariances; whereas the focus on the large $p$ problem is still at a relatively nascent stage, which we review below.

A simple and traditional large $p$ approach is the linear model of coregionalization \citep{goulard1992linear}, which leads to quite restrictive smoothness properties of the component processes, in that the smoothness of any component defaults to that of the roughest \citep{genton15}. More flexibly, 
\cite{dey_graphical_2022} introduced a notion of \emph{stitching} processes together at specific spatial locations, where conditional independence is enforced between different processes at other locations using an iterative proportional scaling approach \citep{speed1986gaussian}. 
Similarly, \cite{peruzzi2024} uses the joint dependence at a reference set of locations to extend the process to the entire spatial domain.  
A potential disadvantage of reference set-based approaches is that estimates, as well as out-of-sample prediction, may depend on the reference set used for construction, including its ordering, i.e., the reference set becomes part of the model construction. If all model parameters are held fixed but two different reference sets are used, then the  predictive distributions at a new location are generally different. \cite{peruzzi2026partial} imposes sparsity \emph{post hoc} via the graphical LASSO \citep{friedman2008sparse} after full spatial estimation of the cross-covariance in a \emph{two-stage} plug-in approach. 
\cite{mitchell_l_krock_modeling_2023} directly model a sparse precision matrix for latent basis coefficients. 
In this approach, one does not model the covariance functions parametrically, making covariance properties like process smoothness and tail decay unavailable, and prediction at unobserved spatial locations is not considered. 

One suitable framework for introducing sparsity is the notion of process-level conditional independence, as introduced by  \citet{brillinger1996remarks} and \citet{dahlhaus2000graphical}, which essentially amounts to sparsity in the inverse spectral density matrix. Approaches such as \citet{dey_graphical_2022} attempt to model this sparsity from the \emph{spatial domain}. In contrast, we develop a multivariate modeling approach in the \emph{spectral domain}, directly targeting a sparse inverse spectral density, which results in a reference-free construction. Further, we use penalized likelihood inference to jointly estimate spatial dependence and sparsity in process dependence, avoiding a two-stage plug-in approach \citep[e.g.,][]{peruzzi2026partial}. Moreover, our modeling approach separately parameterizes the multivariate dependence, modeled via a sparse graphical model, and the process-specific covariance parameters. 
Our approach also enables process-level conditional independence for processes with different range parameters, as well as processes with different classes of covariance functions, a noted gap in \cite{peruzzi2026partial}. 
Inference proceeds via a random Fourier features approach \citep{rahimi2007, emery2018continuous, emery2016improved, miller2022bayesian}, and a computationally-efficient stochastic approximation expectation--maximization (SAEM) scheme \citep{delyon1999convergence} for penalized maximum likelihood inference is developed. 

In simulations, we consider both when the processes are observed at the same locations (the \emph{colocated} setting) and when there are at least some locations not observed for all processes (the \emph{misaligned} setting). 
Asymptotic results suggest that in the colocated setting, leveraging additional processes for prediction of a single process may not yield improvements in prediction performance \citep{zhang2015doesn, lim2022conditions}. 
However, many practical settings may have misaligned observations. While works in highly multivariate spatial processes have traditionally focused on the colocated setting \citep{peruzzi2024, dey_graphical_2022, mitchell_l_krock_modeling_2023},  
we demonstrate improved performance under misalignment.
In addition, we examine quantities that may be of interest for the combined processes (e.g., the sum of the processes). For example, practitioners may be interested in aerosol column mass, obtained by summing the weighted mixing ratios of aerosol components across atmospheric layers, or the total toxic load, calculated as the sum of toxicity levels for different chemicals at each location.
In this case, accurately capturing dependence between variables is necessary for appropriate uncertainty quantification of the predictions. 
We show improvements over existing approaches in terms of estimation, prediction accuracy, and predictive uncertainty quantification. 

We demonstrate this methodology in an application to the MERRA-2 aerosol data set \citep{gelaro2017modern}. 
We find substantial improvements in prediction for our approach in the misaligned setting. Such misalignment can arise when MERRA-2 fields are combined with observations at different locations or spatial resolutions \citep{christensen2019bayesian,zhang2024high}. This demonstrates the benefits of modeling the underlying multivariate spatial process directly, rather than relying on independent univariate analyses.

\section{Preliminaries and Prior Works}

Let $\{\mb{Y}^{\circ}(s) \in \mathbb{R}^p, s \in \mathbb{R}^d\}$ be a multivariate spatial process with output dimension $p$ and input dimension $d$. We refer to the entries $\{Y_j^{\circ}(s) \in \mathbb{R}, s \in \mathbb{R}^d\}$ of $\mb{Y}^{\circ}(s)$ for $j=1, \dots, p$ as individual or marginal processes. 
A principal consideration for Gaussian processes is the covariance structure. 
In particular, if we assume $\mathbb{E}[\mb{Y}^{\circ}(s)] = \mb{\mu}(s)$ for some $\mb{\mu}(s) \in \mathbb{R}^p$, and the processes are second-order, the multivariate covariance function is: \begin{align*}
\mathbb{E}\left[(\mb{Y}^{\circ}(s) - \mb{\mu}(s)) \left(\mb{Y}^{\circ}(t)-\mb{\mu}(t)\right)^\top \right] &= \mb{C}(s, t;\mb{\theta}),
\end{align*}depending on parameters $\mb{\theta}$. 
The $\mathbb{R}^{p\times p}$-valued covariance function $\mb{C}(s, t; \mb{\theta})$ describes dependence both between and within processes at two locations.
Let $\mb{C}(s,t;\mb{\theta}) = [C_{jk}(s,t; \mb{\theta})]_{j,k=1}^p$, and we refer to $C_{jj}(s,t; \mb{\theta})$ as the marginal covariance function of process $j$ and $C_{jk}(s,t; \mb{\theta})$ as the cross-covariance function between processes $j$ and $k$. 
We say that $\{\mb{Y}^{\circ}(s)\}$ is second-order stationary (or henceforth, stationary) if $\mathbb{E}[\mb{Y}^{\circ}(s)] = \mb{\mu}$ and $\mb{C}(s, t; \mb{\theta}) = \mb{C}(t-s; \mb{\theta})$, so that the covariance function $\mb{C}(\cdot, \cdot; \mb{\theta})$ only depends on the lag between $t$ and $s$ through $t-s$ \citep{Cressie1993}. 
Throughout, we assume stationarity and that $\mb{\mu} = \mb{0}_p$. 

In the stationary case, many multivariate covariances admit a spectral representation: \begin{align*}
    \mb{C}(h; \mb{\theta})&= \int_{\mathbb{R}^d} e^{\I h^\top x } \mb{f}(x;\mb{\theta}) dx,
\end{align*}where $\mb{f}(x; \mb{\theta}) = [f_{jk}(x; \mb{\theta})]_{j,k=1}^p$ is a matrix-valued spectral density and $\I$ is the imaginary unit.
If $\mb{f}(x;\mb{\theta})$ exists, the validity of the covariance $\mb{C}(h;\mb{\theta})$ may be shown by ensuring $\mb{f}(x; \mb{\theta})$ is Hermitian and positive semidefinite for all $x \in \mathbb{R}^d$; see, for example, Section 2.1 of \cite{yarger2023multivariate}. 
The spectral density representation is a special case of multivariate Bochner's theorem, ensuring multivariate validity when spectral densities exist. In this paper, we present an approach that accommodates any stationary marginal covariance function with a spectral density, including mixed families of processes.

A key motivation for our work is the notion of process-level conditional independence \citep{dahlhaus2000graphical}. 
Let $\mb{Y}^{\circ}_{-jk}(s) \in \mathbb{R}^{p-2}$ be all entries of $\mb{Y}^{\circ}(s)$ except marginal processes $j$ and $k$.
\cite{dahlhaus2000graphical} defines \emph{process-level} conditional independence for fixed $j\neq k$ as $Y^\circ_j(s_1) \independent Y^\circ_k(s_2) \mid \{\mb{Y}^{\circ}_{-jk}(s), s \in \mathbb{R}^d\}$ for all $s_1, s_2 \in \mathbb{R}^d$, and shows that this process-level conditional independence holds whenever the $j,k$ entry of the inverse spectral density matrix $\mb{f}(x; \mb{\theta})^{-1}$ is $0$ for all $x \in \mathbb{R}^d$. 
While conceptually appealing, one cannot practically condition on all $\{\mb{Y}^{\circ}_{-jk}(s), s \in \mathbb{R}^d\}$ for a direct spatial domain construction. 
However, by directly modeling in the spectral domain, process-level conditional independence enables computational tractability when the conditional independence graph of the processes is sparse. The key bridge between our spectral construction to the spatial domain estimation and computation is provided by \emph{random Fourier features} \citep[see, e.g.,][]{rahimi2007}.

\subsection{Existing Approaches for Highly Multivariate Cross-covariances}
Although already briefly summarized in Section~\ref{sec:intro}, we review relevant approaches for modeling highly multivariate spatial data with a bit more technical detail. First, \citet{cressie2016multivariate} construct a multivariate covariance one process at a time. Each subsequent variable depends on earlier variables across space, with a separate covariance for its remaining unexplained variation. 
The paper further uses a potentially-sparse directed acyclic graph to express dependence among variables. 
Alternatively, the stitching approach of \citet{dey_graphical_2022} uses a finite set of locations $\mathcal{S}$ of size $|\mathcal{S}|$, referred to as reference or stitching locations, and a desired multivariate covariance function $\mb{C}(h; \mb{\theta})$. They construct $p^2$ covariance matrices of size $|\mathcal{S}| \times |\mathcal{S}|$, denoted by $M_{jk}(\mathcal{S}, \mathcal{S})$, for the reference locations such that conditional independence holds where desired for the spatial domain, and the covariances and cross-covariances match with $M_{jk}(\mathcal{S}, \mathcal{S}) = C_{jk}(\mathcal{S}, \mathcal{S})$. 
The multivariate covariance is defined by: \begin{align}
    M_{jk}(s_1, s_2) &= C_{jj}(s_1, \mathcal{S}) C_{jj}(\mathcal{S}, \mathcal{S})^{-1} M_{jk}(\mathcal{S}, \mathcal{S})C_{kk}(\mathcal{S}, \mathcal{S})^{-1} C_{kk}(\mathcal{S}, s_2).\label{eq:dey}
\end{align}
Essentially, the processes are stitched together at the reference locations $\mathcal{S}$, then independently extended to all locations. This approach encourages sparsity in the inverse cross-spectral density matrix through a sparse underlying graph $G$. However, the out-of-sample predictive variance computed in this manner at a new location $s^*_{n+1}$ is dependent on the choice of reference set $\mathcal{S}$. The sample locations and reference locations $\mathcal{S}$ may or may not be the same.
\cite{dey_graphical_2022} propose a chromatic Gibbs sampler for Bayesian estimation. 

More recently, \cite{peruzzi2024} proposed a similar approach with a few notable differences. Most importantly, they introduce an efficient likelihood estimation scheme that is computationally scalable to the setting where both $n$ and $p$ are large. 
The proposed model is similar to \cite{dey_graphical_2022} in \eqref{eq:dey}, using reference locations $\mathcal{S}$ with the choice of: $ M_{jk}(\mathcal{S}, \mathcal{S}) = \mb{L}_j\mb{L}_k^\top$, where $\mb{L}_j$ is the lower-triangular Cholesky decomposition of $C_{jj}(\mathcal{S}, \mathcal{S})$. 
Therefore, the model has the same dependence on the reference location set $\mathcal{S}$ or its ordering.
\cite{peruzzi2026partial} examines spectral density structures that induce process-level conditional independence without a reference set. However, they note that their proposed framework does not accommodate general multivariate Mat\'ern models with process-specific scale parameters. Whether these parameters can encode process-level conditional independence remains an open question that this paper addresses. For inference, \cite{peruzzi2026partial} uses a two-stage plug-in approach.

\cite{mitchell_l_krock_modeling_2023} is conceptually the most similar to our work, in that they model directly in the frequency domain, obviating the need for a reference set. The $p\times p$ inverse spectral density matrix $\mb{f}(x; \mb{\theta})^{-1}$ is estimated directly and nonparametrically across $x$ using the graphical LASSO \citep{friedman2008sparse} and the fused graphical LASSO \citep{danaher2014joint} penalties. 
This handles nonstationarities in highly multivariate data efficiently. 
On the other hand, process-level conditional independence is not specifically encouraged, in favor of piecewise conditional independence in the spectral domain.

\subsection{Approximating Cross-covariances by Random Fourier Features}

We outline the model construction and key ideas of our paper by illustrating the approximation of a multivariate cross-covariance following \cite{emery2016improved}, where a full derivation is provided.
We drop $\mb{\theta}$ in the $\mb{f}(x; \mb{\theta})$ notation for simplified development. 
Applying $\exp({\I h^\top x}) = \cos(h^\top x ) + \I \sin(h^\top x)$, the overall goal is to approximate the integral: \begin{align}\mathbb{E}\left[\mb{Y}^{\circ}(s)\mb{Y}^{\circ}(s+h)^\top\right] = \int_{\mathbb{R}^d}\left\{ \cos(h^\top x) \Re[\mb{f}(x)] - \sin(h^\top x )\Im[\mb{f}(x)]\right\}dx,\label{eq:cos_sin}\end{align} 
using a set of random features in the spectral domain, where $\Re(z)$ and $\Im(z)$ are the real and imaginary parts of $z$, respectively.
First consider a proposal probability density function $g(x)$ with $g(x)>0$ wherever $\mb{f}$ is nonzero, and simulate $X_{\ell, j} \sim g(\cdot)$, $\psi_{\ell, j}\sim \mathrm{U}[0, 2\pi]$ for $j=1, \dots, p$ and $\ell = 1, \dots, L$ independently. 
Here, $L$ indicates the size of the Fourier feature approximation and $\mathrm{U}$ denotes the uniform density. 
\cite{emery2016improved} then construct: \begin{align}
    \mb{Y}^{\circ}(s) &= \frac{1}{\sqrt{L}}\sum_{\ell = 1}^L \sum_{j=1}^p \mb{a}_{j}(X_{\ell, j})\cos(\langle X_{\ell, j}, s\rangle + \psi_{\ell,j}) + \mb{b}_{j}(X_{\ell, j})\sin(\langle X_{\ell, j}, s\rangle + \psi_{\ell,j}).\label{eq:emery}
\end{align}
The two terms of \eqref{eq:emery} correspond to those in \eqref{eq:cos_sin}. 
With particular choices of $\mb{a}_j(x)\in \mathbb{R}^{p}$ and $\mb{b}_j(x)\in \mathbb{R}^{p}$, \cite{emery2016improved} show that \eqref{eq:emery} has covariance \eqref{eq:cos_sin} in expectation. 
The vectors $\mb{a}_j(x)$ and $\mb{b}_j(x)$ encode both the dependence relations between processes and information about the individual spectral densities. 
In particular, we let $\mb{a}_j(x) $ and $\mb{b}_j(x)$ denote the $j$-th columns of the matrices $\mb{A}(x)\in \mathbb{R}^{p\times p}$ and $\mb{B}(x)\in \mathbb{R}^{p\times p}$, which in turn are the real and imaginary Cholesky components of the (weighted) spectral density matrix: \begin{align}
    \frac{2\mb{f}(x)}{g(x)} &= (\mb{A}(x) + \I \mb{B}(x))(\mb{A}(x) + \I \mb{B}(x))^*,\label{eq:cholesky}
\end{align}
where $\mb{M}^*$ denotes the conjugate transpose of the matrix $\mb{M}$. 
The density $g(\cdot)$ is thus a proposal in the spirit of importance sampling in order to target the matrix-valued integral $\int_{\mathbb{R}^d} e^{\I \langle x,  h\rangle}\mb{f}(x)dx$. \cite{emery2018continuous} extend this work to nonstationary models, and \cite{arroyo_algorithm_2021} describe the computational algorithm in more detail. 
This approach for approximate simulation has an advantage over exact circulant embedding approaches \citep[e.g.,][]{moreva_fast_2018} by providing simulations at arbitrary spatial locations. 

In this paper's context, the algorithm's issue is the computational cost of $O(Lp^4 + Lp^2n)$ floating point operations, where $O(\cdot)$ denotes the general order, and $n$ is the number of simulation locations.
This is prohibitive for large $p$, needing the Cholesky decomposition of the dense $p\times p$ matrix $2\mb{f}(X_{\ell, j})/g(X_{\ell, j})$, for each of the $Lp$ summands in \eqref{eq:emery}.

\section{Simulation and Likelihood via Fourier Features}

\subsection{Simulating Highly Multivariate Processes}\label{sec:simulation_methodology}
We first propose an efficient simulation approach for highly multivariate GPs using random Fourier features.
Our framework follows a similar, but subtly different construction to \cite{emery2016improved}. 
Let $X_\ell$ and $\psi_\ell$ be simulated as $X_{\ell, j}$ and $\psi_{\ell, j}$ in \eqref{eq:emery}, the matrices $\mb{A}(x)$ and $\mb{B}(x)$ be given by \eqref{eq:cholesky}, and each $\mb{Z}_\ell \overset{iid}{\sim} \mathcal{N}(\mb{0}_p, \mb{I}_p)$.
Our approach then constructs: 
\begin{align*}
    \mb{Y}^{\circ}(s) &= \frac{1}{\sqrt{L}} \sum_{\ell = 1}^L \mb{A}(X_{\ell}) \mb{Z}_{\ell}\cos(\langle s, X_{\ell} \rangle + \psi_\ell) + \frac{1}{\sqrt{L}}\sum_{\ell = 1}^L \mb{B}(X_{\ell}) \mb{Z}_{\ell}\sin(\langle s, X_{\ell} \rangle + \psi_\ell).
\end{align*}
This construction replaces a sum over the columns of $\mb{A}(X_{\ell, j})$ and $\mb{B}(X_{\ell, j})$ in \eqref{eq:emery} with the matrix-vector products $\mb{A}(X_\ell) \mb{Z}_\ell$ and $\mb{B}(X_{\ell}) \mb{Z}_\ell$. 
Na\"ively, this reduces the number of Choleskys needed from $Lp$ in \eqref{eq:emery} to $L$. 
Importantly, we need not construct $\mb{Y}^{\circ}(s)$ using $\mb{A}(X_{\ell})$ and $\mb{B}(X_{\ell})$ directly.
Instead, we may backsolve against the Cholesky of the matrix $g(X_{\ell})\mb{f}(X_{\ell})^{-1}/2$, where $\mb{f}(x)^{-1}$ is potentially sparse, encoding process-level conditional independence.

We thus formulate $\mb{f}(x)$ so that sparsity is directly interpretable. 
In particular, we choose: \begin{align}
    \mb{f}(x)&= \mb{h}(x)\mb{\sigma}\mb{h}(x)^\top,\quad \mb{h}(x)=\textrm{diag}\left(f_j(x), j=1, \dots, p\right)^{\frac{1}{2}},\label{eq:simplified_form}
\end{align}for a positive-definite matrix $\mb{\sigma} \in \mathbb{R}^{p\times p}$ and unit-integral marginal   spectral density functions $\{f_j(x), j=1, \dots, p\}$. 
Since $\int_{\mathbb R^d} f_j(x)\,dx=1,$ all marginal process variances reside in $\mb{\sigma}$, and there is no scaling ambiguity between $\mb h(x)$ and $\mb{\sigma}$.
In this setting, the Cholesky decomposition of $\mb{f}(x)^{-1}$ may be computed directly from the Cholesky of $\mb{\sigma}^{-1}$. 
Simulation thus requires computing a Cholesky decomposition {\it once}. 
Specifying process-level conditional independence is simple through zeros in $\mb{\sigma}^{-1}$, and validity is immediate \citep{yarger2023multivariate}. 
For the Mat\'ern model, this spectrally-separable model is similar to the parsimonious model of \cite{gneiting2010matern}.
\cite{terres2018bayesian}, \cite{bolin_multivariate_nodate}, \cite{yarger2023multivariate}, \cite{guinness2014multivariate}, and \cite{peruzzi2026partial} propose similar spectrally-separable constructions of multivariate models. 
We give a detailed algorithm in Supplementary Section~\ref{app:ex_sim}, where we construct $\mb{W}_\ell:=\mb{A}(X_\ell)\mb{Z}_\ell  \sim \mathcal{N}(\mb{0}_p, 2\mb{f}(X_\ell)/g(X_\ell))$. A key advance of the present work is to move from spectral domain \emph{process construction} to spatial domain \emph{penalized likelihood estimation} using random Fourier features as a bridge from the two domains.

\subsection{Likelihood and Penalty for Highly Multivariate Processes}\label{sec:marginal_lik}
We next model the observed data by adding a nugget to the spatial process $\mb{Y}^\circ(s)$.
Defining the coefficients as $c_{i, \ell}= L^{-1/2}\cos(\langle s_i, X_\ell\rangle + \psi_{\ell})$, we express the observation at spatial location $s_i$ as $\mb{Y}(s_i) = \sum_{\ell=1}^L c_{i,\ell} \mb{W}_\ell + \mb{\epsilon}_i$, where $\mb{\epsilon}_i\sim \mathcal{N}(\mb{0}_p,\mb{d})$ represents the nugget vector with $\mb{d} = \mathrm{diag}(d_1, \dots, d_p)$ (Equation~\eqref{eq:simplified_form} is real-valued so $\mb{B}(x) = \mb{0}_{p\times p}$). 
With the exception of Proposition~\ref{prop:marginal_Y}, we condition on the realized features $\{(X_\ell,\psi_\ell)\}$ from Section~\ref{sec:simulation_methodology} throughout all subsequent likelihood and expectation evaluations, leading to the following marginal log-likelihood of $\mb{Y}$ after integrating out $\mb{W}$:
\begin{align*}
    \mathcal{L}(\mb{\theta}\mid\mb{Y}) &= -\frac{1}{2} \left\{np\log(2\pi) + \log\left(\det \left[\mb{\Sigma}_L(\mb{\theta}) + \mb{D}(\mb{\theta})\right]\right) + \mb{Y}^\top \left[\mb{\Sigma}_L(\mb{\theta}) + \mb{D}(\mb{\theta})\right]^{-1} \mb{Y}\right\},
\end{align*}
where $[\mb{\Sigma}_L(\mb{\theta})]_{ij} = \sum_{\ell=1}^{L} c_{i,\ell}c_{j,\ell}\,\mb{\Sigma}_{\mb{W}_\ell}$ with $\mb{\Sigma}_{\mb{W}_\ell} = 2\mb{f}(X_\ell)/g(X_\ell) \in \mathbb{R}^{p\times p}$,  $\mb{D}(\mb{\theta})$ is an $np \times np$ diagonal matrix with entries $\{d_{j}\}_{j=1}^p$ appropriately repeated along the diagonal, and $\mb{Y}=\{\mb{Y}(s_i)\}_{i=1}^n \in \mathbb{R}^{np}$. 
Here, $\mb{\theta}$ includes $\mb{\sigma}^{-1}$, $\mb{d}$, and any marginal covariance parameters in $\mb{h}(x)$.
For $L\ll n$, the Woodbury identity can reduce computational cost.

Even with suitable computational complexity, numerical optimization of the marginal log-likelihood is complicated by the high dimensionality of the parameter space, as $\mb{\sigma}^{-1}$ contains $p(p+1)/2 = O(p^2)$ parameters. 
To tackle this problem, we add a graphical LASSO penalty \citep{friedman2008sparse} to the log-likelihood. 
In particular, consider: 
\begin{align*}
    \mathcal{L}_P(\mb{\theta} \mid \mb{Y}) = \mathcal{L}(\mb{\theta} \mid \mb{Y}) - \lambda \sum_{1 \le j \neq k \le p} |\eta_{jk}| = \mathcal{L}(\mb{\theta} \mid \mb{Y}) - P(\mb{\theta}),
\end{align*}where $\mb{\sigma}^{-1} = [\eta_{jk}]_{j,k=1}^p$ and $\lambda >0$ is a tuning parameter. 
In this case, $\eta_{jk} = 0$ directly corresponds to process-level conditional independence for processes $j$ and $k$.

An immediate challenge is that standard graphical LASSO methodology is not directly applicable as $\mb{W}$ is unobserved. 
The matrix $\mb{\sigma}$, defined in the spectral domain, does not necessarily correspond to the covariance matrix of $\mb{Y}(s)$, and the marginal precision of $\mb{Y}$ may be dense even when $\mb{\sigma}^{-1}$ is sparse.
To address estimation with this penalty, we propose an SAEM algorithm  \citep{wei_tanner_mcem_1990, delyon1999convergence} in the following section that bridges the spectral domain construction and spatial domain estimation.

\section{Inference Using Stochastic Approximation EM}

For simplicity, the following derivations use colocated
observations. Under misalignment, sums are restricted to observed entries and $n$ is replaced by the process-specific sample sizes $n_j$. 
We formulate the penalized complete-data log-likelihood as follows:
\begin{align}
    &\mathcal{L}_P(\mb{\theta} \mid \mb{Y}, \mb{W}) = \log(f(\mb{Y} \mid \mb{W}, \mb{\theta})) + \log(f(\mb{W} \mid \mb{\theta}))- \lambda \sum_{1 \le j \neq k \le p}|\eta_{jk}|\nonumber\\
    &= C
    -\frac{n}{2}\log|\mb{d}|
    -\frac{1}{2}\sum_{i=1}^{n}
    \left(\mb{Y}(s_i)-\sum_{\ell=1}^{L}c_{i,\ell}\mb{W}_\ell\right)^\top
    \mb{d}^{-1}
    \left(\mb{Y}(s_i)-\sum_{\ell=1}^{L}c_{i,\ell}\mb{W}_\ell\right)-\frac{L}{2}\log|\mb{\sigma}|\nonumber\\
    &\; \; 
    -\sum_{\ell=1}^{L}\log|\mb{h}(X_\ell)|
    -\frac{1}{4}\sum_{\ell=1}^{L}g(X_\ell)
    \mb{W}_\ell^\top\mb{h}(X_\ell)^{-1}
    \mb{\sigma}^{-1}\mb{h}(X_\ell)^{-1}\mb{W}_\ell
    -\lambda\sum_{1\leq j\neq k\leq p}|\eta_{jk}|,\label{eq:comp_loglik}
\end{align}
where $C$ is a constant independent of $\mb{\theta}$. At step $m$, the E step of an EM approach requires: $${Q}\left(\mb{\theta} \mid \hat{\mb{\theta}}^{(m-1)}\right) = \mathbb{E}\left[ \mathcal{L}_P(\mb{\theta} \mid \mb{Y}, \mb{W}) \mid \mb{Y}, \hat{\mb{\theta}}^{(m-1)}\right],$$
where $\hat{\mb{\theta}}^{(m-1)}$ are the estimates from the previous iteration, while the M step maximizes ${Q}(\mb{\theta} \mid \hat{\mb{\theta}}^{(m-1)})$ over the parameters $\mb{\theta}$ to form the new estimate $\hat{\mb{\theta}}^{(m)}$.
However, computing the multivariate normal conditional distribution $\mb{W} \mid \mb{Y}, \hat{\mb{\theta}}^{(m-1)}$ directly requires solving against a matrix of size $np$, the size of the original computational problem. 
To ameliorate this challenge, we propose a Gibbs sampler-based E step, detailed next. 

\subsection{Markov Chain Monte Carlo E step}
For a fixed EM step $m$, the Monte Carlo E step proceeds as follows. For a Monte Carlo sample of size $T$, we simulate $\mb{W}^{(\tau)}\sim \mb{W}\mid\mb{Y},\hat{\mb{\theta}}^{(m-1)}$ for $\tau=1,2,\dots,T$, suppressing the dependence of $\mb{W}^{(\tau)}$ on $m$ for simplicity.
The estimate of ${Q}(\mb{\theta} \mid \hat{\mb{\theta}}^{(m-1)})$ is then: \begin{align*}
        \hat{Q}\left(\mb{\theta} \mid \hat{\mb{\theta}}^{(m-1)}\right) &= \frac{1}{T} \sum_{\tau = 1}^T \mathcal{L}_P\left(\mb{\theta} \mid \mb{Y}, \mb{W}^{(\tau)}\right).
    \end{align*} 
To generate samples of $\mb{W}^{(\tau)}$ efficiently, we use Gibbs sampling. At the $m$-th iteration, we initialize the chain at the final Gibbs state from iteration $m-1$ and then perform $T$ sequential Gibbs sweeps.
After setting an initial $\mb{W}^{(0)}$,
in each step $\tau$, we update each $\mb{W}_\ell$ sequentially: for $\ell = 1, \dots, L$, we simulate $\mb{W}_\ell^{(\tau)} \mid \mb{W}_{-\ell}^{(\tau, prev)}, \mb{Y}$ where $ \mb{W}_{-\ell}^{(\tau, prev)} = \{\mb{W}_{\ell^\prime}^{(\tau)},  \ell^\prime < \ell\}\cup \{\mb{W}_{\ell^\prime}^{(\tau-1)},  \ell^\prime >\ell\}$ is the most recent value for all other $\ell^\prime \neq \ell$.
This process continues for $\tau = 1, \dots, T$. 
Samples from the distribution $\mb{W}_\ell^{(\tau)} \mid \mb{W}_{-\ell}^{(\tau, prev)}, \mb{Y}$ can be generated based on the Cholesky decomposition of size $p$ via the Woodbury matrix identity. 
We detail the form of the multivariate Gaussian conditional distribution for $\mb{W}_{\ell}^{(\tau)}$ in Supplementary Section~\ref{app:conditional}.

\subsection{Conditional Maximization Steps}

We detail the updates for the parameters $\mb{\sigma}^{-1}$, $\{d_j\}_{j=1}^p$, and $\mb{\iota}$, which denotes the finite-dimensional spatial parameters included in $\mb{h}(x)$. We perform three conditional maximization steps in an alternating fashion \citep{meng1993maximum}:
\begin{enumerate}

\item CM1. (\emph{The precision matrix}).
Since $\mb{Y}(s) \mid \mb{W}, \mb{\theta}$ in Equation~\eqref{eq:comp_loglik} is only a function of the nugget $\mb{d}$, the update for $\mb\sigma$ is:
\begin{align*}
\mb{\sigma}^{(m)}
&=\argmin_{\mb{\sigma}\succ0}\Bigg\{
\frac{L}{2}\log|\mb{\sigma}|
+\frac{1}{4T}\sum_{\tau=1}^T\sum_{\ell=1}^L
g(X_\ell)\,
\mb{W}_\ell^{(\tau)\top}\mb{h}(X_\ell)^{-1}
\mb{\sigma}^{-1}\mb{h}(X_\ell)^{-1}
\mb{W}_\ell^{(\tau)}
+\lambda\sum_{1\le j\neq k\le p}|\eta_{jk}|
\Bigg\},
\end{align*}

which can be recognized as a graphical LASSO problem with an estimated second-moment matrix $\sum_{\ell=1}^L g(X_\ell)\mb{h}(X_\ell)^{-1}{\overline{U}_{m,\ell}}\mb{h}(X_\ell)^{-1}/{(2L)}$ and a penalty parameter $\rho_L=2\lambda/L$, where ${\overline{U}_{m,\ell}}=\sum_{\tau=1}^T {\mb{W}}_{\ell}^{(\tau)}{\mb{W}}_{\ell}^{(\tau) \top}/T$ is the Monte Carlo estimate of the conditional expectation of the complete-data sufficient statistic.

\item CM2. (\emph{The measurement error variance}). Holding other parameters fixed, $\mb{d}$ is updated using the $\log f(\mb{Y}\mid\mb{W},\mb{\theta})$ term. The update for the $j$-th entry of $\mb{d}$ at iteration $m$ is
\begin{align*}
\hat{d}_j^{(m)} = \frac{1}{nT}\sum_{\tau = 1}^T \sum_{i=1}^n
\left(Y_j(s_i) - \sum_{\ell=1}^L c_{i,\ell} W_{\ell,j}^{(\tau)}\right)^2,
\end{align*}
where $W_{\ell,j}^{(\tau)}$ is the $j$-th entry of $\mb{W}_{\ell}^{(\tau)}$.

\item CM3. (\emph{Parameters governing spatial covariance}). The spatial parameters $\mb{\iota}$  governing the individual processes through the spectral densities $\{f_j(x)\}$ are only related to $\log(f(\mb{W}\mid\mb{\theta}))$. Holding $\mb{\iota}_k,\,k\neq j$ fixed and cycling through $j=1,\dots,p$, yields:
\begin{align*}
    \hat{\mb{\iota}}_j^{(m)} = \argmax_{\mb{\iota}_j} \left\{-\sum_{\ell=1}^L \log|\mb{h}(X_\ell)|-\frac{1}{4}\sum_{\ell=1}^L g(X_\ell)\left(\eta_{jj}\frac{[\overline U_{m,\ell}]_{jj}}{f_j(X_\ell)}+2\sum_{k\neq j}\eta_{jk}\frac{[\overline U_{m,\ell}]_{jk}}{\sqrt{f_k(X_\ell)f_j(X_\ell)}}\right)\right\}.
\end{align*}
This coordinatewise approach also accommodates more flexible spatial models, such as those with nonstationarity \citep{emery2018continuous}. 

\end{enumerate}

\subsection{Stochastic Approximation Updates}

To ensure convergence, the MCEM algorithm requires the Monte Carlo sample size $T$ to increase over iterations. One alternative approach is the stochastic approximation EM (SAEM) algorithm, which does not require $T$ to increase with iterations by averaging E step estimates \citep{delyon1999convergence}. We consider the stochastic approximation expectation conditional maximization (SAECM) algorithm. In the $m$-th iteration, let $\mb{\suffstat}_\ell^{(m)}$ and $\hat{\mb d}_{SA}^{(m)}$ denote the stochastic approximation estimates of the complete-data sufficient statistic. Define
\begin{align*}
\hat{Q}_{SA}^{(m)}\left(\mb{\theta} \right) &= (1-\gamma_m)\hat{Q}_{SA}^{(m-1)}\left(\mb{\theta}\right) + \gamma_m\hat{Q}\left(\mb{\theta} \mid \hat{\mb{\theta}}^{(m-1)}\right),\end{align*}
where $\{\gamma_m\}$ is a nonincreasing sequence with $\gamma_1 = 1$. Then, for a set of parameters $\mb\theta=(\mb{\theta}_1,\dots,\mb{\theta}_k)$, SAECM cycles through parameters: 
$
\hat{\mb{\theta}}_j^{(m)}= \argmax_{\mb{\theta}_j}\hat{Q}_{SA}^{(m)}\left(\mb{\theta}_j\right),\; \text{for }j=1,\dots,k    
$
with the rest of $\mb{\theta}$ fixed at $(\hat{\mb{\theta}}_1^{(m)},\dots,\hat{\mb{\theta}}_{j-1}^{(m)},\hat{\mb{\theta}}_{j+1}^{(m-1)},\dots,\hat{\mb{\theta}}_k^{(m-1)})$. 
For $\gamma_m < 1$, the procedure uses information from previous steps to form the new estimates. 
Therefore, we update the sufficient statistic estimates using $\mb{\suffstat}_\ell^{(m)}
=(1-\gamma_m)\mb{\suffstat}_\ell^{(m-1)}
+\gamma_m\overline U_{m,\ell}$, and $\hat{\mb{d}}_{SA}^{(m)} = (1-\gamma_m)\hat{\mb{d}}_{SA}^{(m-1)} + \gamma_m\hat{\mb{d}}^{(m)}$. The remaining parameters are updated by conditional maximization
using $\mb{\suffstat}_\ell^{(m)}$.

\subsection{Prediction at Unobserved Locations}

In order to predict $\mb{Y}(s)$ at a new location $s_*$, we sample the conditional distribution of $(\mb{W}^{(\tau)} \sim \mb{W} \mid \mb{Y}, \hat{\mb{\theta}})$, similar to the E step of the algorithm with the final estimated parameters $\hat{\mb{\theta}}$.
We then use the prediction $\hat{\mb{Y}}(s_*)= L^{-1/2}\sum_{\ell = 1}^L \hat{\mb{W}}_\ell\cos(\langle s_*, X_\ell\rangle + \psi_\ell),$ where $\hat{\mb{W}}_\ell =(1/T) \sum_{\tau = 1}^T\mb{W}_\ell^{(\tau)}$. The predictive variance matrix is formed as:
\begin{align*}
\widehat{\textrm{Var}}\big({\mb Y}(s_*)\mid \mb{Y},\hat{\mb{\theta}}\big)
&=\frac1L\sum_{\ell=1}^{L}\sum_{\ell'=1}^{L}
\bigg[\left(\frac{1}{T-1}\sum_{\tau=1}^{T}\big(\mb W_\ell^{(\tau)}-\hat{\mb W}_\ell\big)\big(\mb W_{\ell'}^{(\tau)}-\hat{\mb W}_{\ell'}\big)^\top\right)\\
&\phantom{=\frac1L\sum_{\ell=1}^{L}\sum_{\ell'=1}^{L}\bigg[} \times\cos(\langle s_*,X_\ell\rangle+\psi_\ell)\cos(\langle s_*,X_{\ell'}\rangle+\psi_{\ell'})\bigg] + \hat{\mb{d}}.
\end{align*}

\subsection{Implementation Details}
\label{subsec:implementation}
We next discuss some algorithmic details.
To select $\rho_L$, information criteria or cross-validation can be used. In our experiments, the Akaike information criterion (AIC) \citep{akaike1998information} tends to select denser precision estimates, whereas the Bayesian information criterion (BIC) \citep{schwarz1978estimating} yields sparser estimates. We select $\rho_L$ among nonzero penalties by BIC and denote the selected fit \emph{Penalized} hereafter. The BIC is defined as
$
\mathrm{BIC}(\rho_L) = -2\mathcal{L}(\hat{\mb\theta}_{\rho_L} \mid \mb{Y}) + k\ln n,
$
where $k=\sum_{1\le i<j\le p}\mathbbm{1}(\hat\eta_{ij}\neq 0)$ counts the selected edges. The remaining spatial parameters shift $\mathrm{BIC}(\rho_L)$ by a common constant and do not affect the selected $\rho_L$. 

Under misalignment, since each process is observed at a different subset of locations, the random feature matrix no longer admits the simple Kronecker structure available under colocated observations.
We handle the misalignment using an observation mask together with the $n\times L$ Fourier feature matrix. 
The required matrix products are evaluated directly from the feature values at the observed locations, without explicitly constructing a full matrix.

\subsection{Computational Complexity}
\label{subsec:complexity}

Let $\bar n$ be the average number of observations over processes.
For each EM iteration, $L$ relevant Cholesky blocks must be computed, so the complexity with precomputed random features, $T$ Gibbs sweeps, and residuals updated after each Gibbs draw is $O(Lp^3+(T+J)Lp^2+TL \bar np)$ plus the cost of the graphical-LASSO update.
Here, $J$ bounds the number of CM3 objective evaluations per process.
The $p^3$ and $p^2$ parts may be reduced appropriately when the sparsity of $\mb{\sigma}^{-1}$ can be exploited.
Each iteration avoids the $O(\bar n^3p^3)$ factorization of the full observation covariance and the potential $O(L^3p^3)$ factorization arising from marginalizing out $\mb{W}$.
Likelihood evaluation conditional on the features costs at most $O(p\bar nL^2+p\bar nL+L^3p^3)$. 
For prediction, generating $T$ posterior draws costs $O\{Lp^3+TLp^2+TL\bar np\}$.

\section{Theoretical Results}\label{sec:theory}

We start by proving in Proposition~\ref{prop:marginal_Y} that the unconditional distribution of $\mb{Y}$ converges to a multivariate Gaussian distribution with the target covariance function as $L\to\infty$, providing asymptotic support for our random Fourier features approximation. Proposition~\ref{prop:set} establishes convergence to the stationary set of the penalized random feature likelihood for fixed $L$. Proposition~\ref{prop:diagonal_convergence} establishes convergence to the stationary set of the exact penalized GP likelihood along suitable sequences $L_k\to\infty$ and $m_k\to\infty$. Explicit assumptions and detailed proofs are deferred to Supplementary Section~\ref{app:proof}.

\begin{proposition}[The unconditional distribution of $\mb{Y}_L$]
    \label{prop:marginal_Y} Consider $\{\mb{Y}_L(s)\}_{s\in \mathbb{R}^d}$ simulated as in Section~\ref{sec:simulation_methodology} with a measurement error with variance $\mb{d}$. Let $[\mb{\Sigma}]_{ij}=\mb{C}(s_i,s_j)\in\mathbb{R}^{p\times p}$ and $\mathbbm{1}(\cdot)$ be the indicator function.
    Then, 
    \begin{enumerate}
        \item[(a)] For each $L \in \{1, 2, \dots\}$, the unconditional moments after averaging over $X_\ell,\psi_\ell,\mb{W}_\ell$ are $\mathbb{E}[\mb{Y}_L(s)] = \mb{0}_p$ and $\operatorname{Cov}(\mb{Y}_L(s_i),\mb{Y}_L(s_j); \mb{\theta})=\mb{C}(s_i,s_j)+\mb{d}\mathbbm{1}\{i=j\}$.
        \item[(b)] For fixed finite collections $s_1,\dots,s_n$, the stacked vector $\mb{Y}_L=(\mb{Y}_L(s_1)^\top,\ldots,\mb{Y}_L(s_n)^\top)^\top$ converges in distribution to $\mathcal{N}(\mb{0},\mb{\Sigma}+\mb{D})$ as $L\to\infty$.
    \end{enumerate}
\end{proposition}
Proposition~\ref{prop:marginal_Y} can be proved using the laws of total expectation and covariance and the multivariate central limit theorem. It demonstrates that our model correctly recovers the multivariate Gaussian distribution with the target covariance $\mb{C}(\cdot, \cdot;\mb{\theta})$ asymptotically.

We next establish the convergence results for the algorithm.
Let $\suffstat^{(m)}$ be the stochastic-approximation iterate of the complete-data sufficient statistic for $\mb{\theta}$ at the $m$-th iteration, and $\hat{\theta}(\suffstat^{(m)})$ denote the fixed point of the CM cycle with $\suffstat$ held fixed. 
To simplify the notation, let $\ellg(\mb{\theta}):= \mathcal{L}(\mb{\theta} \mid \mb{Y})$ denote the marginal random feature log-likelihood under the proposal distribution $g$, and $\ellgP(\mb{\theta}):=\ellg(\mb\theta)-P(\mb\theta)$. 
We denote the stationary set of $\ellgP(\mb{\theta})$ as $\stationary=\bigl\{\mb{\theta}^\star\in\Theta : 0\in\sd^\circ\ellgP(\mb{\theta}^\star)\bigr\}$, where $\partial^\circ$ is the Clarke subdifferential \citep{clarke1990optimization}. For a point $x$ and a nonempty set $A$ in a finite-dimensional Euclidean space, write
$\dist(x,A)=\inf_{a\in A}\|x-a\|$. Matrix components are vectorized when evaluating the norm.

\begin{proposition}[Fixed-$L$ convergence with a Gibbs E step]
\label{prop:set}
Fix $n,L$, and the realized random features.  Under
\cref{ass:model,ass:gibbs,ass:steps,ass:stability,%
ass:contraction,ass:stratum,ass:curvature}, the SAECM algorithm satisfies, almost surely, 
\begin{align}
  \dist\!\left(
    \hat{\mb\theta}(\suffstat^{(m)}),
    \stationary
  \right)
  &\longrightarrow0,
  \label{eq:result-hat}\\
  \dist(\mb\theta^{(m)},\stationary)
  &\longrightarrow 0.
  \label{eq:result-theta}
\end{align}
\end{proposition}
\Cref{prop:set} shows that, for fixed $n$ and $L$, the parameter estimates from our SAECM algorithm with a Gibbs E step approach the stationary set of the random-feature-approximated penalized likelihood as the iterations increase, providing a convergence guarantee for the computational procedure. This is similar to the result of \citet{kuhn_coupling_2004}, who further note that, under additional conditions on the stochastic perturbations, convergence to local minima or saddle points can be avoided.

Next, we present a convergence result for the algorithm based on the random feature approximation to the target covariance function. Let $\ell$ denote the exact GP log-likelihood and $\Theta_P^\star=\{\mb{\theta}\in\Theta: \nabla\ell(\mb{\theta})\in\partial P(\mb{\theta})\}$
its penalized stationary set.

\begin{proposition}[Convergence to the stationary set of the exact penalized GP likelihood]
\label{prop:diagonal_convergence}
Fix $n$, $p$. Suppose \cref{ass:rff,ass:common-compact} hold. Suppose further that, almost surely with respect to the random features, \cref{ass:model,ass:gibbs,ass:steps,ass:stability,ass:contraction,ass:stratum,ass:curvature}
hold for all sufficiently large $L$. Then there exist deterministic sequences $L_k\longrightarrow\infty$ and $m_k\longrightarrow\infty$ such that, almost surely,
\begin{align}
  \dist\left(
    \hat{\mb\theta}_{L_k}(\suffstat_{L_k}^{(m_k)}),
    \Lstar
  \right)
  &\longrightarrow0,
  \label{eq:proposition-conclusion}\\
  \dist\left(
    \mb\theta_{L_k}^{(m_k)},
    \Lstar
  \right)
  &\longrightarrow0.
  \label{eq:actual-proposition-conclusion}
\end{align}
\end{proposition}

To extend \Cref{prop:set} from fixed $L$ to $L\to\infty$, additional regularity and containment assumptions are required (\cref{ass:rff,ass:common-compact}). Then, uniform score convergence can be applied to establish convergence to the stationary set along suitably chosen sequences $L_k\to\infty$ and $m_k\to\infty$, thereby proving \Cref{prop:diagonal_convergence}. It shows that, along these sequences, not only does the covariance approximation converge, but the parameter estimates also approach the stationary set of the exact penalized log-likelihood with the target covariance.

\section{Numerical Experiments}

We consider a simulation design consisting of $p=50$ processes, with a sparse data-generating  $\mb{\sigma}^{-1}$. The data are simulated using the algorithm proposed in Supplementary Section~\ref{app:ex_sim} with $L=100{,}000$, which allows processes from different covariance families, with each marginal process represented by a potentially different covariance function. In particular, we consider three families: Mat\'ern, squared-exponential (SE), and confluent hypergeometric (CH).

The Mat\'ern model \citep{matern1960} has covariance function and spectral density:
\begin{align*}
C_{\mathcal M}(h;\nu,\phi,\sigma)
    &= \sigma\frac{2^{1-\nu}}{\Gamma(\nu)}
       \left(\frac{\lVert h\rVert}{\phi}\right)^\nu
       K_\nu\left(\frac{\lVert h\rVert}{\phi}\right);
&f_{\mathcal M}(x;\nu,\phi,\sigma)
    &= \sigma
       \frac{\Gamma\left(\nu+\frac{d}{2}\right)}
       {\pi^{\frac{d}{2}}\Gamma(\nu)\phi^{2\nu}}
       \left(\phi^{-2}+\lVert x\rVert^2\right)^{-\nu-\frac{d}{2}},
\end{align*}
where $K_\nu(\cdot)$ is the modified Bessel function of the second kind, $\nu$ is a smoothness parameter, $\phi$ is a range parameter, and $\sigma$ is a variance parameter. The SE covariance function can be derived from a Mat\'ern model by letting $\nu\to\infty$ with appropriate range rescaling. Its covariance function and spectral density are 
\begin{align*}
C_{\mathcal G}(h;\phi,\sigma)
    &= \sigma\exp\left(-\frac{\lVert h\rVert^2}{2\phi^2}\right);
&f_{\mathcal G}(x;\phi,\sigma)
    &= \sigma\left(\frac{\phi}{\sqrt{2\pi}}\right)^d
       \exp\left(-\frac{\phi^2\lVert x\rVert^2}{2}\right).
\end{align*}
The CH model \citep{ma2022beyond,yarger2025multivariate} additionally allows flexible control of covariance tail decay. Its covariance function and spectral density are
\begin{align*}
C_{\mathcal{CH}}(h;\nu,\alpha,\beta,\sigma)
    &= \sigma\frac{\Gamma(\nu+\alpha)}{\Gamma(\nu)}
       \mathcal U\left(
           \alpha,1-\nu,\frac{\lVert h\rVert^2}{2\beta^2}
       \right), \\
f_{\mathcal{CH}}(x;\nu,\alpha,\beta,\sigma)
    &= \sigma
       \frac{\Gamma\left(\nu+\frac{d}{2}\right)\beta^d}
       {(2\pi)^{d/2}B(\alpha,\nu)}
       \mathcal U\left(
           \nu+\frac{d}{2},
           1-\alpha+\frac{d}{2},
           \frac{\beta^2\lVert x\rVert^2}{2}
       \right);\; \alpha > \frac{d}{2},
\end{align*}
where $B(a,b)$ is the beta function and $\mathcal U(a,b,z)$ is the confluent hypergeometric function of the second kind. The CH covariance combines Mat\'ern-like smoothness control with flexible polynomial tail behavior that can model either short or long range dependence. The parameter $\nu$ controls smoothness, $\beta$ plays a role similar to that of $\phi$ in the Mat\'ern covariance, and $\alpha$ controls the covariance tail decay and the degree of long range dependence. 

Throughout the simulations, we use a mixed-family setting in which $20$ of the $50$ processes are randomly assigned to the Mat\'ern covariance family, $20$ to the CH covariance family, and $10$ to the SE covariance family. The parameters are randomly drawn from uniform distributions. For the Mat\'ern and SE families, we use the inverse-range parameterization $a=1/\phi$. The Mat\'ern models have $a\sim\mathrm{U}[2.5,3.5]$ and $\nu\sim\mathrm{U}[1.5,2.5]$. The CH models have $\alpha\sim\mathrm{U}[1.25,1.75]$, $\beta\sim\mathrm{U}[0.6,0.9]$, and $\nu\sim\mathrm{U}[1.5,2.5]$, whereas the SE models have $a\sim\mathrm{U}[1.75,2.25]$. There are $n_o=500$ observed locations and $n_t=125$ test locations, resulting in $25{,}000$ values. We take $d=2$, with locations sampled uniformly over $[0,5]\times[0,5]$.

We assume a randomly generated precision structure among the processes. The diagonal entries are set to $(\mb{\sigma}^{-1})_{jj}=2$. For the off-diagonal entries, we consider two separate scenarios: a positive off-diagonal setting, in which $39$ randomly selected upper-triangular entries are set to $0.8$, and a negative off-diagonal setting, in which $48$ randomly selected upper-triangular entries are set to $-0.8$. All remaining off-diagonal entries are set to $0$, and the corresponding lower-triangular entries are determined by symmetry (see Figure~\ref{fig:precision_random}). These settings yield variance ranges of $[0.5,4.34]$ for the positive off-diagonal case and $[0.5,3.41]$ for the negative off-diagonal case. The nugget variance is set to $0.1$ for all processes and is assumed to remain identical across processes during parameter estimation, while the process-specific covariance parameters are estimated separately. Results for a tridiagonal precision matrix are provided in Supplementary Section~\ref{app:sim}.

Previous literature suggests that, for several multivariate approaches, modeling processes independently offers predictive performance similar to cokriging when data are colocated \citep{zhang2015doesn, lim2022conditions}. Because observations of a target process at training locations already provide substantial information, colocated data from correlated processes may offer limited or no improvement to the prediction. Consequently, we also consider a \emph{misaligned} setting where a randomly chosen $40\%$ of the observed locations for each process are missing. In this setting, observations from correlated processes at locations where the target process is unobserved can provide additional predictive information, highlighting the benefit of multivariate modeling over independent process estimation. We evaluate these approaches based on parameter estimation, the predictive performance of individual and summed processes, and the associated uncertainty quantification.

We evaluate our method using both the graphical LASSO without a penalty (Dense) and the graphical LASSO with a penalty selected using the BIC (Penalized). We compare our method with kriging under the assumption that the processes are independent (Ind), the latent model in the inside-out cross-covariance approach (IOX, \citealp{peruzzi2024}), and the linear model of coregionalization (LMC). For IOX, we consider both colocated and misaligned settings using Mat\'ern marginals because the implementation does not support fitting CH marginals, although the IOX construction itself allows different marginal covariance families.

For LMC, full-rank deterministic maximum likelihood estimation is challenging when $p$ is large, and the covariance family of each latent process must be determined because we model processes using multiple covariance families. 
Therefore, given a predetermined number of latent processes, $q$, we construct the loading matrix from a rank-$q$ approximation to the sample covariance matrix and determine the covariance function of each latent process through AIC. Then the spatial parameters and the nugget are estimated through maximum likelihood estimation. 
We set the number of latent processes to $q=10$, which explains around 70\% of the variance in the population covariance matrix, while increasing $q$ to $15$ adds only 6--7 percentage points at greater computational cost.
We omit the basis graphical LASSO approach proposed by \citet{mitchell_l_krock_modeling_2023} from our prediction comparison because the paper does not develop a procedure for prediction at new locations. Their application uses an empirical orthogonal basis defined at the observed locations, and extension of this basis to new locations is neither discussed in the paper nor implemented in the supplied example code. 
Further implementation details of the benchmark methods are given in Supplementary Section~\ref{app:benchmark}.
All results use AMD EPYC 7662 ``Rome'' processors with 128 cores and 256 GB of memory per node. Computer code is available at: \url{https://github.com/bychen713/SSGGP/}.

\subsection{Estimation Results}
\begin{figure}[!ht]\centering\tiny
\begin{tabular}{ccccc}
& \hspace*{-0.8em}\textbf{Correlation} & \hspace*{-0.8em}\textbf{Precision} & \hspace*{-0.8em}\textbf{Penalized} & \hspace*{-0.8em}\textbf{Dense}\\
\simrowlabel{Positive colocated} &
\includegraphics[width=.21\linewidth]{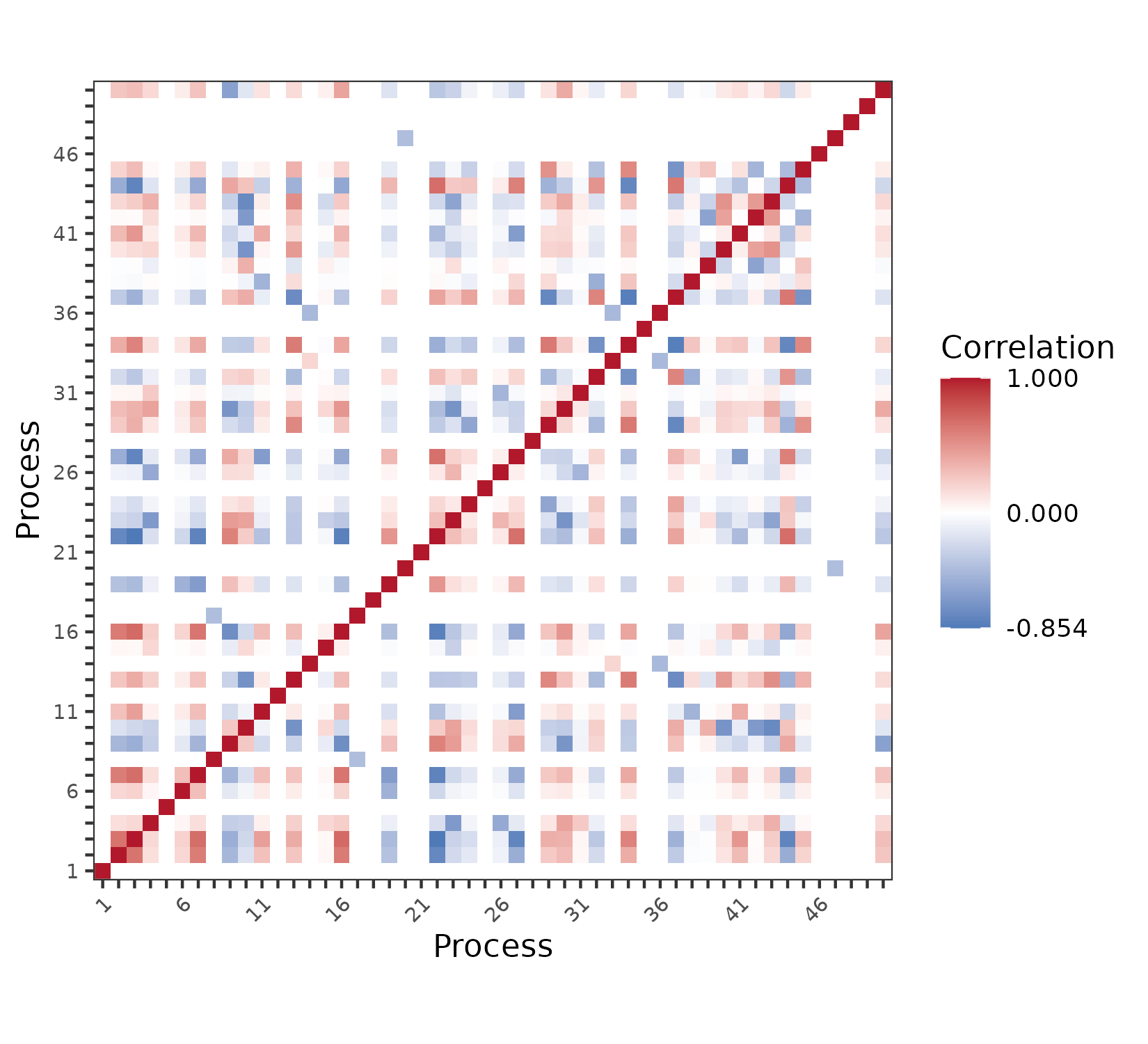} &
\includegraphics[width=.21\linewidth]{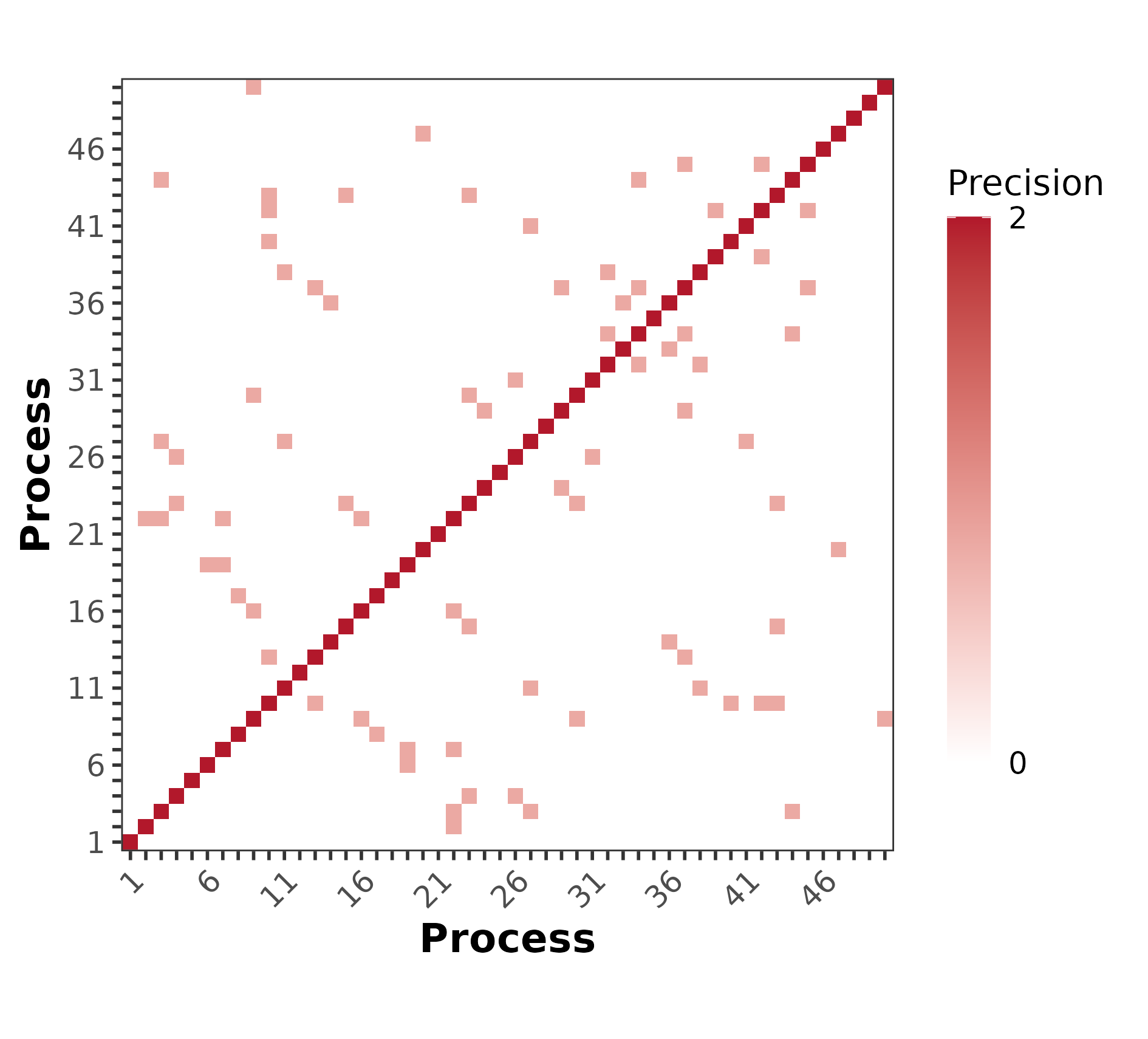} &
\includegraphics[width=.21\linewidth]{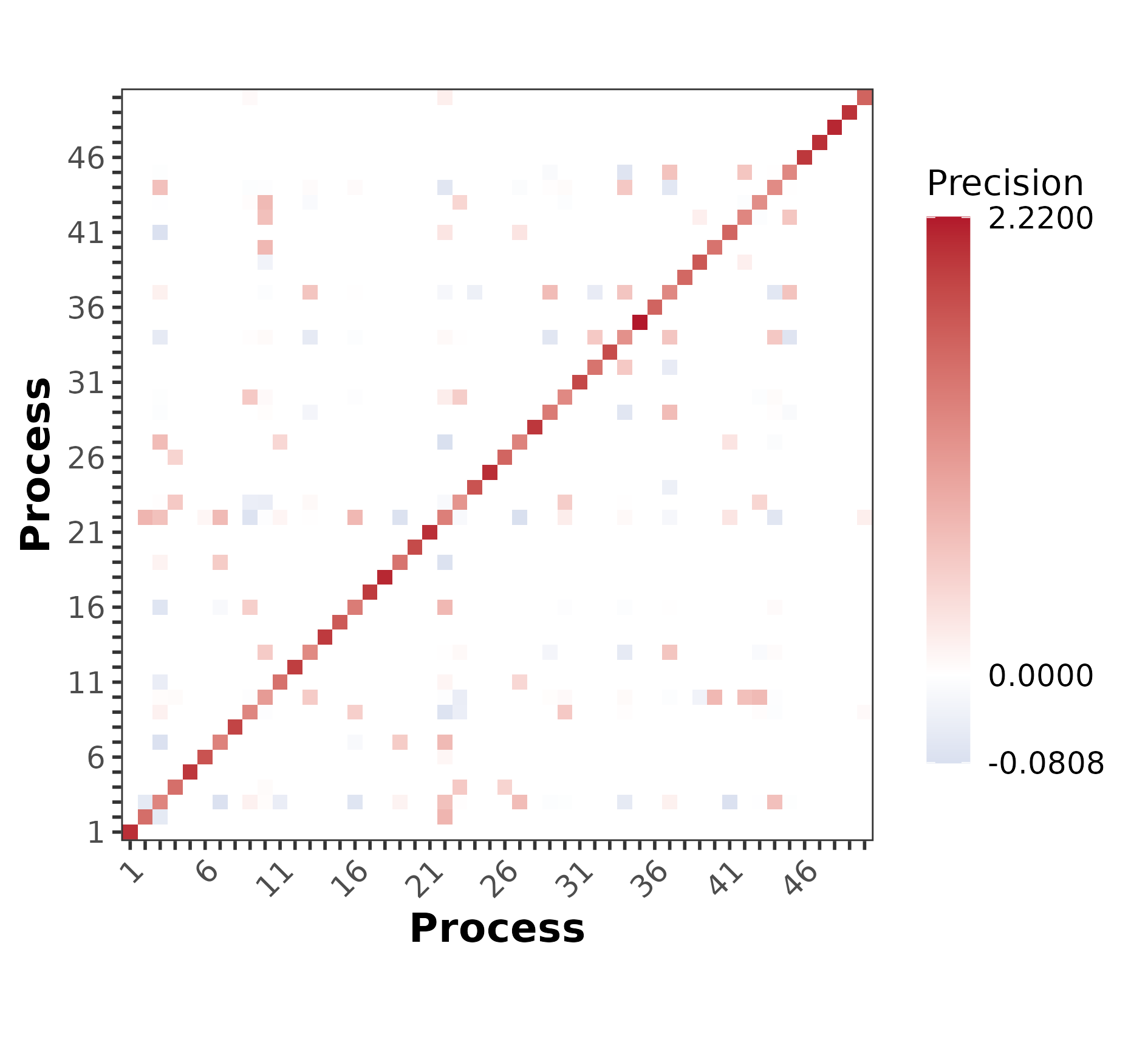} &
\includegraphics[width=.21\linewidth]{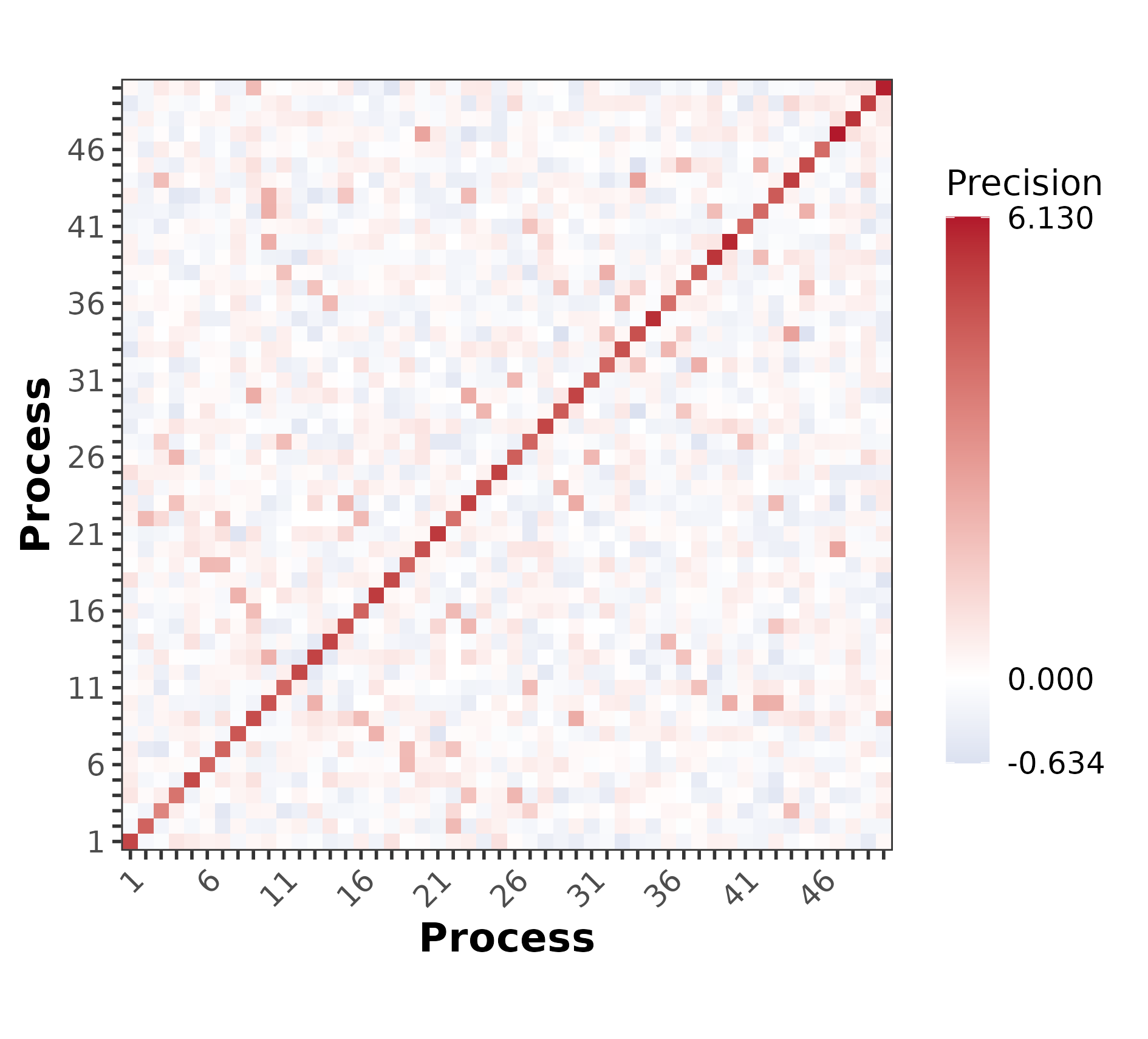} \\
\simrowlabel{Negative colocated} &
\includegraphics[width=.21\linewidth]{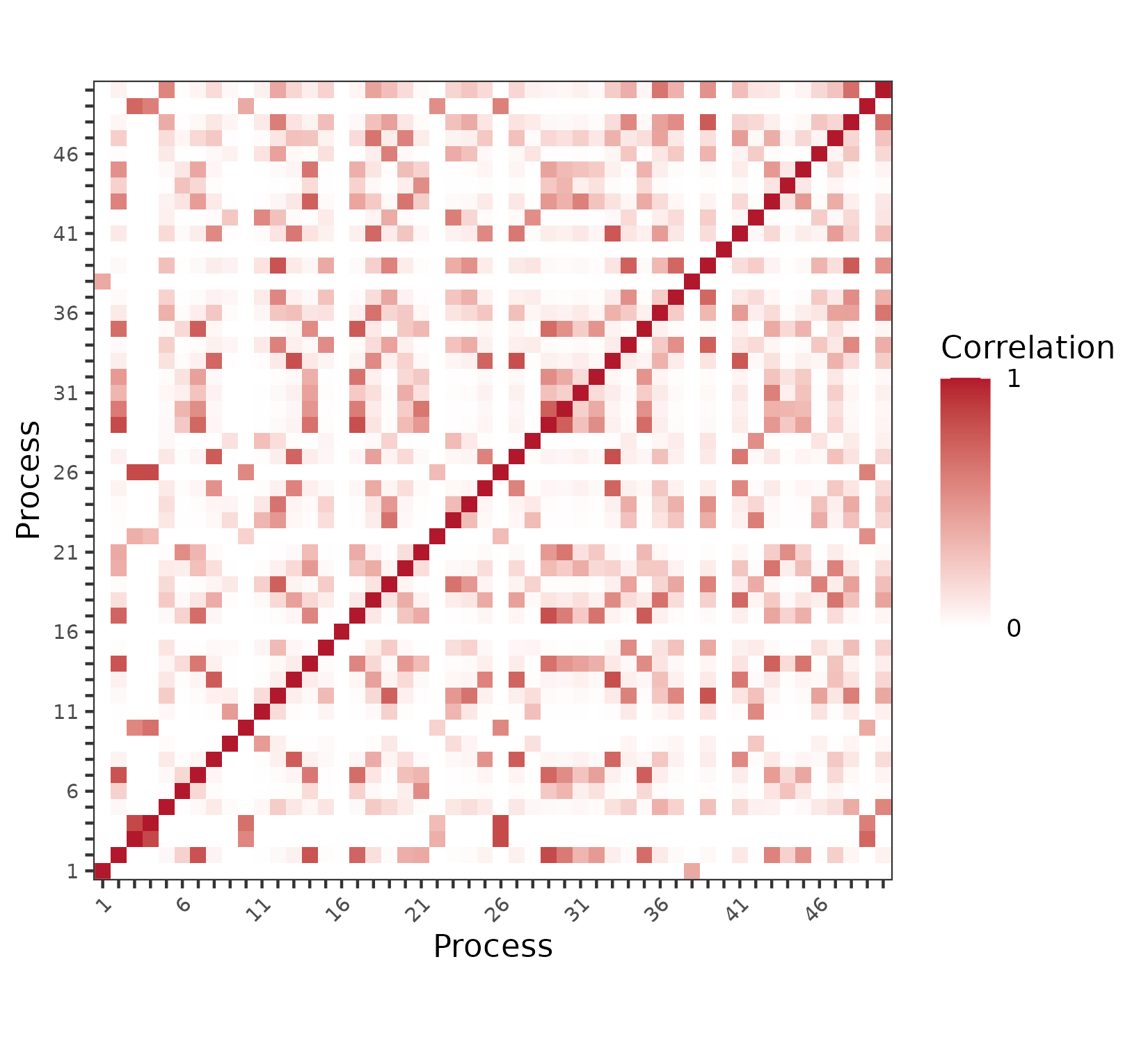} &
\includegraphics[width=.21\linewidth]{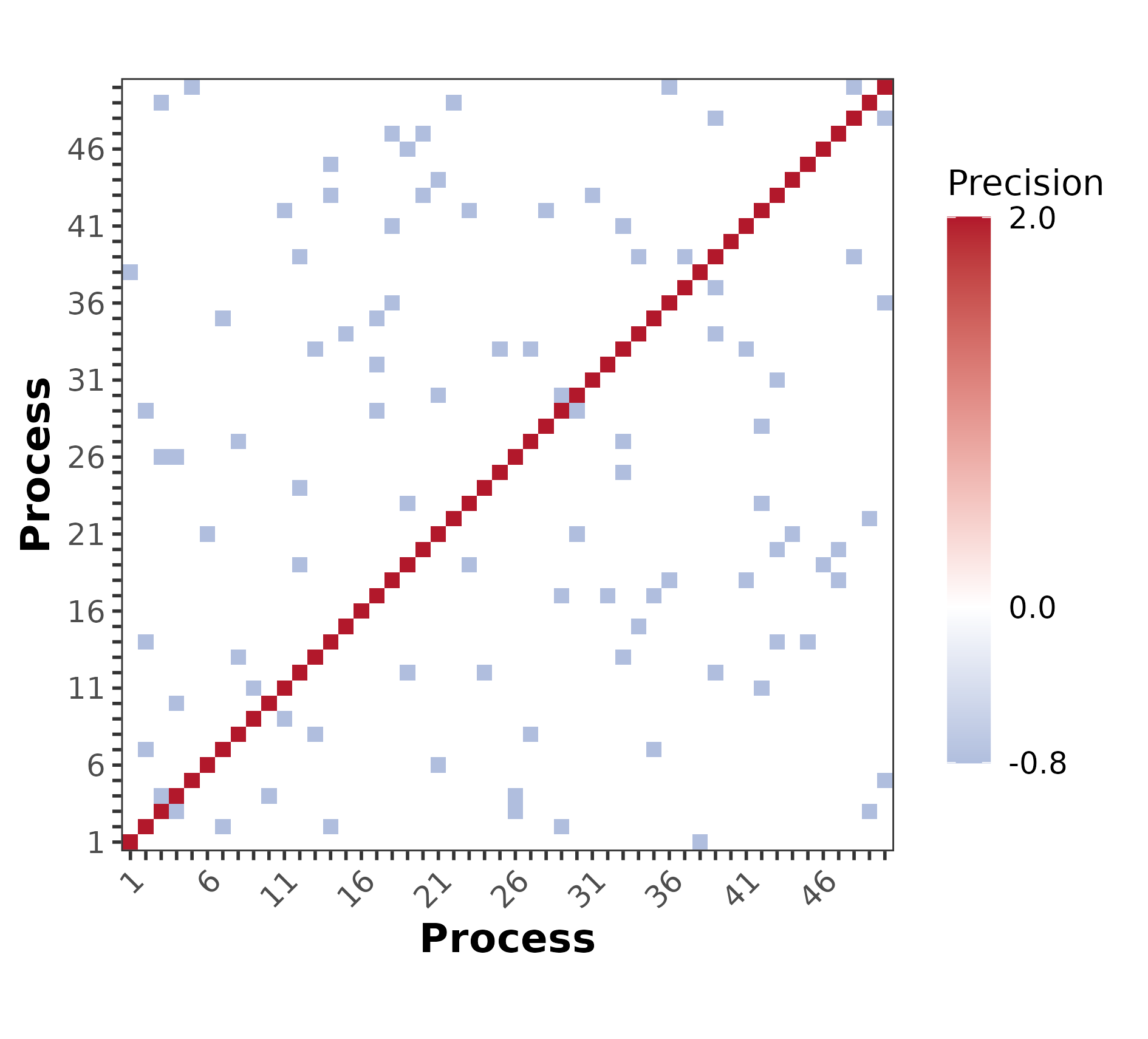} &
\includegraphics[width=.21\linewidth]{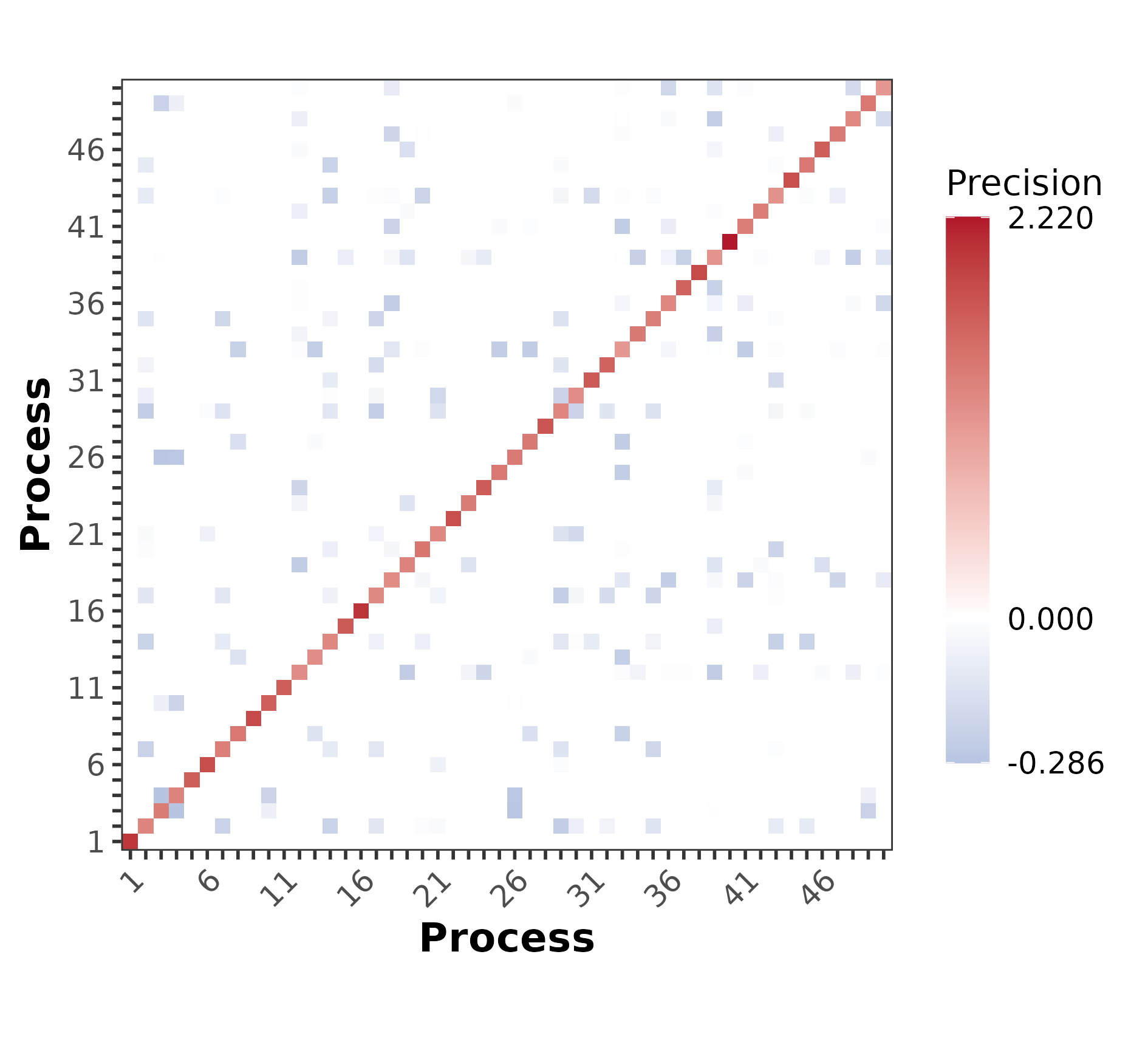} &
\includegraphics[width=.21\linewidth]{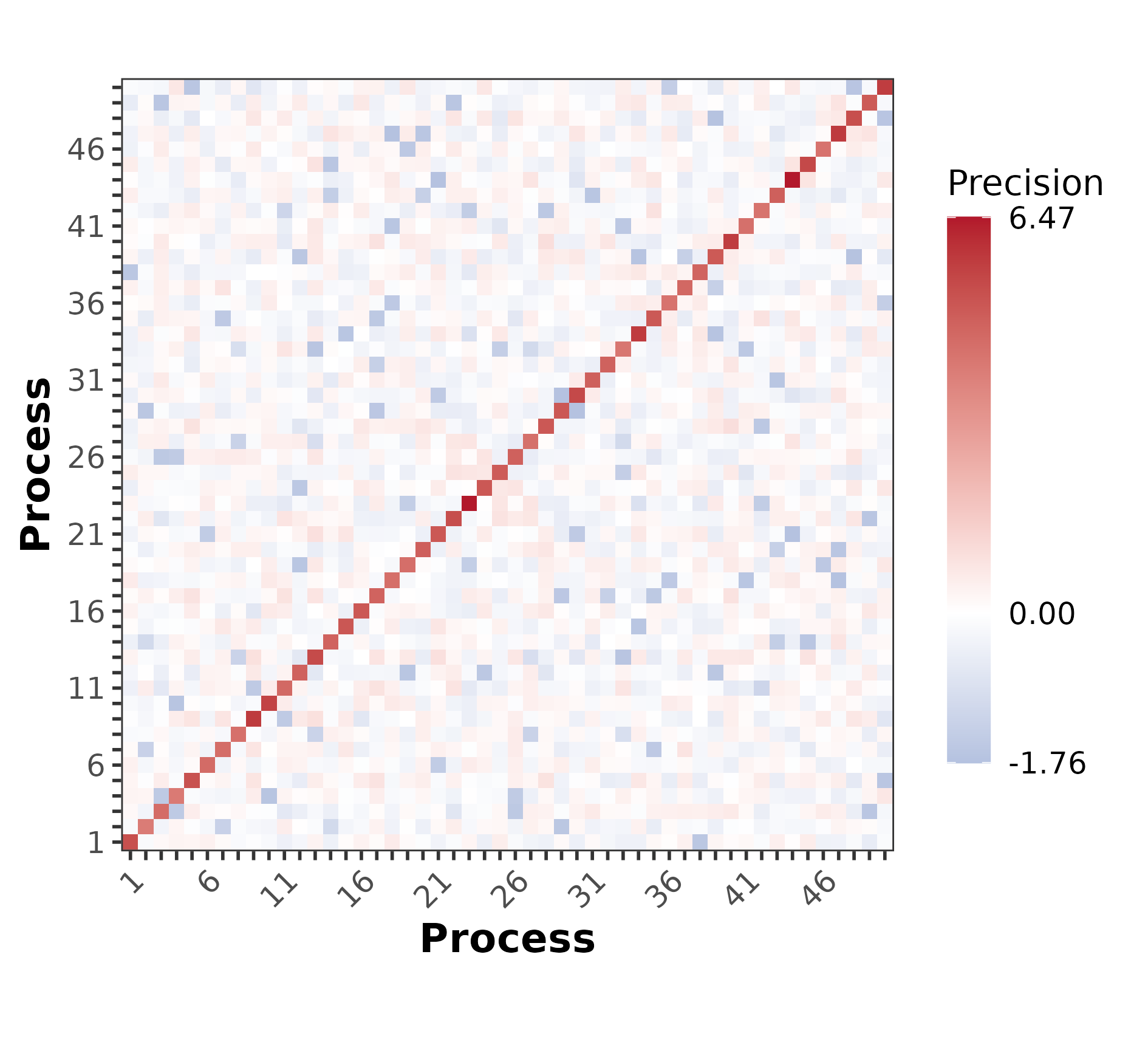}

\end{tabular}
\caption{True correlation and precision matrices and entrywise median precision estimates across 15 replicates for colocated simulations.}
\label{fig:precision_random}
\end{figure}

We present the estimation results in Table~\ref{table:estimation_sim_rand} and Figure~\ref{fig:precision_random}. Our method produces lower bias than the Ind approach for most parameters. Figure~\ref{fig:precision_random} displays the elementwise median of the precision-matrix estimates from the proposed method across the 15 replicates. Both the Dense and Penalized approaches capture the nonzero patterns in the true precision matrix. The results for the misaligned observations are qualitatively similar, although they generally exhibit larger biases and standard deviations (Supplementary Table~\ref{table:estimate_random_misaligned}).

Complete timing results for all settings are provided in Supplementary Table~\ref{table:time_full}. 
SAECM requires a sufficiently large $L$ to approximate the covariance functions well when $p$ is large, and the E step requires repeated $O(p^3)$ computations. 
Furthermore, CM3 requires sequential optimization of the spatial parameters, so both the Penalized and Dense fits take approximately 2--3.5 hours, with  
the Penalized fits slightly faster.

\begin{table}[!ht]
\centering
\small
\setlength{\tabcolsep}{3.5pt}
\renewcommand{\arraystretch}{0.8}
\caption{Parameter estimation for colocated observations. Entries are bias (SD) across 15 replicates, averaged over the processes in each family. Bold indicates the smallest absolute bias.}
\label{table:estimation_sim_rand}
\resizebox{\textwidth}{!}{%
\begin{tabular}{ccc|ccc|ccc}\toprule
&&&\multicolumn{3}{c|}{Positive off-diagonals}&\multicolumn{3}{c}{Negative off-diagonals}\\
Family & Par. & Truth range & Penalized & Dense & Ind & Penalized & Dense & Ind\\\midrule
CH & $\alpha$ & [1.27, 1.73] & \textbf{1.28 (3.25)} & 2.67 (4.78) & 3.30 (4.06) & \textbf{1.04 (3.38)} & 2.51 (4.56) & 3.39 (3.97)
 \\
CH & $\beta$ & [0.60, 0.87] & 0.41 (0.73) & \textbf{0.21 (0.64)} & 0.52 (0.92) & 0.49 (0.76) & \textbf{0.22 (0.59)} & 0.60 (0.95)
 \\
CH & $\nu$ & [1.50, 2.50] & \textbf{0.45 (1.21)} & 0.70 (1.22) & 0.57 (1.31) & \textbf{0.14 (0.99)} & 0.51 (1.09) & 0.39 (1.20)
 \\
CH & $d$ & 0.10 & \textbf{0.00 (0.00)} & 0.00 (0.00) & 0.00 (0.01) & \textbf{0.00 (0.00)} & 0.00 (0.00) & 0.00 (0.01)
 \\
SE & $a$ & [1.76, 2.16] & 0.02 (0.11) & 0.02 (0.11) & \textbf{0.01 (0.11)} & 0.01 (0.11) & 0.01 (0.11) & \textbf{0.00 (0.11)}
 \\
SE & $d$ & 0.10 & 0.00 (0.00) & 0.00 (0.00) & \textbf{0.00 (0.01)} & \textbf{0.00 (0.00)} & 0.00 (0.00) & 0.00 (0.01)
 \\
Mat\'ern & $a$ & [2.52, 3.48] & \textbf{0.24 (1.39)} & 0.57 (1.26) & 0.47 (1.39) & \textbf{-0.16 (1.17)} & 0.49 (1.05) & 0.43 (1.21)
 \\
Mat\'ern & $\nu$ & [1.50, 2.50] & \textbf{0.57 (1.24)} & 0.62 (1.19) & 0.57 (1.28) & \textbf{0.24 (0.89)} & 0.41 (0.87) & 0.38 (0.96)
 \\
Mat\'ern & $d$ & 0.10 & 0.00 (0.00) & 0.00 (0.00) & \textbf{0.00 (0.01)} & \textbf{0.00 (0.00)} & 0.00 (0.00) & 0.00 (0.01)
 \\
\bottomrule\end{tabular}
}
\end{table}

\subsection{Prediction of  Colocated Observations}

We evaluate the $\ell_2$ norm of the residuals, the root mean squared error (RMSE) of the sum of processes, and the coverage of the predictive intervals for both individual processes and the sum of the processes across the 15 replicates. PI lengths refer to predictive interval lengths for individual
processes and their sum. We demonstrate in the simulation that cokriging outperforms the Ind approach in the prediction of the sums. We present these results in Table~\ref{table:prediction_sim_rand_coloc}. All metrics are computed over all test locations and, when calculated per process, averaged across all processes. Letting $n_t$ denote the number of test locations, the average $\ell_2$ norm of the residuals is calculated as:
\begin{equation*}
\text{Norm} = \frac{1}{n_t}\sum_{i=1}^{n_t}\sqrt{\sum_{j=1}^p (Y_{j}(s_i)-\hat{Y}_{j}(s_i))^2}.
\end{equation*}
For the sum of processes, we calculate the RMSE as:
\begin{equation*}
    \label{eq:msesum}
    \text{RMSE}_{\text{sum}} = \left[\frac{1}{n_t}\sum_{i=1}^{n_t}\left(\sum_{j=1}^p Y_{j}(s_i)-\sum_{j=1}^p\hat{Y}_{j}(s_i)\right)^2\right]^{\frac12}. 
\end{equation*}

\begin{table}[!ht]
\centering
\small
\setlength{\tabcolsep}{3.5pt}
\renewcommand{\arraystretch}{0.80}
\caption{Prediction summaries for colocated simulations. Entries are mean (SD) across 15 replicates. Pos/Neg denote positive/negative precision off-diagonals. Bold indicates coverage closest to 95\% and the lowest mean for other metrics.}
\label{table:prediction_sim_rand_coloc}
\begin{tabular}{llccccc}
\toprule
Setting & Metric & Penalized & Dense & Ind & LMC & IOX \\
\midrule
Pos & Norm
    & \textbf{2.66 (0.03)} & 2.67 (0.03) & 2.68 (0.03) & 4.29 (0.15) & 2.70 (0.03) \\
    & Coverage (\%)
    & 94.77 (0.29) & 94.59 (0.28) & 94.72 (0.27) & \textbf{94.91 (0.36)} & 96.33 (0.32) \\
    & Sum RMSE
    & \textbf{2.62 (0.16)} & 2.62 (0.14) & 2.64 (0.16) & 4.11 (0.36) & 2.65 (0.14) \\
    & Sum coverage (\%)
    & 95.52 (1.88) & \textbf{95.09 (1.89)} & 95.89 (1.76) & 94.61 (1.80) & 96.27 (1.88) \\
    & PI length
    & 1.47 (0.02) & \textbf{1.46 (0.02)} & 1.48 (0.02) & 2.39 (0.07) & 1.62 (0.03) \\
    & Sum PI length
    & 10.41 (0.14) & \textbf{10.26 (0.16)} & 10.53 (0.13) & 16.67 (0.53) & 11.14 (0.27) \\
\addlinespace
Neg & Norm
    & \textbf{2.74 (0.04)} & 2.74 (0.04) & 2.77 (0.03) & 4.44 (0.17) & 2.77 (0.03) \\
    & Coverage (\%)
    & 94.79 (0.25) & 94.66 (0.19) & 94.89 (0.26) & \textbf{95.10 (0.27)} & 96.41 (0.20) \\
    & Sum RMSE
    & 3.62 (0.32) & \textbf{3.59 (0.35)} & 3.81 (0.26) & 4.68 (0.43) & 3.69 (0.30) \\
    & Sum coverage (\%)
    & 92.32 (2.20) & \textbf{94.40 (2.42)} & 85.55 (2.92) & 96.64 (2.14) & 97.71 (1.57) \\
    & PI length
    & 1.51 (0.02) & \textbf{1.50 (0.02)} & 1.53 (0.02) & 2.50 (0.09) & 1.66 (0.02) \\
    & Sum PI length
    & 12.41 (0.36) & 13.42 (0.49) & \textbf{10.92 (0.15)} & 19.76 (0.69) & 16.54 (0.68) \\
\bottomrule
\end{tabular}
\end{table}

For individual-process prediction based on colocated observations, our approach outperforms all other methods under both precision matrix settings. The Penalized approach performs slightly better than the Dense approach on average, particularly in prediction norm,  
which could be attributed to the reduced number of false positive edges while retaining true edges. The Ind approach performs similarly as it works with the correctly specified marginal model. Our approach also demonstrates superior performance in the sum RMSE, as it estimates correlation between processes, unlike independent prediction.

The results for prediction uncertainty are also shown in Table~\ref{table:prediction_sim_rand_coloc}. The coverage rates for individual processes are similar across all methods and are all close to the nominal rate. We also report the average predictive interval lengths, with our approach generally providing shorter intervals and coverage close to the nominal level. 
The coverage rate for the sum under cokriging is closer to the nominal rate than that under the Ind approach, especially in the negative off-diagonal setting. 
The Penalized approach also slightly undercovers the sum in the negative off-diagonal setting, possibly because penalization attenuates estimated cross-process dependence.
This difference between cokriging and Ind can be explained by the variance decomposition, $\operatorname{Var}\left(\sum_{j=1}^p Y_j(s)\right) = \sum_{j=1}^p \operatorname{Var}(Y_j(s)) + 2\sum_{k < j} \operatorname{Cov}(Y_j(s), Y_k(s)).$
Because the Ind approach ignores the second term, its predictive intervals are narrower in the negative off-diagonal setting and wider in the positive off-diagonal setting. The pattern becomes clearer in the results of the tridiagonal precision setting in Supplementary Section~\ref{app:sim}. 

Low-rank approximations such as the LMC cannot achieve comparable performance because the true processes have a full-rank dependence structure. Its wide predictive intervals provide  good coverage results but are less meaningful given the significantly larger norm. IOX is implemented under a full-rank specification. However, the current implementation of IOX does not support the CH marginal covariance functions and is therefore misspecified in this setting, which may contribute to its slightly higher norm and RMSE. To separate the effect of this misspecification from the method’s performance, Supplementary Section~\ref{subsec:all_matern} presents a setting where every marginal process has a Mat\'ern covariance function. In both settings, IOX does not estimate the sparsity of the precision matrix and shows slight overcoverage.

\subsection{Prediction of  Misaligned Observations}
For the simulation of misaligned observations, removing $40\%$ of the observations leads to $300$ points per process, resulting in a total of $15{,}000$ observed values across all processes. 
Parameter estimates are obtained as described in Section \ref{subsec:implementation} and are similar to those obtained for the colocated observations (Supplementary Table~\ref{table:estimate_random_misaligned}).
The prediction results are reported in Table~\ref{table:prediction_sim_rand_mis}. 
Overall, our approach achieves the best performance in most measures. 
The improvement over the Ind approach is even greater here because our method can utilize information from correlated processes at locations where the target process is unobserved. 
While LMC has the best coverage rate for the sum of the processes, it has wider predictive intervals and the worst point-prediction performance. 
\begin{table}[!ht]
\centering
\small
\setlength{\tabcolsep}{3.5pt}
\renewcommand{\arraystretch}{0.80}
\caption{Prediction summaries for misaligned simulations. Entries are mean (SD) across 15 replicates. Pos/Neg denote positive/negative precision off-diagonals. Bold indicates coverage closest to 95\% and the lowest mean for other metrics.}
\label{table:prediction_sim_rand_mis}
\begin{tabular}{llccccc}
\toprule
Setting & Metric & Penalized & Dense & Ind & LMC & IOX \\
\midrule
Pos & Norm
    & 2.85 (0.04) & \textbf{2.85 (0.04)} & 2.90 (0.04) & 4.48 (0.15) & 2.94 (0.04) \\
    & Coverage (\%)
    & \textbf{94.87 (0.31)} & 94.23 (0.33) & 94.63 (0.31) & 94.63 (0.29) & 98.18 (0.24) \\
    & Sum RMSE
    & \textbf{2.82 (0.18)} & 2.82 (0.17) & 2.85 (0.19) & 4.25 (0.40) & 2.89 (0.19) \\
    & Sum coverage (\%)
    & 95.41 (2.31) & \textbf{95.04 (1.92)} & 95.68 (2.09) & \textbf{95.04 (2.52)} & 98.40 (1.45) \\
    & PI length
    & 1.58 (0.02) & \textbf{1.54 (0.02)} & 1.59 (0.02) & 2.47 (0.08) & 1.99 (0.04) \\
    & Sum PI length
    & 11.21 (0.16) & \textbf{10.80 (0.20)} & 11.38 (0.15) & 17.11 (0.53) & 13.54 (0.37) \\
\addlinespace
Neg & Norm
    & 2.97 (0.04) & \textbf{2.94 (0.04)} & 3.04 (0.04) & 4.65 (0.16) & 3.03 (0.05) \\
    & Coverage (\%)
    & \textbf{94.86 (0.23)} & 94.54 (0.25) & 94.74 (0.30) & 94.80 (0.37) & 98.15 (0.23) \\
    & Sum RMSE
    & 4.03 (0.32) & \textbf{3.85 (0.33)} & 4.36 (0.32) & 5.21 (0.41) & 4.11 (0.42) \\
    & Sum coverage (\%)
    & 88.96 (3.03) & 93.33 (3.34) & 83.63 (3.02) & \textbf{95.41 (2.68)} & 98.99 (1.53) \\
    & PI length
    & 1.64 (0.02) & \textbf{1.59 (0.02)} & 1.66 (0.02) & 2.58 (0.09) & 2.04 (0.03) \\
    & Sum PI length
    & 12.76 (0.25) & 14.09 (0.48) & \textbf{11.95 (0.16)} & 20.81 (0.93) & 21.38 (0.79) \\
\bottomrule
\end{tabular}
\end{table}
\subsection{Sparsity Recovery}\label{sec:sparse_rec}
\begin{figure}[!t]\centering\tiny
\includegraphics[width=.95\linewidth, height=6cm]{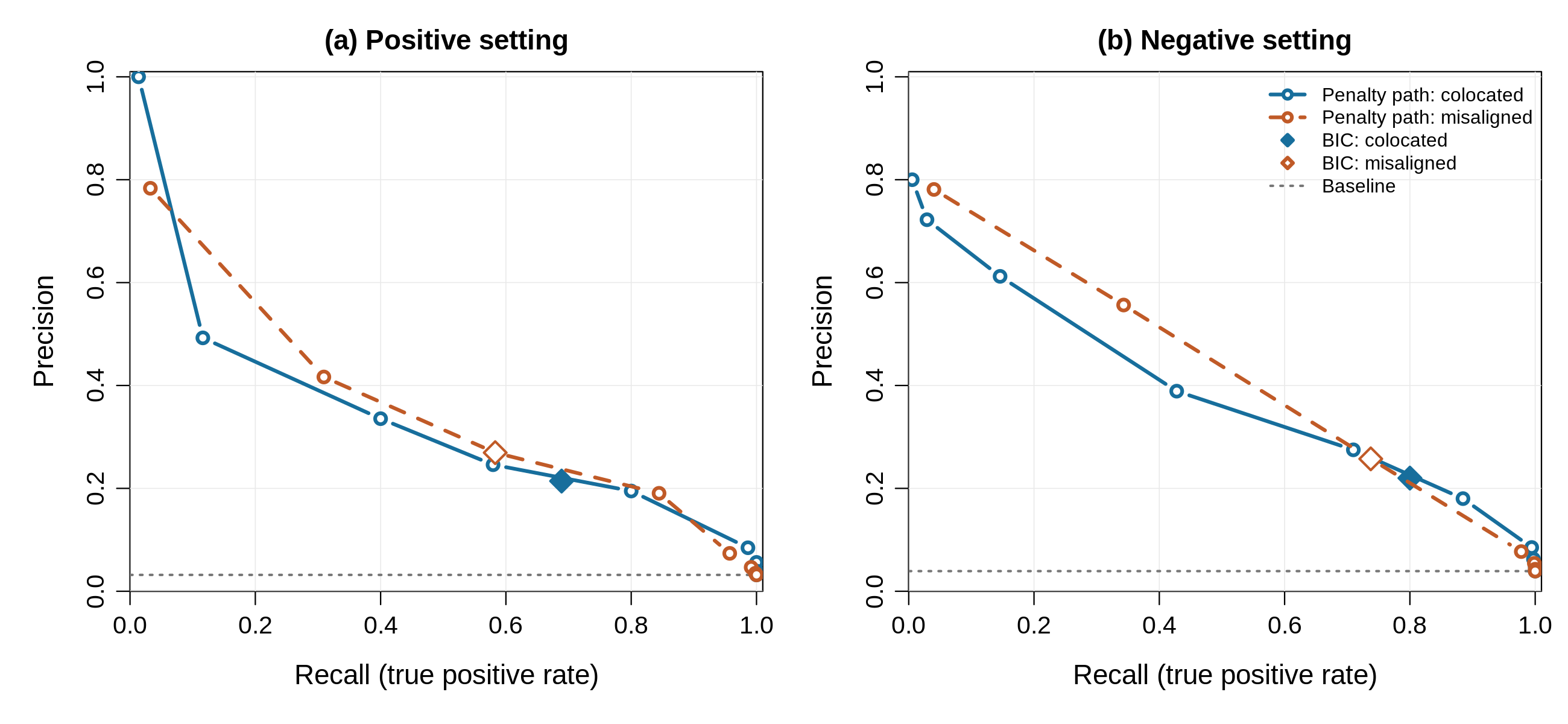} 
\caption{Precision--recall curves for precision-matrix edge recovery in colocated and misaligned simulations, averaged across 15 replicates.}
\label{fig:sparsity}
\end{figure}

Although our results have mainly focused on prediction and estimation performances so far, graph selection is a useful outcome of our methodology, and we evaluate the selection results using the precision-recall curves for the estimated precision matrix in Figure~\ref{fig:sparsity}. The Dense approach is a baseline that does not perform selection and thus has recall one. The Penalized approach retains a large proportion of the true edges and substantially reduces the false positive rate (FPR) relative to the Dense approach (Supplementary Table~\ref{table:sparsity}). However, the figure shows a trade-off between precision and recall across penalties. When the penalty is large, only a few off-diagonal entries with strong signals are selected, resulting in low recall and high precision. As the penalty decreases, more off-diagonal entries are selected, including false positives. The resulting loss of precision is exacerbated by the imbalance between true and absent edges, since only $39$ of the $1{,}225$ possible edges are nonzero in the positive off-diagonal sign setting, and only $48$ are nonzero in the negative off-diagonal sign setting. At the BIC-selected penalties, precision remains low, suggesting that likelihood-based selection may retain spurious edges alongside true dependencies that improve model fit. The LMC and IOX approaches do not directly perform selection, so their results are excluded.

\section{Analysis of MERRA-2 Data}

We consider the aerosol data set from the Modern-Era Retrospective Analysis for Research and Applications, Version 2 (MERRA-2) for our data analysis. MERRA-2 is a global atmospheric reanalysis developed by NASA's Global Modeling and Assimilation Office that incorporates aerosol modeling and assimilates aerosol optical depth observations, with data available from 1980 onward \citep{gelaro2017modern}. It combines satellite observations of aerosol optical depth with simulated aerosol transport and physical processes, providing output on a regular grid with a resolution of $0.625^{\circ}$ longitude by $0.5^{\circ}$ latitude. 

\begin{figure}[!ht]
  \centering
  \includegraphics[width=.98\textwidth, height=6cm]{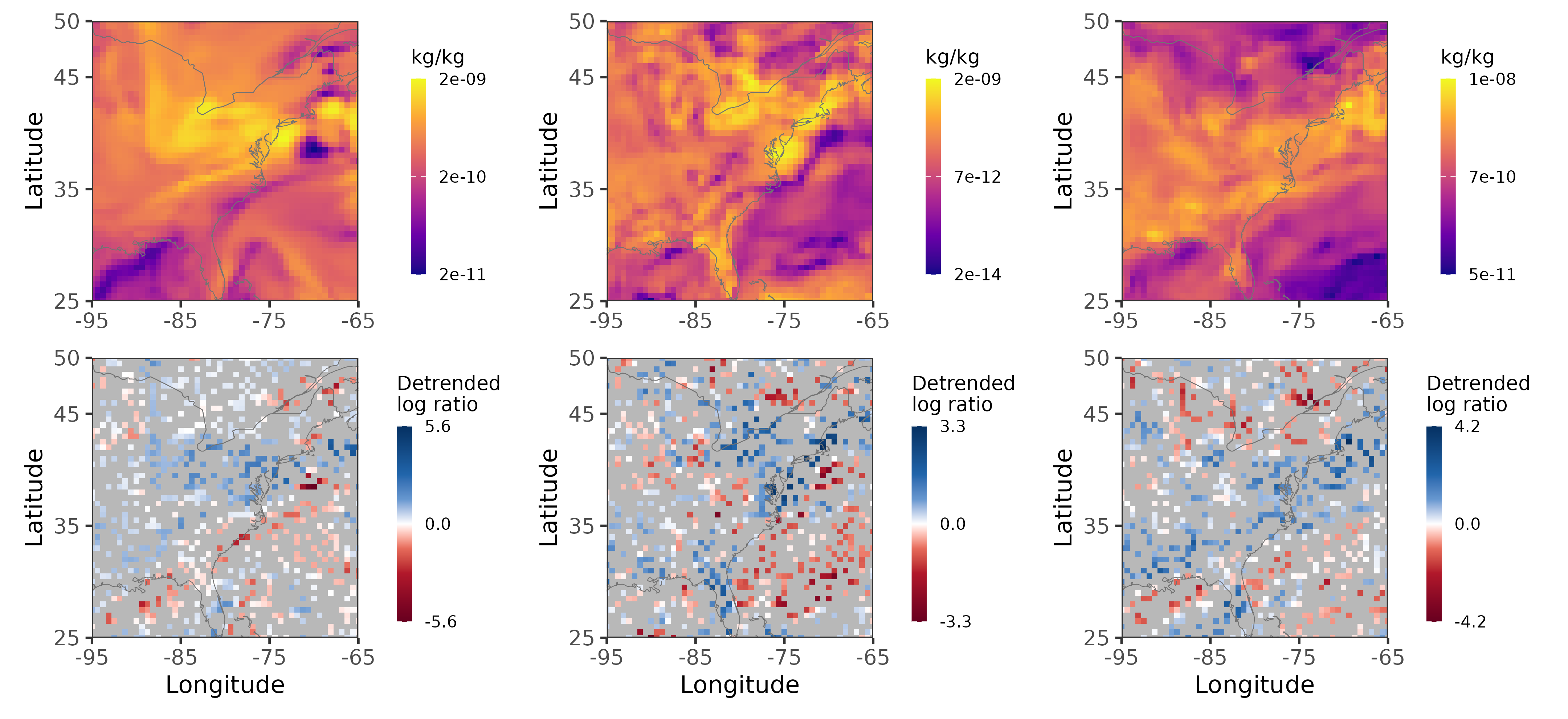}
\caption{MERRA-2 data for colocated fold 1. Left to right: dust \texttt{DU001} at 506 hPa, sea salt \texttt{SS005} at 843 hPa, and sulfate \texttt{SO4} at 843 hPa. Top: mixing ratios at all 2,499 original grid cells. Bottom: preprocessed values at the 700 sampled locations.}
\label{fig:merra-maps}
\end{figure}

We use the three-hourly instantaneous aerosol mixing-ratio product on native model levels, M2I3NVAER version 5.12.4 \citep{GMAO2015MERRA2}. Ten aerosol components, three dust sizes, two sea-salt sizes, hydrophobic and hydrophilic black carbon, hydrophobic and hydrophilic organic carbon, and sulfate, are selected from four vertical layers numbered 50, 52, 55, and 62, with regional-mean mid-layer pressures of approximately 506, 582, 688, and 843 hPa. We treat each component--layer combination as a distinct spatial process, resulting in $p=40$ processes. We focus on the region bounded by longitudes $95^{\circ}$W and $65^{\circ}$W and latitudes $25^{\circ}$N and $50^{\circ}$N at 00 UTC on 15 July 2022, yielding 2,499 grid cells with complete data. 
For each process and fold, we apply a logarithmic transformation using an offset equal to half its smallest positive observed value and subtract a spatially varying mean estimated from training observations in the fold by local linear regression with a bandwidth of 1,000 kilometers. 
We then divide the residuals by their process-specific standard deviations computed from the same observations.
The first, twentieth, and fortieth processes, corresponding to \texttt{DU001} at layer 50, \texttt{SS005} at layer 62, and \texttt{SO4} at layer 62, are plotted in Figure~\ref{fig:merra-maps}, with the resulting processed values shown below each map.

For prediction assessment, we construct three folds, each containing $n=500$ training locations and 200 non-overlapping held-out test locations. In the colocated design, all 40 processes are observed at each training location. In the misaligned design, we mask 40\% of the training responses separately for each process, leaving $300$ observed values per process. 
The nugget is estimated separately for each process, and other implementation details are described in Supplementary Section~\ref{app:data}.

Figure~\ref{fig:merra-colocated-precision} shows the entrywise median of precision matrices from the three data folds. There is a clear pattern of negative off-diagonal entries within the dust variables and among the black carbon and the organic carbon, especially at corresponding vertical levels. \texttt{DU001} and \texttt{DU003} represent the finest and intermediate-sized dust bins, respectively, and their negative precision entries suggest positive partial associations, possibly reflecting shared emission and transport. \texttt{DU005}, the coarsest dust bin, shows weaker associations with the other dust bins, possibly because larger particles settle faster \citep{Maring2003dust}. Several black and organic carbon variables also show positive partial associations, consistent with shared combustion sources \citep{Bond2004carbon,randles2017}.

The estimation results are shown in Supplementary Section~\ref{app:data} and Figure~\ref{fig:merra-colocated-precision}. In Figure~\ref{fig:merra-colocated-precision}, the Penalized estimate is on a different scale from the Dense estimate, but both capture the main precision structure among the dust and carbon variables.
\Cref{table:merra-prediction} shows the prediction results for all approaches. When the data are colocated, all approaches have similar individual-process coverage and sum RMSE, although LMC has a larger prediction norm. The Ind approach has substantially lower coverage for the sum of the processes. This is expected because several variables are positively correlated, and ignoring their dependence results in underestimation of the predictive variance of the sum. When the data are misaligned, IOX, Penalized, and Dense have similar prediction errors and outperform the Ind approach in individual-process prediction, with IOX providing sum coverage closest to 95\%. LMC has the largest prediction norm but a lower sum RMSE than Ind. Prediction norms increase under misalignment for all methods, whereas sum RMSE is essentially unchanged for Penalized and Dense.

\begin{figure}[!ht]\centering\small
  \begin{tabular}{@{}ccc@{}}
    \textbf{Sample precision} & \textbf{Penalized} & \textbf{Dense}\\
    \includegraphics[width=.3\linewidth]{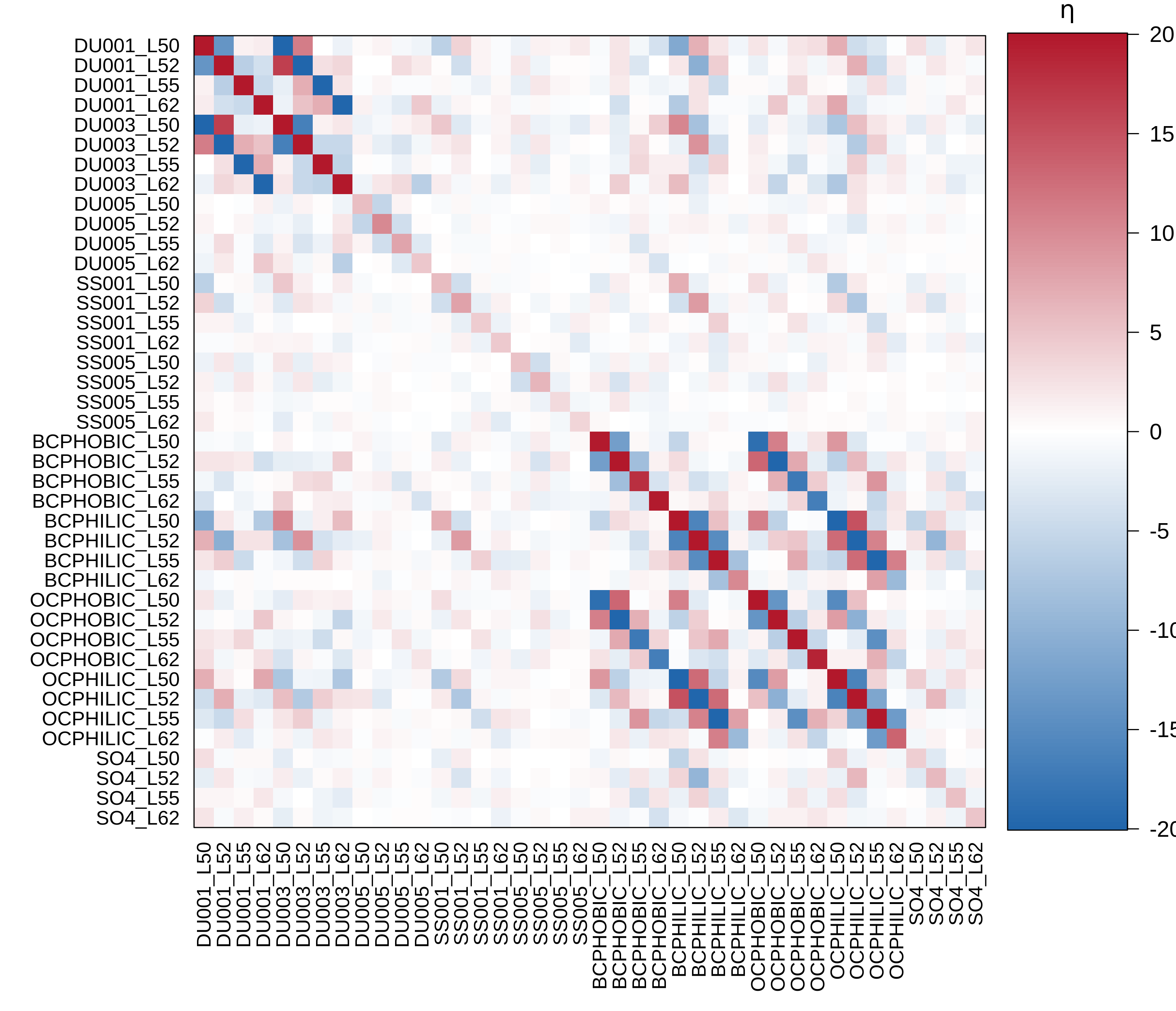} &
    \includegraphics[width=.3\linewidth]{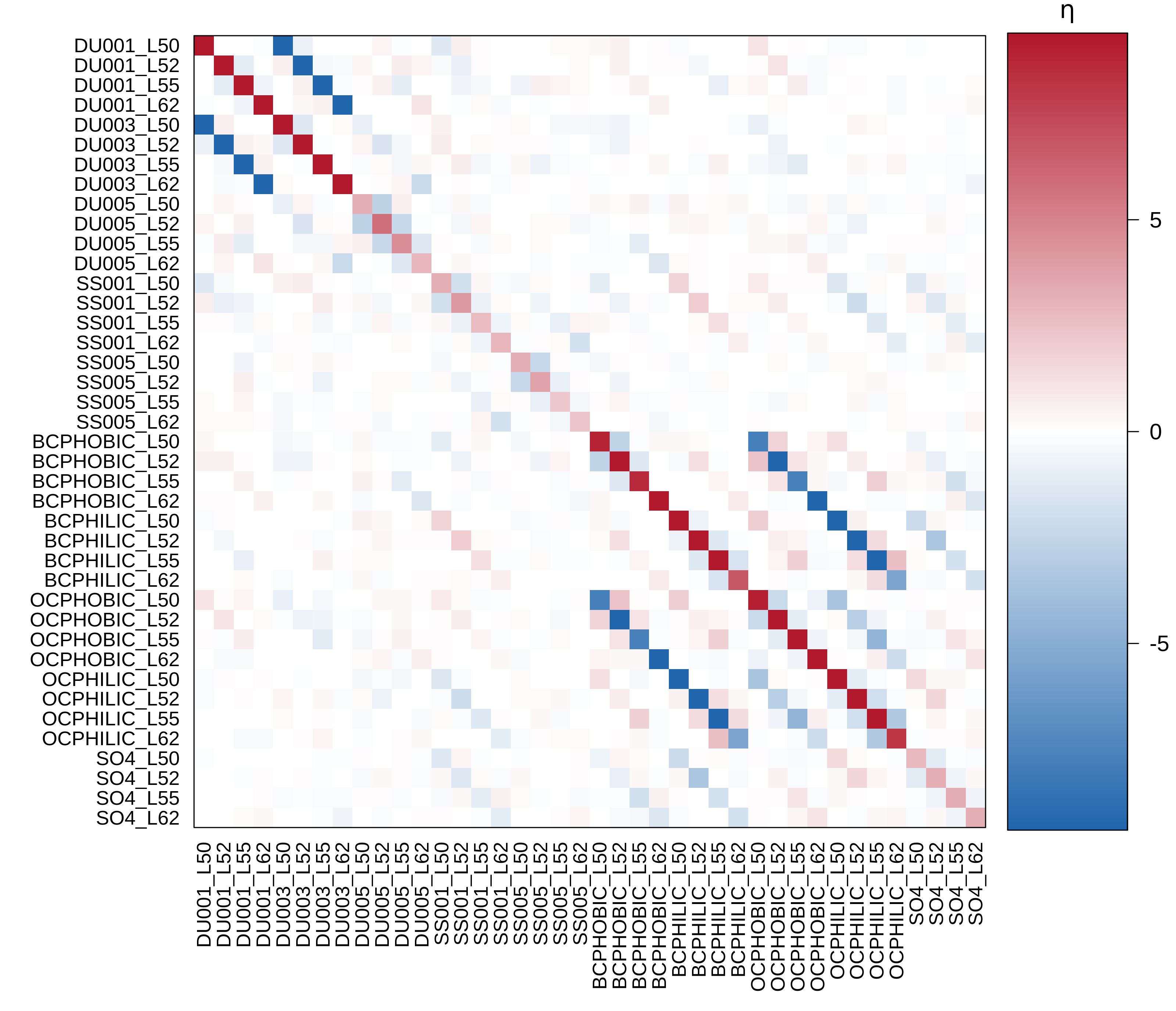} &
    \includegraphics[width=.3\linewidth]{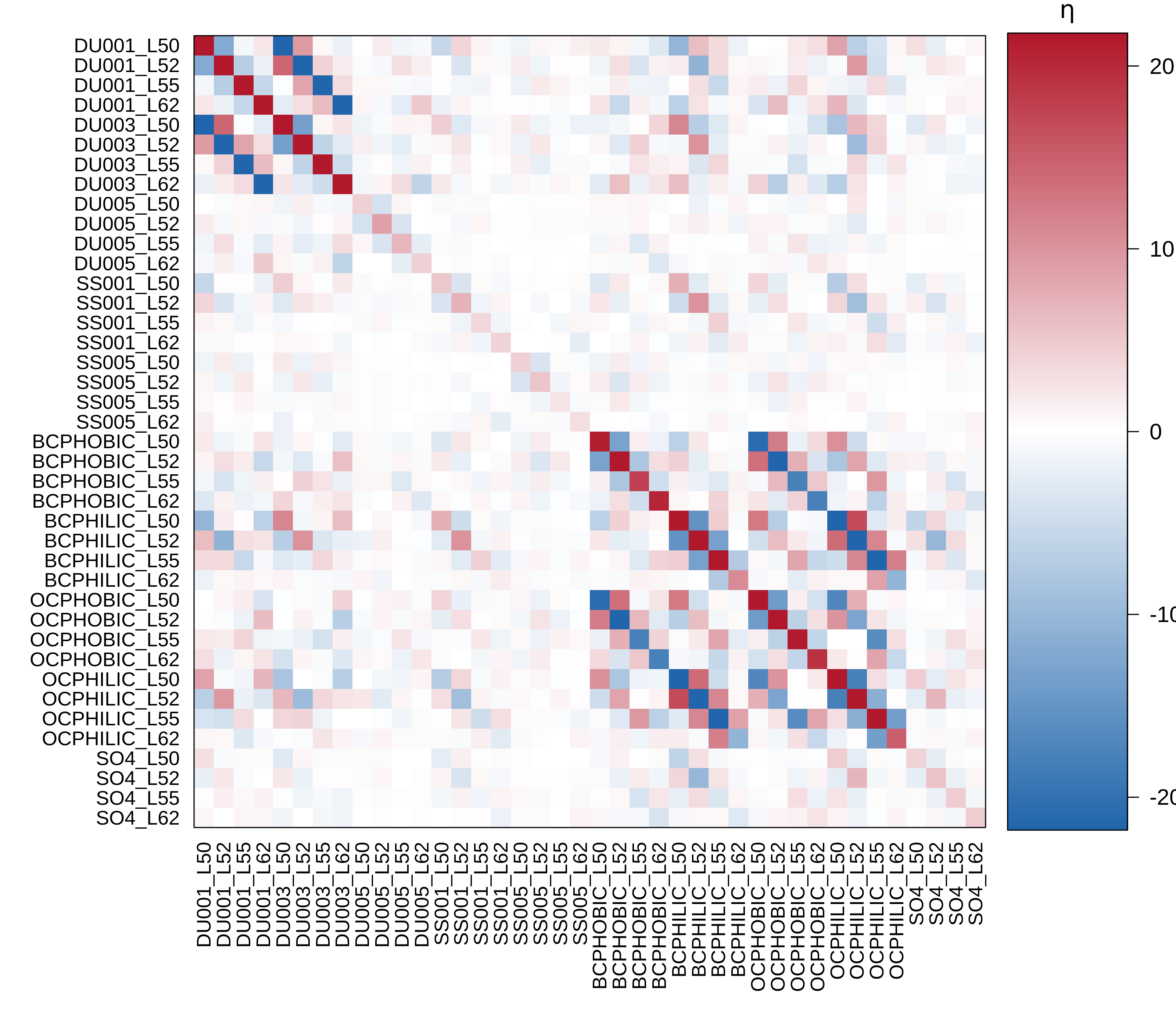}
  \end{tabular}
    \caption{MERRA-2 sample precision matrices and SAECM precision estimates from colocated observations, shown as entrywise medians across three folds.}
\label{fig:merra-colocated-precision}
\end{figure}

\begin{table}[!ht]
\centering
\small
\setlength{\tabcolsep}{1.5pt}
\renewcommand{\arraystretch}{0.8}
\caption{MERRA-2 prediction results for colocated and misaligned observations. Entries are mean (SD) across three folds. Bold indicates coverage closest to 95\% and the lowest mean for other metrics.}
\label{table:merra-prediction}
\begin{tabular}{llccccc}
\toprule
Design & Metric & Penalized & Dense & Ind & LMC ($q=10$) & IOX \\
\midrule
Colocated & Norm
& 2.65 (0.12) & 2.66 (0.12) & \textbf{2.56 (0.12)} & 3.55 (0.15) & 2.61 (0.12) \\
& Coverage (\%)
& 93.51 (0.79) & 93.71 (0.83) & 93.67 (1.08) & 93.75 (1.31) & \textbf{94.70 (0.84)} \\
& Sum RMSE
& 8.00 (1.19) & 8.01 (1.22) & 7.53 (1.11) & \textbf{7.43 (0.92)} & 7.66 (1.12) \\
& Sum coverage (\%)
& \textbf{94.83 (2.08)} & 94.67 (1.89) & 71.00 (3.28) & 93.67 (0.76) & 94.50 (1.73) \\
& PI
& 1.69 (0.09) & 1.72 (0.10) & \textbf{1.68 (0.14)} & 2.27 (0.09) & 1.83 (0.12) \\
& PI sum
& 29.29 (1.20) & 29.67 (1.73) & \textbf{10.73 (0.88)} & 26.11 (2.38) & 28.02 (2.59) \\
\midrule
Misaligned & Norm
& 3.11 (0.09) & \textbf{3.08 (0.10)} & 3.38 (0.10) & 3.94 (0.10) & 3.13 (0.24) \\
& Coverage (\%)
& 93.48 (0.85) & 93.55 (0.85) & 93.20 (0.83) & 93.17 (1.11) & \textbf{95.58 (0.59)} \\
& Sum RMSE
& 7.84 (1.31) & \textbf{7.83 (1.36)} & 8.92 (1.49) & 8.33 (1.21) & 8.09 (1.27) \\
& Sum coverage (\%)
& 89.50 (1.32) & 92.67 (1.04) & 70.00 (2.18) & 92.17 (1.76) & \textbf{94.83 (0.58)} \\
& PI
& 1.98 (0.14) & \textbf{1.96 (0.13)} & 2.09 (0.12) & 2.43 (0.11) & 2.24 (0.10) \\
& PI sum
& 22.96 (1.59) & 25.69 (2.07) & \textbf{13.61 (0.76)} & 26.76 (2.89) & 30.87 (2.55) \\
\bottomrule
\end{tabular}
\end{table}

\section{Discussion}
We introduce a new penalized likelihood estimation methodology for highly multivariate Gaussian processes that defines the model in the spectral domain, where process-level conditional independence is most naturally interpretable, while inference can proceed in the spatial domain using random Fourier features as the key connecting bridge. Unlike previous works using spatial domain constructions \citep[e.g.,][]{dey_graphical_2022, peruzzi2024}, our approach does not require a reference set. Also unlike previous works using spectral constructions \citep[e.g.,][]{terres2018bayesian}, our approach obviates the need for Whittle likelihood-based estimation, and targets maximizing the penalized observed data likelihood.

By separating the covariance parameters and the spatial parameters in the spectral domain, our covariance function is valid by construction and easily allows different marginal covariance families with spectral densities in finite-dimensional parametric families.
Furthermore, the combination of the SAECM algorithm with the graphical LASSO method provides sparsity in the precision estimate, establishing process-level conditional independence estimation. 
Our construction allows process-specific Mat\'ern marginal scales while retaining this graph interpretation, at the expense of cross-covariances that are generally not Mat\'ern, thereby partially addressing the modeling challenge noted by \cite{peruzzi2026partial}.

Our simulation and data analyses demonstrate comparable or better individual-process predictive performance relative to benchmarks under colocated observations, with more pronounced improvements under misalignment, particularly under heterogeneous marginal covariance families. The consideration of process dependence provides additional information when the data are not fully observed, and further improves prediction and inference regarding the functionals of the process. This approach can be readily applied to environmental studies including atmospheric science \citep{gelaro2017modern} or oceanography \citep{argo_argo_2023}.

Although we work with stationary covariance functions for demonstrations, future work can consider extensions to nonstationary \citep{emery2018continuous, ton_spatial_2018} or spatio-temporal covariance functions \citep{allard_fully_2022}. Extension to stationary multivariate spatio-temporal processes is conceptually straightforward, provided suitable joint spatio-temporal marginal spectral densities are available. Accommodating frequency-dependent partial coherence \citep{mitchell_l_krock_modeling_2023} is another promising direction.

\spacingset{1.8} 
\section*{Supplementary Material}
The Supplementary Material contains mathematical proofs, additional technical details and numerical results.

\section*{Data Availability Statement}
The MERRA-2 data analyzed in this study are available from NASA's Goddard Earth Sciences Data and Information Services Center under product M2I3NVAER, version 5.12.4 (doi:10.5067/LTVB4GPCOTK2). Code for the simulations and data analysis is available at \url{https://github.com/bychen713/SSGGP}.

\begingroup
\fontsize{12pt}{11pt}\selectfont 
\bibliographystyle{apalike}
\bibliography{mGP.bib}
\endgroup

\normalsize

\clearpage
\spacingset{1.8}
\begin{center}
{\LARGE \textbf{Supplementary Material to}
\\\emph{Penalized Likelihood Inference in Highly Multivariate Sparse Spectral Graphical Gaussian Processes}}
\end{center}

\setcounter{section}{0}
\setcounter{equation}{0}
\setcounter{figure}{0}
\setcounter{table}{0}
\setcounter{page}{1}
\renewcommand{\thepage}{S.\arabic{page}}

\renewcommand{\thesection}{S.\arabic{section}}
\renewcommand{\theequation}{S\arabic{equation}}
\renewcommand{\thefigure}{S\arabic{figure}}
\renewcommand{\thetable}{S\arabic{table}}
\renewcommand{\theHsection}{S.\arabic{section}}
\renewcommand{\theHequation}{S\arabic{equation}}
\renewcommand{\theHfigure}{S\arabic{figure}}
\renewcommand{\theHtable}{S\arabic{table}}

\renewcommand{\thealgocf}{S\arabic{algocf}} 

\section{Example Simulation Details}\label{app:ex_sim}

\begin{figure}[!th]
    \centering
    \includegraphics[width = .9\textwidth]{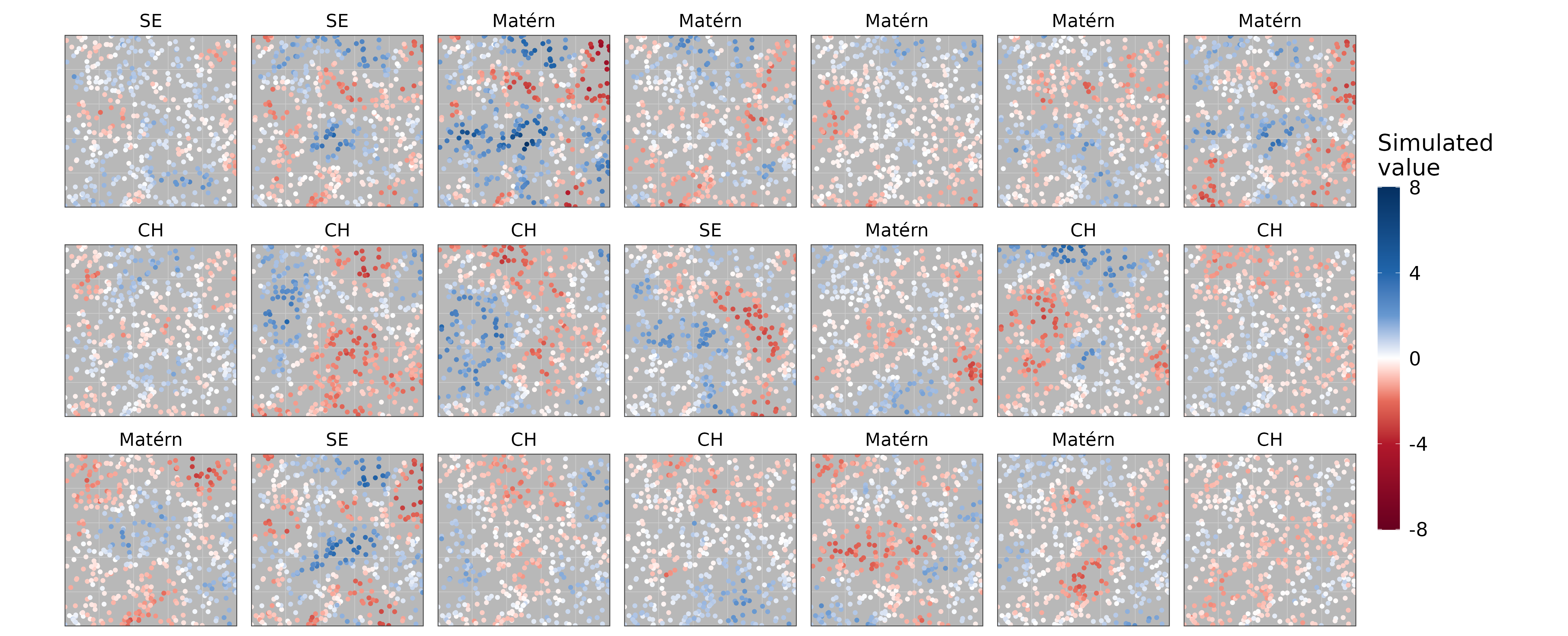}
    \caption{Simulated fields for the first 21 processes in replicate 1 of the random-precision setting with positive off-diagonal entries.}    \label{fig:visual}
\end{figure}

We summarize the described approach in Algorithm~\ref{sim:ours} for $\mb{f}(x)$ as in \eqref{eq:simplified_form}, which constructs $\mb{W}_\ell \sim \mathcal{N}(\mb{0}_p, 2\mb{f}(X_\ell)/g(X_\ell))$. 
We provide an example in Figure~\ref{fig:visual}.

\begin{algorithm}
 \KwData{$L$, $\mb{s}\in \mathbb{R}^{n\times d}$, $\mb{\sigma}^{-1} \in \mathbb{R}^{p\times p}$, $\{f_{j}(x)\}_{j=1}^p$, $g(x)$.}
 \KwResult{Simulated multivariate process $\mb{Y}(\mb{s}) \in \mathbb{R}^{n \times p}$ at locations $\mb{s}$.}

 Initialize $\mb{Y}(\mb{s})$'s entries to $0$, and compute the Cholesky factor of $\mb{\sigma}^{-1} = \mb{\Lambda}\mb{\Lambda}^\top $.
 
 \For{$\ell = 1, 2, \dots, L$}{
    Simulate $X_\ell \sim g(x)$, $\psi_\ell \sim \textrm{U}[0, 2\pi]$, and $\mb{Z}_\ell \sim \mathcal{N}(\mb{0}_p, \mb{I}_p)$.

    Compute $\mb{M}_\ell = \left(g(X_\ell)/2\right)^{\frac{1}{2}}\mb{\Lambda}^\top\mb{h}(X_\ell)^{-1}  =\mb{\Lambda}^\top \textrm{diag}\left(\sqrt{\frac{g(X_{\ell})}{2f_j(X_{\ell})}}, j=1, \dots, p\right)$.

    Compute $\mb{W}_\ell$ by solving $\mb{M}_\ell \mb{W}_\ell = \mb{Z}_\ell$.

    \For{$i=1, \dots, n$}{

    Set $\mb{Y}(s_i) = \mb{Y}(s_i) + c_{i,\ell} \mb{W}_\ell$, where $c_{i,\ell} = \frac{1}{\sqrt{L}} \cos(\langle X_{\ell},  s_i\rangle + \psi_\ell)$.
    }
 }
 \caption{Simulation of $\mb{Y}(\mb{s})$ for $\mb{f}(x) \in \mathbb{R}^{p\times p}$ as in \eqref{eq:simplified_form}. \vspace{-5mm}}\label{sim:ours}
\end{algorithm}
 
The computational bottlenecks are the Cholesky decomposition of $\mb{\sigma}^{-1}$ which is $O(p^3)$ for dense $\mb{\sigma}^{-1}$, 
the creation of $\{\mb{M}_\ell\}_{\ell=1}^L$ and $\{\mb{W}_\ell\}_{\ell=1}^L$ by backsolving which is $O(Lp^2)$, and the inner loop of constructing $\mb{Y}(\mb{s})$ has overall complexity $O(Lpn + Ldn)$.
Furthermore, the cubic and quadratic terms of $p$ can be effectively reduced with sparsity, greatly reducing the computational complexity for large $p$.

\section{Marginal Likelihood Calculation}\label{app:marginal_likelihood}

In the following, we suppress the notation on $\mb{\theta}$, using the matrices $\mb{\Sigma}_L$, $\mb{D}$, and:
\begin{align*}
    \mb{\Sigma}_{\mb{W}} &= \textrm{block-diagonal}(\mb{\Sigma}_{\mb{W}_{\ell^\prime}},\ell^\prime =1, \dots, L)\in \mathbb{R}^{
   Lp\times Lp}, &\mb{A} &=[c_{i,\ell}]_{i=1,\ell=1}^{n,L} \otimes \mb{I}_p \in \mathbb{R}^{np\times Lp},
\end{align*}
where $c_{i, \ell}= L^{-1/2}\cos(\langle s_i, X_\ell\rangle + \psi_{\ell})$.

\subsection{Likelihood calculation without simulated \texorpdfstring{$\mb{W}$}{W}}
By integrating out $\mb{W}$, the random feature covariance can be expressed:
$
\mb{\Sigma}_L = \mb{A}\mb{\Sigma}_{\mb{W}}\mb{A}^\top
$.
Then we can apply the Woodbury identity for the likelihood calculation:
\begin{align*}
    \big[\mb A\mb\Sigma_{\mb W}\mb A^\top+\mb D\big]^{-1}
&=\mb D^{-1}-\mb D^{-1}\mb A\big(\mb\Sigma_{\mb W}^{-1}+\mb A^\top\mb D^{-1}\mb A\big)^{-1}\mb A^\top\mb D^{-1}.\\
\det\big[\mb A\mb\Sigma_{\mb W}\mb A^\top+\mb D\big]
&=\det\big[\mb\Sigma_{\mb W}^{-1}+\mb A^\top\mb D^{-1}\mb A\big]\det\big[\mb\Sigma_{\mb W}\big]\det\big[\mb D\big].
\end{align*}
The calculation is dominated by the Cholesky factorization of the $Lp\times Lp$ matrix
$\mb{\Sigma}_{\mb{W}}^{-1}+\mb{A}^\top\mb{D}^{-1}\mb{A}$, which requires $O(L^3 p^3)$.

\subsection{Likelihood calculation with simulated \texorpdfstring{$\mb{W}$}{W}}

Directly following our approach for simulation above for $\mb{f}(x) \in \mathbb{R}^{p\times p}$, we take $
    \mb{\Phi}
    = \left[c_{i,\ell} \mb{W}_\ell\right]_{i=1, \ell=1}^{n, L} \in \mathbb{R}^{np \times L}$. 
The Monte Carlo estimate of the marginal log-likelihood function based on $\mb{\Phi}(\mb{\theta})$ is then:
\begin{align*}
    \tilde{\mathcal{L}}(\mb{\theta} \mid \mb{Y}) &= -\frac{1}{2} \left\{np\log(2\pi) + \log\left(\textrm{det} \left[\tilde{\mb{\Sigma}}_L(\mb{\theta}) + \mb{D}\right]\right) +\mb{Y}^\top (\tilde{\mb{\Sigma}}_L(\mb{\theta}) + \mb{D})^{-1} \mb{Y}\right\},
\end{align*}where $\tilde{\mb{\Sigma}}_L(\mb{\theta}) = \mb{\Phi}(\mb{\theta})  \mb{\Phi}(\mb{\theta})^\top$ is a Monte Carlo estimate of the covariance $\mb{\Sigma}_L(\mb{\theta})$, obtained by simulating the latent variables $\mb{W}=\{\mb{W}_\ell\}$ rather than integrating them out analytically, and
$\mathbb{E}_{\mb{W}}[\tilde{\mb{\Sigma}}_L(\mb{\theta};\mb{W})] = \mb{\Sigma}_L(\mb{\theta})$.
To compute the likelihood, we need to find the determinant of $\tilde{\mb{\Sigma}}_L + \mb{D}$ and solve against $\tilde{\mb{\Sigma}}_L + \mb{D}$. 
The Woodbury matrix identity and the matrix determinant lemma, respectively, yield:
\begin{align}\begin{split}
    \left[\tilde{\mb{\Sigma}}_L+ \mb{D}\right]^{-1}&=     \left[\mb{\Phi}\mb{\Phi}^\top + \mb{D}\right]^{-1} = \mb{D}^{-1} - \mb{D}^{-1}\mb{\Phi}(\mb{I}_{L}  + \mb{\Phi}^\top \mb{D}^{-1} \mb{\Phi})^{-1}\mb{\Phi}^\top \mb{D}^{-1},\\  
    \textrm{det}\left[\tilde{\mb{\Sigma}}_L + \mb{D}\right] &= \textrm{det}\left[\mb{\Phi}\mb{\Phi}^\top + \mb{D}\right]
    =\textrm{det}\left[\mb{I}_{L} + \mb{\Phi}^\top \mb{D}^{-1}\mb{\Phi}\right]\textrm{det}\left[\mb{D}\right].
    \end{split}\label{eq:woodbury}
\end{align}
This reduces the necessary Cholesky decomposition from size $np$ to $L$.
We summarize the overall algorithm for likelihood evaluation in Algorithm~\ref{sim:likeli}.
The computational complexity for simulation is $O(p^3 + Lp^2 + Lpn + Lnd)$, but the likelihood evaluation has additional terms. 
In particular, forming the matrix $\mb{I}_L + \mb{\Phi}^\top \mb{D}^{-1} \mb{\Phi}$ has complexity $O(L^2 p + L^2n)$, and computing its Cholesky factor $\mb{\Psi}$ has complexity $O(L^3)$. 
The additional complexity is thus directly controllable by the Fourier approximation size $L$, and remains less (in $p$) than the $O(Lp^3)$ in \cite{mitchell_l_krock_modeling_2023} for estimation. 

\begin{algorithm}[!h]
\DontPrintSemicolon
 \KwData{$\mb{Y}(\mb{s}) \in \mathbb{R}^{n\times p}$, $L$, $\mb{s}\in \mathbb{R}^{n\times d}$, $\mb{\sigma}^{-1} \in \mathbb{R}^{p\times p}$, $\{f_{j}(x)\}_{j=1}^p$, $\mb{d} = \textrm{diag}(\{d_j\}_{j=1}^p)$, $g(x)$.}
 \KwResult{Likelihood evaluation for parameters $\mb{\sigma}^{-1}$, $\{d_j\}_{j=1}^p$, and those in $\{f_{j}(x)\}_{j=1}^p$. }

Compute the Cholesky factor of $\mb{\sigma}^{-1} = \mb{\Lambda}\mb{\Lambda}^\top $.

Compute $\mb{W}_\ell$ and $c_{i, \ell}$ as in Algorithm~\ref{sim:ours} for all $\ell = 1$, $2$, $\dots$, $L$.

 \For{$\ell_1 = 1, 2, \dots, L$}{
 \For{$\ell_2 = 1, 2, \dots, L$}{
    Compute $(\mb{I}_L + \mb{\Phi}^\top \mb{D}^{-1} \mb{\Phi})_{\ell_1, \ell_2} = \mathbbm{1}(\ell_1 = \ell_2)+  \mb{W}_{\ell_1}^\top \mb{d}^{-1} \mb{W}_{\ell_2} \sum_{i=1}^n c_{i, \ell_1} c_{i, \ell_2}$.
 }
 }
 Compute the Cholesky factor $\mb{I}_L + \mb{\Phi}^\top \mb{D}^{-1} \mb{\Phi} = \mb{\Psi} \mb{\Psi}^\top$. 
 
 Compute the likelihood using $\mb{\Psi}$ and \eqref{eq:woodbury}. 

 \caption{Likelihood evaluation when $\mb{f}(x)$ has form \eqref{eq:simplified_form}.}
 \label{sim:likeli}
\end{algorithm}

This approach also lessens the memory burden of the likelihood evaluation, which would be $O(p^2n^2)$ for the $np \times np$ matrix $\tilde{\mb{\Sigma}}_L + \mb{D}$.
Our proposed methodology for computation of the likelihood circumvents direct formation of $\tilde{\mb{\Sigma}}_L + \mb{D}$ and has a memory burden of size $O(p^2 + L^2 + Lp + Ln + np)$. 
We outline the sources for these terms: the $p^2$ for $\mb{\sigma}^{-1}$ and $\mb{\Lambda}$, $L^2$ for $\mb{I}_L + \mb{\Phi}^\top \mb{D}^{-1} \mb{\Phi}$ and $\mb{\Psi}$, $Lp$ for $\{\mb{W}_\ell\}_{\ell=1}^L$, $Ln$ for $\{c_{i, \ell}\}$, and $np$ for $\mb{Y}$. 
The largest terms are again directly controllable by $L$ or by sparsity in $\mb{\sigma}^{-1}$.

\section{Conditional Distribution for Gibbs Sampling}\label{app:conditional}

First, we introduce some notation. 
With $\otimes$ being the standard Kronecker product, define \begin{align*}
    \mb{A}_\ell &=[c_{i,\ell}]_{i=1}^n \otimes \mb{I}_p \in \mathbb{R}^{np\times p}, & \mb{W}_{-\ell}&= [\mb{W}_{\ell^\prime}]_{\ell^\prime\neq \ell}^L \in \mathbb{R}^{(L-1)p},\\
    \mb{A}_{-\ell} &= [c_{i,\ell^\prime}]_{i=1, \ell^\prime\neq \ell}^{n, L}\otimes \mb{I}_p \in \mathbb{R}^{np \times (L-1)p},& 
    \mb{D} &=\mb{I}_n \otimes \mb{d}.
\end{align*}
With this notation, we have the following result.

\begin{proposition}\normalfont\label{prop:cond}
The distribution of $\mb{W}_\ell \mid \mb{W}_{-\ell}, \mb{Y}, \mb{\theta}$ is multivariate Gaussian with:
    \begin{align*}
        \mathbb{E}[\mb{W}_\ell \mid \mb{W}_{-\ell}, \mb{Y},  \mb{\theta}] &= \mb{\Sigma}_{\mb{W}_\ell} \left[  \mb{I}_p  - \mb{A}_\ell^\top \mb{D}^{-1} \mb{A}_\ell (\mb{\Sigma_{\mb{W}_\ell}}^{-1} + \mb{A}_\ell^\top \mb{D}^{-1} \mb{A}_\ell)^{-1}\right]  \\ 
        &~~~~~~~~~~~~~~~\left[\mb{A}_\ell^\top \mb{D}^{-1} (\mb{Y} - \mb{A}_{-\ell}\mb{W}_{-\ell})\right],\\
           \textrm{Var}[\mb{W}_\ell \mid \mb{W}_{-\ell}, \mb{Y},  \mb{\theta}] &=
 (\mb{\Sigma}_{\mb{W}_\ell}^{-1} + \mb{A}_\ell^\top \mb{D}^{-1} \mb{A}_\ell)^{-1}.
    \end{align*}
\end{proposition}

Therefore, we can simulate from this distribution using a matrix decomposition of dimension $p$. 
We study the computation in a bit more detail. Notice that \begin{align*}
    \mb{A}_\ell^\top \mb{D}^{-1} \mb{A}_\ell &= ([c_{i,\ell}]_{i=1}^n \otimes \mb{I}_p)^\top (\mb{I}_n \otimes \mb{d}^{-1}) ([c_{i,\ell}]_{i=1}^n \otimes \mb{I}_p) =\left[ \sum_{i=1}^n c_{i,\ell}^2\right] \mb{d}^{-1},
\end{align*}a $p\times p$ diagonal matrix whose computation only has a computational complexity of $O(np)$.
Furthermore, the matrices $\mb{\Sigma}_{\mb{W}_\ell}^{-1}$ (which can be formed directly from $\mb{\sigma}^{-1}$) and $\mb{\Sigma}_{\mb{W}_\ell}^{-1} + \mb{A}_\ell^\top \mb{D}^{-1} \mb{A}_\ell$ have the same sparsity as $\mb{\sigma}^{-1}$. 
Thus, process-level conditional independence relations result in computational improvements even within the Gibbs sampling algorithm of the E step.
We note that similar expressions for the conditional expectation may be formed, though we have found the above most computationally efficient in practice. The proof is as follows.

\begin{proof}[Proof of Proposition~\ref{prop:cond}]
    Let \begin{align*}
 \mb{\Sigma}_{\mb{W}_{-\ell}} &= \textrm{block-diagonal}(\mb{\Sigma}_{\mb{W}_{\ell^\prime}}, \ell^\prime\neq \ell) \in \mathbb{R}^{
   (L-1)p\times (L-1)p}  
   \end{align*}
    We have the joint distribution \begin{align*}
   \begin{bmatrix} \mb{W}_\ell \\ \mb{W}_{-\ell} \\ \mb{Y} \end{bmatrix} \mid \mb{\theta} &\sim  \mathcal{N}\left(\begin{bmatrix} \mb{0} \\ \mb{0} \\ \mb{0}\end{bmatrix}, \begin{bmatrix} \mb{\Sigma}_{\mb{W}_\ell}  & \mb{0} & \mb{\Sigma}_{\mb{W}_\ell} \mb{A}_\ell^\top\\
   \mb{0} &\mb{\Sigma}_{\mb{W}_{-\ell}} & \mb{\Sigma}_{\mb{W}_{-\ell}}\mb{A}_{-\ell}^\top \\
   \mb{A}_\ell \mb{\Sigma}_{\mb{W}_\ell}&\mb{A}_{-\ell} \mb{\Sigma}_{\mb{W}_{-\ell}} & \mb{A}\mb{\Sigma}_{\mb{W}}\mb{A}^\top + \mb{D}\end{bmatrix}\right).
\end{align*}
We begin with the variance.
Let \begin{align*}
    \mb{\Omega} = \begin{bmatrix}
   \mb{\Sigma}_{\mb{W}_{-\ell}} & \mb{\Sigma}_{\mb{W}_{-\ell}}\mb{A}_{-\ell}^\top \\
   \mb{A}_{-\ell} \mb{\Sigma}_{\mb{W}_{-\ell}} & \mb{A}\mb{\Sigma}_{\mb{W}}\mb{A}^\top + \mb{D}\end{bmatrix}^{-1} = \begin{bmatrix}
       \mb{\Omega}_{11} & \mb{\Omega}_{21}^\top \\
       \mb{\Omega}_{21} & \mb{\Omega}_{22}
   \end{bmatrix}.
\end{align*}
Based on properties of the multivariate normal distribution, 
\begin{align*}
    \textrm{Var}[\mb{W}_\ell \mid \mb{W}_{-\ell}, \mb{Y},  \mb{\theta}] &= \mb{\Sigma}_{\mb{W}_\ell} - \begin{pmatrix}\mb{0} & \mb{\Sigma}_{\mb{W}_\ell} \mb{A}_\ell^\top \end{pmatrix}\mb{\Omega}
\begin{pmatrix}\mb{0} \\ \mb{A}_\ell\mb{\Sigma}_{\mb{W}_\ell}  \end{pmatrix} =\mb{\Sigma}_{\mb{W}_\ell} -  \mb{\Sigma}_{\mb{W}_\ell} \mb{A}_\ell^\top \mb{\Omega}_{22}
 \mb{A}_\ell\mb{\Sigma}_{\mb{W}_\ell}.
\end{align*}
We next compute $\mb{\Omega}_{22}$ using the Schur complement:
\begin{align*}
    \mb{\Omega}_{22} &= \left[\mb{A}\mb{\Sigma}_{\mb{W}}\mb{A}^\top + \mb{D} - \mb{A}_{-\ell} \mb{\Sigma}_{\mb{W}_{-\ell}} \mb{\Sigma}_{\mb{W}_{-\ell}}^{-1} \mb
    {\Sigma}_{\mb{W}_{-\ell}}\mb{A}^\top_{-\ell}\right]^{-1} \\
&=\left[\mb{A}\mb{\Sigma}_{\mb{W}}\mb{A}^\top + \mb{D} - \mb{A}_{-\ell} \mb{\Sigma}_{\mb{W}_{-\ell}}\mb{A}^\top_{-\ell}\right]^{-1}.
\end{align*}
We may decompose 
    $\mb{A}\mb{\Sigma}_{\mb{W}}\mb{A}^\top  
    = \mb{A}_\ell \mb{\Sigma}_{\mb{W}_\ell}\mb{A}_\ell^\top + \mb{A}_{-\ell} \mb{\Sigma}_{\mb{W}_{-\ell}}\mb{A}_{-\ell}^\top$, which results in $    \mb{\Omega}_{22} = \left[\mb{A}_\ell \mb{\Sigma}_{\mb{W}_\ell}\mb{A}_\ell^\top + \mb{D}\right]^{-1}$.
This leads to the variance \begin{align*}
 \textrm{Var}[\mb{W}_\ell \mid \mb{W}_{-\ell}, \mb{Y},  \mb{\theta}] &=
\mb{\Sigma}_{\mb{W}_\ell} -  \mb{\Sigma}_{\mb{W}_\ell} \mb{A}_\ell^\top \left[\mb{A}_\ell \mb{\Sigma}_{\mb{W}_\ell}\mb{A}_\ell^\top + \mb{D}\right]^{-1}
 \mb{A}_\ell\mb{\Sigma}_{\mb{W}_\ell} \\
 &= 
 \left[\mb{\Sigma}_{\mb{W}_\ell}^{-1} + \mb{A}_\ell^\top \mb{D}^{-1} \mb{A}_\ell\right]^{-1},
\end{align*}
where the last line follows from the Woodbury matrix identity.

The expectation follows in a similar manner:\begin{align*}
    \mathbb{E}[\mb{W}_\ell \mid \mb{W}_{-\ell}, \mb{Y},  \mb{\theta}] &= \begin{pmatrix}\mb{0} & \mb{\Sigma}_{\mb{W}_\ell} \mb{A}_\ell^\top \end{pmatrix} \mb{\Omega}\begin{pmatrix} \mb{W}_{-\ell} \\ \mb{Y}\end{pmatrix} \\
    &= \mb{\Sigma}_{\mb{W}_\ell} \mb{A}_\ell^\top \mb{\Omega}_{21}
     \mb{W}_{-\ell} + \mb{\Sigma}_{\mb{W}_\ell} \mb{A}_\ell^\top \mb{\Omega}_{22}
    ^\top \mb{Y}.
\end{align*}
Using blockwise inversion, we have $
    \mb{\Omega}_{21} = -\mb{\Omega}_{22} \mb{A}_{-\ell}\mb{\Sigma}_{\mb{W}_{-\ell}}\mb{\Sigma}_{\mb{W}_{-\ell}}^{-1} =-\mb{\Omega}_{22} \mb{A}_{-\ell}$, which gives \begin{align*}
    \mathbb{E}[\mb{W}_\ell \mid \mb{W}_{-\ell}, \mb{Y},  \mb{\theta}] &= \mb{\Sigma}_{\mb{W}_\ell} \mb{A}_\ell^\top\mb{\Omega}_{22}\left[\mb{Y} - \mb{A}_{-\ell}\mb{W}_{-\ell}\right] \\
    &= \mb{\Sigma}_{\mb{W}_\ell} \mb{A}_\ell^\top\left[\mb{A}_\ell \mb{\Sigma}_{\mb{W}_\ell}\mb{A}_\ell^\top + \mb{D}\right]^{-1}\left[\mb{Y} - \mb{A}_{-\ell}\mb{W}_{-\ell}\right].
    \end{align*}
    Applying the Woodbury matrix identity: \begin{align*}
        \left[\mb{A}_\ell \mb{\Sigma}_{\mb{W}_\ell}\mb{A}_\ell^\top + \mb{D}\right]^{-1} &= \mb{D}^{-1} - \mb{D}^{-1}\mb{A}_\ell \left[\mb{\Sigma}_{\mb{W}_\ell}^{-1} + \mb{A}_\ell^\top\mb{D}^{-1}\mb{A}_\ell\right]^{-1}\mb{A}_{\ell}^\top \mb{D}^{-1},
    \end{align*}the conditional expectation is \begin{align*}
         \mathbb{E}[\mb{W}_\ell \mid \mb{W}_{-\ell}, \mb{Y},  \mb{\theta}] &= \mb{\Sigma}_{\mb{W}_\ell} \mb{A}_\ell^\top\left[\mb{D}^{-1} - \mb{D}^{-1}\mb{A}_\ell \left[\mb{\Sigma}_{\mb{W}_\ell}^{-1} + \mb{A}_\ell^\top\mb{D}^{-1}\mb{A}_\ell\right]^{-1}\mb{A}_{\ell}^\top \mb{D}^{-1}\right] \\
         &~~~~~~~~~~~~~~~\left[\mb{Y} - \mb{A}_{-\ell}\mb{W}_{-\ell}\right].
    \end{align*}
    The expression in the proposition then follows from factoring and rearranging.
\end{proof}

\section{Proofs of Propositions}\label{app:proof}
\subsection{Proof of \texorpdfstring{\Cref{prop:marginal_Y}}{prop1}} 
Let $\mb{\theta}_{\textrm{prop}}$ be the parameters of the proposal distribution in the importance sampling.
We have the following hierarchical structure:
\begin{align*}
X_\ell&\overset{i.i.d.}{\sim} g(\cdot\mid\mb{\theta}_{\textrm{prop}}),\\
\mb{W}_\ell&\mid X_\ell,\mb{\theta}_{\textrm{prop}},\mb{\theta}\sim  \mathcal{N}\left(0,2\frac{\mb{f}(X_\ell)}{g(X_\ell)}\right)= \mathcal{N}\left(\mb{0}_p,\mb{\Sigma}_{\mb{W}_\ell}\right),\\
\mb{Y}_L(s)&\mid\mb{W},\mb{d},\{X_\ell,\psi_\ell\}_{\ell=1}^L\sim  \mathcal{N}\left(\sum_\ell c_\ell{\mb{W}_\ell},\mb{d}\right),
\end{align*}
where $c_\ell = \frac{1}{\sqrt{L}}\cos(\langle s,X_\ell\rangle + \psi_\ell)$ and $\psi_\ell\overset{i.i.d.}{\sim} U[0,2\pi]$.
In matrix form:
\begin{align*}
    \mb{W}&\mid \{X_\ell\}_{\ell=1}^L,\mb{\theta}_{\textrm{prop}},\mb{\theta} \sim \mathcal{N}(\mb{0},\mb{\Sigma}_{\mb{W}}), \\
    \mb{Y}_L&\mid\mb{W},\mb{d},\{X_\ell,\psi_\ell\}_{\ell=1}^L \sim \mathcal{N}(\mb{AW}, \mb{D}),
\end{align*}
where $\mb D=\mb I_n\otimes\mb d$, and $\mb{\Sigma}_{\mb{W}}=\textrm{blockdiag}\{\mb{\Sigma}_{\mb{W}_\ell}\}$. By integrating out $\mb{W}$, we have
\begin{align*}
    \mb{Y}_L\mid \mb{d},\{X_\ell,\psi_\ell\}_{\ell=1}^L \sim \mathcal{N}(\mb{0}, \mb{A\Sigma_W A}^\top + \mb{D}).
\end{align*}
The first two moments of $\mb{Y}_L$ can be calculated by the law of total expectation and variance:
\begin{align*}
    \mathbb E[\mb Y]
    &=
    \mathbb E_{\{X_\ell,\psi_\ell\}_{\ell=1}^L}
    \left[
    \mathbb E\left[
    \mb Y\mid\{X_\ell,\psi_\ell\}_{\ell=1}^L
    \right]
    \right]
    =\mb 0,\\
    \operatorname{Var}(\mb Y)
    &=
    \operatorname{Var}_{\{X_\ell,\psi_\ell\}_{\ell=1}^L}
    \left(
    \mathbb E\left[
    \mb Y\mid\{X_\ell,\psi_\ell\}_{\ell=1}^L
    \right]
    \right)+
    \mathbb E_{\{X_\ell,\psi_\ell\}_{\ell=1}^L}
    \left[
    \operatorname{Var}\left(
    \mb Y\mid\{X_\ell,\psi_\ell\}_{\ell=1}^L
    \right)
    \right]\\
    &=
    \mb 0+
    \mathbb E_{\{X_\ell,\psi_\ell\}_{\ell=1}^L}
    \left[
    \mb A\mb\Sigma_{\mb W}\mb A^\top+\mb D
    \right].
\end{align*}
Each $p\times p$ block of $\mathbb{E}_{\{X_\ell,\psi_\ell\}_{\ell=1}^L}[\mb{A\Sigma_W A}^\top]$ is:
\begin{align*}
    \sum_\ell \,&\mathbb{E}\left[\frac{2}{L}\cos(\langle s_1,X_\ell\rangle+\psi_\ell)\cos(\langle s_2,X_\ell\rangle+\psi_\ell)\frac{\mb{f}(X_\ell)}{g(X_\ell)}\right]\\
    &=\frac{1}{L}\sum_\ell \mathbb{E}_{X_\ell}\left[\mathbb{E}_{\psi_\ell}\left[\left\{\cos(\langle s_1-s_2,X_\ell\rangle)+\cos(\langle s_1+s_2,X_\ell\rangle+2\psi_\ell)\right\}\frac{\mb{f}(X_\ell)}{g(X_\ell)}\mid X_\ell\right]\right]\\
    &=\frac{1}{L}\sum_\ell \mathbb{E}_{X_\ell}\left[\cos(\langle s_1-s_2,X_\ell\rangle)\frac{\mb{f}(X_\ell)}{g(X_\ell)}\right]\\
    &=\mb{C}(s_1,s_2).
\end{align*}
Express $\mb{Y}_L(s)=\sum_\ell c_\ell(s)\mb{W}_\ell + \mb{\epsilon}=\frac{1}{\sqrt{L}}\sum_\ell \cos(\langle s,X_\ell\rangle+\psi_\ell)\mb{W}_\ell + \mb{\epsilon} \equiv \frac{1}{\sqrt{L}}\sum_\ell \mb{V}_\ell(s) + \mb{\epsilon}$. 
Since $\E\|\mb W_\ell\|^2=2\operatorname{tr}(\mb\sigma)<\infty$ and the cosine factors are bounded, the finite-dimensional summands have finite second moments.
The i.i.d. random vectors $\mb{V}_\ell(s)$ are not normally distributed, but by the multivariate central limit theorem, $\mb{Y}_L\to \mathcal{N}(\mb{0},\mb{\Sigma+D})$ as $L\to \infty$.

\subsection{Preliminaries for the proofs of Propositions~\ref{prop:set} and~\ref{prop:diagonal_convergence}}
With $\mb\theta=(\precision,\mb d,\mb\iota)$, the penalized complete-data log-likelihood has an exponential form
$$
\mathcal{L}_P(\mb\theta\mid\mb Y,\mb W)
=-\psi(\mb\theta)
+\langle U(\mb Y,\mb W),\phi(\mb\theta)\rangle
-P(\mb\theta),
$$
where $U(\mb Y,\mb W)$ is the complete-data sufficient statistic, $\phi(\mb\theta)=\bigl(\phi_{\precision},\phi_d\bigr)$ and $\psi(\mb\theta)$ are known functions of $\mb\theta$.
\begin{align*}
U(\mb Y,\mb W)
  &=\Bigl(\bigl(\mb W_\ell^{\vphantom{\top}} \mb W_\ell^\top\bigr)_{\ell=1}^{L},\;
   \Bigl(\textstyle\sum_{i=1}^{n}
     \bigl(Y_j(s_i)-\sum_{\ell=1}^{L}c_{i,\ell}W_{\ell j}\bigr)^{2}
   \Bigr)_{j=1}^{p}\Bigr)
  =\bigl(U_{\precision},\,U_d\bigr),\\
  \phi(\mb{\theta})
  &=\Bigl(
   \Bigl(-\tfrac{g(X_\ell)}{4}\,
     \mb h_{\mb{\iota}}(X_\ell)^{-1}\,\precision\,\mb h_{\mb{\iota}}(X_\ell)^{-1}
   \Bigr)_{\ell=1}^{L},\;
   \Bigl(-\tfrac{1}{2d_j}\Bigr)_{j=1}^{p}
  \Bigr),\\
  \psi(\mb{\theta})
  &=\frac{n}{2}\log\lvert\mb d\rvert
   \;-\;\frac{L}{2}\log\det\precision
   \;+\;\sum_{\ell=1}^{L}\log\bigl\lvert\mb h_{\mb{\iota}}(X_\ell)\bigr\rvert,
\end{align*}
For a sufficient statistic $\suffstat$, let $\mathcal{L}_P(\mb\theta\mid u)$ be the complete-data log-likelihood with $U(\mb Y,\mb W)$ replaced by $u$. Parameter ${\mb{\iota}}$ enters $f(\mb Y,\mb W;\mb{\theta})$ only through $\mb h_{\mb{\iota}}(X_\ell)$ in $\phi_{\precision}$ and $\psi$, sharing $U_{\precision}$ and has no sufficient statistic of its own. Additionally, let $\bar{\suffstat}(\mb{\theta})=\E_\theta[U(\mb{Y},\mb{W})\mid \mb{Y}]$ be the conditional mean of the complete-data sufficient statistic given the observed data, and $  \overline U_m= \frac1B\sum_{b=1}^B U(\mb Y,\mb W_{m,b})$ its sample estimate. For notational simplicity, we write the block averages without burn-in. The same convergence arguments apply when a fixed number $B_0<B$ of initial sweeps is discarded and the averages are taken over the remaining $B-B_0$ states. 
Here, $\mb{\theta}'=\mathcal{T}(\suffstat,\mb{\theta})$ is the updated parameter after one iteration, $\mathcal{T}(u,\hat{\mb\theta}(u))=\hat{\mb\theta}(u)$. 

In the Gibbs algorithm, let $\pi_{\mb\theta}=p_g(w\mid\mb Y,\mb\theta)$ be the fixed-$L$ posterior distribution of the latent values, so $\bar{\suffstat}(\mb{\theta})=\pi_{\mb\theta}U$, and let $\mathsf P_{\mb\theta}$ denote the Markov kernel of one complete Gibbs scan targeting $\pi_{\mb\theta}$, so
\begin{align}
  \mb W_{m,0}
  &=\mb W_{m-1,B},\notag\\
  \mb W_{m,b}\mid\mathcal F_{m,b-1}
  &\sim
  \mathsf P_{\mb\theta^{(m-1)}}(\mb W_{m,b-1},\cdot),
  \qquad b=1,\ldots,B,
\end{align}
where $\{\mathcal F_{m,b}\}$ denote the filtration of the algorithm recording the initialization and algorithmic history through the $b$-th Gibbs sweep of iteration $m$. Write $\mathcal F_{m-1}=\mathcal F_{m,0}$. The SAECM recursion is
\begin{align}
  \suffstat^{(m)}
  &=(1-\gamma_m)\suffstat^{(m-1)}+\gamma_m\overline U_m,
  \notag\\
  \mb\theta^{(m)}
  &=\mathcal{T}(\suffstat^{(m)},\mb\theta^{(m-1)}).
\end{align}

\subsection{Proof of \texorpdfstring{\Cref{prop:set}}{prop2}}

\subsubsection{Assumptions}
Let $\Theta$ denote the open interior parameter space and let
$\mathcal U$ denote the open domain of admissible averaged sufficient
statistics. The latter is a domain for the SA averages. Individual raw
statistics $U(\mb Y,w)$ need not belong to $\mathcal U$. Define a graphical-lasso stratum to be a parameter region with fixed zeros and fixed signs of the nonzero precision entries, and the derivatives on a stratum use its free coordinates.

\begin{assumption}[Fixed-$L$ regularity]
\label{ass:model}
Assume $\psi$, $\phi$, and $\bar{\suffstat}$ are $C^2$ on $\Theta$, and $\mathcal{T}$ and $\hat{\mb\theta}$ corresponding to each graphical-LASSO stratum in $\Theta$ are $C^2$. For every compact $K\subset\Theta$,
$p_g(\mb Y,w;\cdot)$ is $C^1$, and
$$
  \int
  \sup_{\mb\theta\in K}
  \norm{\nabla_{\mb\theta}p_g(\mb Y,w;\mb\theta)}
  \,dw
  <\infty.
$$
Hence Fisher's identity applies.
\end{assumption}

\begin{assumption}[Moment bound for the Gibbs E step]
\label{ass:gibbs}
Let $\pi_{\mb\theta}(dw)=p_g(w\mid \mb Y;\mb\theta)\,dw$
denote the finite-$L$ posterior distribution of $W=(W_1^\top,\ldots,W_L^\top)^\top$, and let $\mathsf P_{\mb\theta}$ be the transition kernel obtained by one
complete Gibbs sweep, $\pi_{\mb\theta}\mathsf P_{\mb\theta}=\pi_{\mb\theta}.$ If one stochastic E step performs $B$ complete Gibbs sweeps, denote its
endpoint kernel by $ \Pi_{\mb\theta}=\mathsf P_{\mb\theta}^{B}.$

Set
$
  V(w)=1+\lVert w\rVert^2.
$
For every compact set $K\subseteq\Theta$, there exist constants
$\kappa_K\in(0,1)$ and $b_K<\infty$ such that
$\left(\Pi_{\mb\theta} V^2\right)(w) = \mathbb E_{{\mb{\theta}}}
\!\left[V(W^+)^2\mid W_0=w\right]
\le \kappa_KV(w)^2+b_K.$ The initial Gibbs state is deterministic or satisfies
$\mathbb E\!\left[V^2(W_{0,B})\right]<\infty.$
\end{assumption}

\begin{assumption}[Step sizes]
\label{ass:steps}
The gains satisfy $0<\gamma_m\leq1$, are nonincreasing, and
\begin{equation}
  \gamma_m\to0,\qquad
  \sum_m\gamma_m=\infty,\qquad
  \sum_m\gamma_m^2<\infty,\qquad
  \frac{\gamma_{m-q}}{\gamma_m}\to1
  \quad\text{for every fixed }q.
  \label{eq:gains}
\end{equation}
\end{assumption}

\begin{assumption}[Almost-sure compactness]
\label{ass:stability}
Almost surely, there exist compact sets $K_u\subset\mathcal U$ and
$K_\theta\subset\Theta$ such that
$$
  \overline{\{(u^{(m)},\mb\theta^{(m)}):m\geq0\}}
  \subseteq K_u\times K_\theta.
$$
\end{assumption}

\begin{assumption}[Finite-cycle contraction]
\label{ass:contraction}
The selected CM updates are exact, well-defined, and interior in their
respective free-coordinate domains. For every compact
$K_{u}\subset\mathcal U$ and $K_\theta\subset\Theta$, there
exist a compact $\widetilde K_\theta\subset\Theta$, $q\ge1$, and
$\kappa_{\mathcal{T}}\in(0,1)$ such that
$$
K_\theta\subset\widetilde K_\theta,
\qquad
\mathcal{T}(K_{u}\times\widetilde K_\theta)
\subset\widetilde K_\theta,
$$
and, writing $\mathcal{T}_\suffstat(\mb\theta)=\mathcal{T}(u,\mb\theta)$,
$$
\|\mathcal{T}_u^q(\mb\theta)-\hat{\mb\theta}(u)\|
\le
\kappa_{\mathcal{T}}\|\mb\theta-\hat{\mb\theta}(u)\|,
\qquad
(u,\mb\theta)\in
K_{u}\times\widetilde K_\theta.
$$
Moreover, $\mathcal{T}$, $\hat{\mb\theta}$, and $\bar{\suffstat}$ are Lipschitz on
$K_{u}\times\widetilde K_\theta$,
$K_{u}$, and $\widetilde K_\theta$, respectively, with
finite constants $C_{\mathcal{T}}$, $C_{\hat\theta}$, and $C_{\bar u}$.
\end{assumption}

\begin{assumption}[Eventual graphical-LASSO pattern stability]
\label{ass:stratum}
Let $\mathcal M_\zeta$ denote the parameter region with a fixed zero
pattern and fixed signs for the nonzero entries of $\mb{\sigma}^{-1}$, and let $\mathcal{L}_{\suffstat}:=\{\suffstat : \hat{\mb{\theta}}(\suffstat)\in\stationary\}$. For each stratum,
let $K_{\zeta,L}$ be a deterministic compact set contained in the
deterministic open set $\mathcal N_{\zeta,L}\subset\mathcal U$. Almost
surely, there exist $\zeta$ and $m_0<\infty$ such that
$$
  K_{\mathrm{tail},L}
  :=\overline{\{u^{(m)}:m\ge m_0\}}
  \subseteq K_{\zeta,L}
  \subset\mathcal N_{\zeta,L}.
$$
For every sufficiently large $m$, both $\mb\theta^{(m)}$ and
$\hat{\mb\theta}(u^{(m)})$ belong to $\mathcal M_\zeta$, and
$\hat{\mb\theta}(u)\in\mathcal M_\zeta$ throughout
$\mathcal N_{\zeta,L}$. On this neighborhood,
$V_P(u):=-\ell_{g,P}(\hat{\mb\theta}(u))$ is $C^2$, with bounded
first and second derivatives on a neighborhood of $K_{\zeta,L}$. Define $
  \mathcal Z_L
  :=K_{\mathrm{tail},L}\cap\mathcal L_{\suffstat}.
$
\end{assumption}

\begin{assumption}[Local curvature and inactive-coordinate optimality]
\label{ass:curvature}
Let $F$ denote the free coordinates of the eventual graphical LASSO
pattern. For every $u\in\mathcal N_{\zeta,L}$, the Hessian
$\mathcal H(u)
=
\nabla^2_{\mb\theta_F\mb\theta_F}
\mathcal{L}_P(\hat{\mb\theta}(u)\mid u)$
is negative semidefinite. Moreover, whenever
$\nabla_{\mb\theta_F}\ell_{g,P}(\hat{\mb\theta}(u))=0$,
every inactive off-diagonal precision entry satisfies
$\left|
\frac{\partial}{\partial\eta_{jk}}
\ell_g(\hat{\mb\theta}(u))
\right|
\le2\lambda.
$
\end{assumption}

\subsubsection{Lemmas}
The SAECM update is decomposed as follows:
\begin{equation}
  \suffstat^{(m)}
  =
  \suffstat^{(m-1)}
  +\gamma_m
  \{
    h(\suffstat^{(m-1)})
    +\varepsilon_m
    +R_m
  \},
  \label{eq:final-recursion}
\end{equation}
where $h(\suffstat)=\bar{\suffstat}(\hat{\mb\theta}(\suffstat))-\suffstat$ is the mean field, $\varepsilon_m=\overline U_m-\bar{\suffstat}(\mb\theta^{(m-1)})$ is the error term, and $R_m = \bar{\suffstat}(\mb\theta^{(m-1)})-
\bar{\suffstat}(\hat{\mb\theta}(\suffstat^{(m-1)}))$ is the tracking remainder. Then, \Cref{lem:tracking} shows the tracking remainder is summable, \Cref{lem:markov-noise} shows the noise can be controlled, and \Cref{lem:Lyapunov} provides the mean field descent identity. Together these imply convergence to the stationary set.

The proofs of Lemmas~\ref{lem:tracking} and~\ref{lem:markov-noise} rely on conditional-moment and martingale-difference bounds for the Gibbs noise. 
These bounds have deterministic constants only on a fixed compact set depending on the entire realized path by \cref{ass:stability}. 
Therefore, it is required to create a localized process on a countable deterministic compact exhaustion, which does not depend on the future event to avoid breaking the martingale-difference property.
Then, almost surely, such localization has no exit, and hence coincides with the original recursion.

Let $K_{\suffstat,j}\times K_{\theta,j}$, $j\geq1$, be a deterministic compact exhaustion of $\mathcal U\times\Theta$, with each member contained in the interior of the next.  
Define the exit time
$$
  \tau_j
  =
  \inf\{m\geq1:
    (\suffstat^{(m)},\mb\theta^{(m)})
    \notin K_{\suffstat,j}\times K_{\theta,j}\}.
$$
For each $j$, define a localized process that follows the original
recursion until $\tau_j$. 
After an attempted exit, hold $(\suffstat,\mb\theta)$ fixed and continue the Gibbs endpoints under the corresponding fixed kernel, so $(\suffstat,\mb\theta)$ remains in the compact set. 
When $\{\tau_j=\infty\}$, the extension coincides with the original recursion.
By \cref{ass:stability}, $
  \Prob\!\left(
    \bigcup_{j\geq1}
    \{(\suffstat^{(m)},\mb\theta^{(m)})
      \in K_{\suffstat,j}\times K_{\theta,j}
      \text{ for all }m\}
  \right)=1.
$
Thus, it is enough to prove the proposition for each localized process.

Fix a localization index $j$, write $\tau=\tau_j$, $K_{\suffstat}=K_{\suffstat,j}$ and $K=K_{\theta,j}$, and let all quantities in \Cref{lem:tracking,lem:markov-noise} refer to its localized process. 

\begin{lemma}[Localized moments and finite-cycle tracking]
\label{lem:tracking}
Define
\begin{equation*}
\begin{aligned}
\Delta_m&=\norm{\suffstat^{(m)}-\suffstat^{(m-1)}},\qquad
e_m^\tau=\mathbf 1_{\{m<\tau\}}
\norm{\mb\theta^{(m)}-\hat{\mb\theta}(\suffstat^{(m)})}.
\end{aligned}
\end{equation*}
Then, almost surely,
\begin{equation}
\begin{aligned}
&\Delta_m\to0,\qquad
\sum_m\gamma_m\Delta_m<\infty,\\
&e_m^\tau\to0,\qquad
\sum_m\gamma_m e_{m-1}^\tau<\infty.
\end{aligned}
\label{eq:small-increments}
\end{equation}
In addition,
\begin{equation}
\begin{aligned}
&\sup_m\E V^2(\mb W_{m,B})<\infty,\qquad
\max_{1\leq b\leq B}\sup_m
\E\norm{U(\mb Y,\mb W_{m,b})}^2<\infty,\\
&\sum_m\gamma_m
\norm{\mb\theta^{(m)}-\mb\theta^{(m-1)}}
V(\mb W_{m,B})<\infty
\quad\text{almost surely}.
\end{aligned}
\label{eq:weighted-control}
\end{equation}
\end{lemma}

\begin{proof}
Taking expectation over the past, the inequality in \cref{ass:gibbs} gives
$$
  \E\!\left[V^2(\mb W_{m,B})\right]
  \leq \kappa_K \E\left[V^2(\mb W_{m-1,B})\right]+b_K.
$$
Iterating this scalar inequality gives
$$
  \sup_m\E V^2(\mb W_{m,B})
  \leq
  \E V^2(\mb W_{0,B})+\frac{b_K}{1-\kappa_K}
  <\infty.
$$
The finite Gibbs steps have bounded coefficients and fourth moments,
so the second moment of $\overline U_m$ is bounded.
By Markov inequality and \cref{ass:steps},
$$
  \sum_m\Prob(\gamma_m\norm{\overline U_m}>\varepsilon)
  \leq
  \varepsilon^{-2}\sum_m\gamma_m^2
  \E\norm{\overline U_m}^2
  <\infty,
$$
and therefore $\gamma_m\norm{\overline U_m}\to0$ almost surely
by Borel--Cantelli lemma.
Next, we plug in the SAEM update into $\Delta_m$ and apply triangle inequality:
$$
  \Delta_m
  \leq\gamma_m\|\overline U_m-\suffstat^{(m-1)}\|
  \leq\gamma_m\|\overline U_m\|
  +\gamma_m\|\suffstat^{(m-1)}\|\to0,
$$
by localization and boundedness of $\suffstat^{(m-1)}$. Moreover,
$$
  \sum_m\gamma_m\Delta_m
  \leq
  \sum_m\gamma_m^2
  \left\{
    \norm{\overline U_m}
    +\sup_{u\in K_{\suffstat}}\norm u
  \right\}.
$$
The expectation of the right hand side is finite, proving
\eqref{eq:small-increments}.

Tracking the frozen-statistic CM fixed point.
Compare $q$ actual CM cycles with $q$ cycles at the frozen
statistic and use the Lipschitz bounds in \cref{ass:contraction}
$$
  e_{m+q}^\tau
  \leq\kappa_{\mathcal{T}} e_m^\tau
  +C\sum_{k=m+1}^{m+q}\Delta_k,
$$
Since $\Delta_m\to0$ and $\sum_m\gamma_m\Delta_m<\infty$,
$e_m^\tau\to0,\,\sum_m\gamma_m e_{m-1}^\tau<\infty$
by geometric summation.

Weighted parameter movement.
The final bound is needed in Lemma~2 because a change in the Gibbs kernel
is multiplied there by the possibly unbounded endpoint $\mb W_{m,B}$.
Since
$$
  \norm{\mb\theta^{(m)}-\mb\theta^{(m-1)}}
  \leq e_m^\tau+e_{m-1}^\tau+C_{\hat\theta}\Delta_m,
$$
square the tracking recursion and choose the Young-inequality constant so
that the coefficient of $(e_m^\tau)^2$ is strictly below one.
Using $\E\Delta_m^2=O(\gamma_m^2)$ and the fixed-lag gain condition
then gives, for some $r<1$,
$\E(e_m^\tau)^2=O(\gamma_m^2)+O(r^m)$.
Consequently,
$\E\norm{\mb\theta^{(m)}-\mb\theta^{(m-1)}}^2
=O(\gamma_m^2)+O(r^m)$.
By Cauchy--Schwarz and the uniform bound on
$\E V^2(\mb W_{m,B})$,
$$
\begin{aligned}
  \E\left[
    \norm{\mb\theta^{(m)}-\mb\theta^{(m-1)}}
    V(\mb W_{m,B})
  \right]
  &\leq
  \E\left[
    \norm{\mb\theta^{(m)}-\mb\theta^{(m-1)}}^2
  \right]^{1/2}
  \E\left[V^2(\mb W_{m,B})\right]^{1/2}\\
  &\leq C\{\gamma_m+r^{m/2}\}.
\end{aligned}
$$
Therefore,
$$
  \sum_m\gamma_m
  \E\left[
    \norm{\mb\theta^{(m)}-\mb\theta^{(m-1)}}
    V(\mb W_{m,B})
  \right]
  <\infty.
$$
Tonelli's theorem now yields
$$
  \sum_m\gamma_m
  \norm{\mb\theta^{(m)}-\mb\theta^{(m-1)}}
  V(\mb W_{m,B})
  <\infty,
  \qquad\text{almost surely}.
$$
\end{proof}

\begin{lemma}[Persistent-Gibbs noise control]
\label{lem:markov-noise}
Almost surely,
\begin{align}
  \sum_m\gamma_m\varepsilon_m
  &\quad\text{converges},
  \label{eq:noise-convergence}\\
  \sum_m\gamma_m^2\norm{\varepsilon_m}^2
  &<\infty.
  \label{eq:noise-squares}
\end{align}
Moreover, for every bounded globally Lipschitz matrix function $A$,
\begin{equation}
  \sum_m\gamma_m A(\suffstat^{(m-1)})\varepsilon_m
  \quad\text{converges almost surely}.
  \label{eq:predictable-noise}
\end{equation}
\end{lemma}

\begin{proof}
Starting a retained block at $W_0=w$, let $\mu_{\mb\theta}=\mathbb E_{\pi_{\mb\theta}}[\mb W]$ denote the posterior mean. Define
$$
  \mathcal{B}_{\mb\theta}(w)
  =
  \E_{\mb\theta}\left[
    \frac1B\sum_{b=1}^B U(\mb Y,W_b)\mid W_0=w
  \right]
  -\bar{\suffstat}(\mb\theta),
$$
which is a quadratic polynomial. If we let $z=w-\mu_{\mb\theta}$,
then one scalar component can be written as
$\mathcal{B}_{\mb\theta}(w)
=c_{\mb\theta}+b_{\mb\theta}^{\top}z
+z^{\top}A_{\mb\theta}z$.
The endpoint of the Gibbs block satisfies
$$
  W_B-\mu_{\mb\theta}
  =J_{\mb\theta}z+\epsilon_{\mb\theta},
  \qquad
  \epsilon_{\mb\theta}\sim \mathcal{N}(0,\Omega_{\mb\theta}),
$$
where $J_{\mb\theta}=G_{\mb\theta}^{B}$ and $G_{\mb\theta}$
is the affine transition of conditional Gaussian.
Since the transition maps quadratic functions to quadratic functions,
we define
$\chi_{\mb\theta}(w)
=\beta_{\mb\theta}^{\top}z+z^{\top}S_{\mb\theta}z$
satisfying the Poisson equation
\begin{equation}
  \chi_{\mb\theta}-\Pi_{\mb\theta}\chi_{\mb\theta}
  =\mathcal{B}_{\mb\theta}.
  \label{eq:poisson-equation}
\end{equation}
Matching the linear and quadratic coefficients gives
$$
  (I-J_{\mb\theta})^\top\beta_{\mb\theta}=b_{\mb\theta},
  \qquad
  S_{\mb\theta}-J_{\mb\theta}^{\top}S_{\mb\theta}J_{\mb\theta}
  =A_{\mb\theta}.
$$
Substituting $w=\mu_{\mb\theta}+tv$ into the drift inequality in
\cref{ass:gibbs}, dividing by $t^4$, and letting $t\to\infty$
give $\norm{J_{\mb\theta}}_{\mathrm{op}}\leq\kappa_K^{1/4}<1$,
so the two coefficient equations have unique solutions. Here, $\|\cdot\|_{\mathrm{op}}$ denotes the operator norm induced by the Euclidean norm. The constant term agrees because $\mathcal{B}_{\mb\theta}$ is centered,
implying
$c_{\mb\theta}=-\operatorname{tr}(S_{\mb\theta}\Omega_{\mb\theta})$.

For a vector-valued function $f$, define
$\norm{f}_V=\sup_w\norm{f(w)}/V(w)$.
$\mu_{\mb\theta}$, $J_{\mb\theta}$, $b_{\mb\theta}$, and
$A_{\mb\theta}$ are Lipschitz in $\mb\theta$ on the localization compact.
Since $\sup_{\mb\theta}\norm{J_{\mb\theta}}_{\mathrm{op}}<1$,
$\beta_{\mb\theta}$ and $S_{\mb\theta}$ are uniformly bounded and Lipschitz.
Since $V$ dominates quadratic growth, there is a uniform
$C_\chi<\infty$ such that
\begin{align}
  \norm{\chi_{\mb\theta}}_V
  +\norm{\Pi_{\mb\theta}\chi_{\mb\theta}}_V
  &\leq C_\chi,
  \label{eq:g-bound}\\
  \norm{\chi_{\mb\theta}-\chi_{\mb\vartheta}}_V
  &\leq C_\chi\norm{\mb\theta-\mb\vartheta}.
  \label{eq:g-Lipschitz}
\end{align}
The vector-valued result follows by applying this construction component-wise.

Let $\chi_m=\chi_{\mb\theta^{(m-1)}}$,
\begin{equation}
\begin{aligned}
  \varepsilon_m
  &=
  \underbrace{
    D_m+\chi_m(\mb W_{m,B})
    -\Pi_{\mb\theta^{(m-1)}}\chi_m(\mb W_{m-1,B})
  }_{M_m}+
  \underbrace{
    \chi_m(\mb W_{m-1,B})-\chi_m(\mb W_{m,B})
  }_{C_m},
\end{aligned}
\label{eq:noise-decomposition}
\end{equation}
where
$D_m=\overline U_m-\E[\overline U_m\mid\mathcal F_{m-1}]$.
The first term $M_m$ is a martingale difference.
The uniform bound $\sup_m\E\norm{\overline U_m}^2<\infty$,
together with \eqref{eq:g-bound}, gives it a uniformly bounded second moment.
Hence $\sum_m\gamma_mM_m$ converges almost surely
by the martingale convergence theorem.
The same series also converges after multiplying a bounded predictable
matrix $A(\suffstat^{(m-1)})$.

To handle $C_m$, we let $B_m=\gamma_m A(\suffstat^{(m-1)})$.
Summation by parts yields
\begin{align}
  \sum_{m=1}^{N}B_m C_m
  &=
  B_1\chi_1(W_{0,B})-B_N\chi_N(W_{N,B})\notag\\
  &\quad+
  \sum_{m=1}^{N-1}
  \left\{
    B_{m+1}\chi_{m+1}(\mb W_{m,B})
    -B_m\chi_m(\mb W_{m,B})
  \right\}.
  \label{eq:summation-by-parts}
\end{align}
$B_N\chi_N(W_{N,B})\to0$ since $B_N$ is bounded
and by \eqref{eq:g-bound},
$\norm{B_N\chi_N(W_{N,B})}\leq C\gamma_NV(W_{N,B})$,
and $\sum_N\gamma_N^2\E V^2(W_{N,B})<\infty$,
then Borel--Cantelli applies.

For the internal difference in \eqref{eq:summation-by-parts},
boundedness and Lipschitz continuity of $A$, together with
\eqref{eq:g-Lipschitz}, give
$$
\begin{aligned}
  &\norm{
    B_{m+1}\chi_{m+1}(\mb W_{m,B})
    -B_m\chi_m(\mb W_{m,B})
  }\\
  &\qquad\leq
  C\Big[
    |\gamma_{m+1}-\gamma_m|
    +\gamma_m\Delta_m
    +\gamma_m\norm{\mb\theta^{(m)}-\mb\theta^{(m-1)}}
  \Big]V(\mb W_{m,B}).
\end{aligned}
$$
The first series converges since $\gamma_m$ is nonincreasing
and $\E V(W_{m,B})$ is bounded,
$$
\begin{aligned}
  \E\sum_m|\gamma_{m+1}-\gamma_m|V(W_{m,B})
  &\leq
  \sup_m\E V(W_{m,B})
  \sum_m(\gamma_m-\gamma_{m+1})\\
  &=\gamma_1\sup_m\E V(W_{m,B})<\infty.
\end{aligned}
$$
The second series converges by applying Cauchy--Schwarz
and the result of Lemma~1.
$$
\begin{aligned}
  \E\sum_m\gamma_m\Delta_m V(W_{m,B})
  &\leq
  \sum_m\gamma_m
  \bigl(\E\Delta_m^2\bigr)^{1/2}
  \bigl(\E V^2(W_{m,B})\bigr)^{1/2}\\
  &\leq C\sum_m\gamma_m^2<\infty.
\end{aligned}
$$
The third series is finite almost surely by \eqref{eq:weighted-control}.
Hence
$$
  \sum_m
  \norm{
    B_{m+1}\chi_{m+1}(W_{m,B})
    -B_m\chi_m(W_{m,B})
  }
  <\infty.
$$
Together with $B_N\chi_N(W_{N,B})\to0$,
we have $\sum_m B_mC_m$ converges almost surely,
which proves \eqref{eq:noise-convergence}
and \eqref{eq:predictable-noise}.

Finally, \Cref{lem:tracking} and the boundedness of
$\bar{\suffstat}$ on the localization compact give
$$
  \sup_m\E\norm{\varepsilon_m}^2
  \leq
  2\sup_m\E\norm{\overline U_m}^2
  +2\sup_{\mb\theta\in K}\norm{\bar{\suffstat}(\mb\theta)}^2
  <\infty.
$$
Consequently,
$$
  \E\left[\sum_m\gamma_m^2\norm{\varepsilon_m}^2\right]
  =
  \sum_m\gamma_m^2\E\norm{\varepsilon_m}^2
  \leq C\sum_m\gamma_m^2<\infty,
$$
so, $\sum_m\gamma_m^2\norm{\varepsilon_m}^2<\infty$ almost surely,
proving \eqref{eq:noise-squares}.
\end{proof}

\begin{lemma}[Lyapunov calculation]
\label{lem:Lyapunov}    
On the stabilized stratum,
\begin{equation}
  \inner{\nabla V_P(\suffstat)}{h(\suffstat)}
  \leq0,
  \label{eq:Lyapunov-inequality}
\end{equation}
with equality if and only if
$\hat{\mb\theta}(\suffstat)\in\stationary$.
\end{lemma}

\begin{proof}
Work in the free coordinates $F$. On the identified stratum, every
inactive coordinate of $\hat{\mb\theta}(\suffstat)$ is identically zero,
so the composite $V_P=-\ellgP\circ\hat{\mb\theta}$ is differentiated
only through the free coordinates. Let
$B(\suffstat)=D_{\suffstat}\hat{\mb\theta}_F(\suffstat)$,
$G(\suffstat)=D_{\mb\theta_F}\phi(\hat{\mb\theta}(\suffstat))$.
The fixed-point condition is
\begin{equation}
  -\nabla_{\mb\theta_F}\psi(\hat{\mb\theta})
  +G(\suffstat)^\top\suffstat
  -\nabla_{\mb\theta_F}P(\hat{\mb\theta})
  =0.
  \label{eq:CM-FOC}
\end{equation}
Differentiate in $\suffstat$,
\begin{equation}
  \mathcal H(\suffstat)B(\suffstat)
  +G(\suffstat)^\top
  =0.
  \label{eq:HB}
\end{equation}
The Fisher identity and \eqref{eq:CM-FOC} give
\begin{equation}
  \nabla_{\mb\theta_F}\ellgP(\hat{\mb\theta})
  =G(\suffstat)^\top h(\suffstat).
  \label{eq:observed-score}
\end{equation}
Consequently,
$\nabla V_P(\suffstat)
=-B^\top\nabla_{\mb\theta_F}\ellgP(\hat{\mb\theta})
=B^\top\mathcal H B h$,
and hence
$$
  \inner{\nabla V_P(\suffstat)}{h(\suffstat)}
  =(Bh)^\top\mathcal H(Bh)\leq0.
$$
As $\mathcal H$ is symmetric and negative semidefinite, equality
holds if and only if $\mathcal H Bh=0$. By \eqref{eq:HB}, this is
equivalent to $G^\top h=0$, and hence
$\nabla_{\mb\theta_F}\ell_{g,P}(\hat{\mb\theta})=0$
by \eqref{eq:observed-score}.
The inactive-coordinate condition in \cref{ass:curvature}
then gives full Clarke stationarity.
Conversely, full Clarke stationarity gives $G^\top h=0$,
hence $\mathcal H Bh=0$, and therefore equality.
\end{proof}

\subsubsection{Final proof}

\begin{proof}[Proof of \Cref{prop:set}]
Since $\nabla V_P^\top$ is bounded and Lipschitz on
$K_{\zeta,L}$, let $A_\zeta$ be a bounded, globally Lipschitz
extension that agrees with $\nabla V_P^\top$ on $K_{\zeta,L}$.
Apply \Cref{lem:tracking,lem:markov-noise} to every deterministic
localization and stratum. Since there are countably many localizations
and finitely many strata, the resulting estimates hold on a common
probability-one event.

Fix a realization in this event. By \cref{ass:stability}, some
localization has $\tau=\infty$, and by \cref{ass:stratum}, the
trajectory eventually remains in some stratum $\zeta$. For such
localization, the original and localized process agree at every
iteration.

On the stabilized tail, write the recursion as
\eqref{eq:final-recursion}. By \Cref{lem:tracking},
$$
  \sum_m\gamma_m\norm{R_m}
  \leq C_{\bar u}\sum_m\gamma_m e_{m-1}^{\tau}
  <\infty.
$$
By \Cref{lem:markov-noise},
$\sum_m\gamma_m\varepsilon_m$ converges,
$\sum_m\gamma_m^2\norm{\varepsilon_m}^2<\infty$.
Applying the weighted conclusion of that lemma to $A_\zeta$, which
equals $\nabla V_P^\top$ on $K_{\zeta,L}$, gives
$\sum_m\gamma_m
\nabla V_P(\suffstat^{(m-1)})^\top\varepsilon_m$ converges.
By \Cref{lem:tracking},
$\norm{\suffstat^{(m)}-\suffstat^{(m-1)}}\to0$,
so the Taylor-expansion segments eventually remain in the region
where $V_P$ is $C^2$. Define
$q(u)=-\inner{\nabla V_P(u)}{h(u)}$.
By \Cref{lem:Lyapunov}, $q\geq0$, and its zero set on
$K_{\mathrm{tail},L}$ is $\mathcal Z_L$.
Taylor expansion of $V_P$ along \eqref{eq:final-recursion} gives
$$
\begin{aligned}
  V_P(\suffstat^{(m)})-V_P(\suffstat^{(m-1)})
  &=-\gamma_m q(\suffstat^{(m-1)})\\
  &\quad+\gamma_m
  \nabla V_P(\suffstat^{(m-1)})^\top\varepsilon_m
  +\gamma_m
  \nabla V_P(\suffstat^{(m-1)})^\top R_m\\
  &\quad+O\!\left(
    \gamma_m^2
    \{1+\norm{\varepsilon_m}^2+\norm{R_m}^2\}
  \right).
\end{aligned}
$$
The preceding bounds make every perturbation series on the right
convergent. Since $V_P$ is bounded on the compact trajectory,
$\sum_m\gamma_mq(\suffstat^{(m-1)})<\infty$.
This, together with $\sum_m\gamma_m=\infty$ implies
$\liminf_m q(\suffstat^{(m)})=0$.
If $q(\suffstat^{(m)})$ did not converge to zero, there would be infinitely many crossings between two fixed positive levels of $q$, each contributing a fixed positive amount to the sum by uniform continuity of $q$, boundedness of $h$, and the preceding noise bounds, contradicting its finiteness. Hence $q(\suffstat^{(m)})\to0$. By compactness and continuity,
every limit point belongs to $\mathcal Z_L$, so
$\mathcal Z_L\neq\varnothing$ and $\dist(\suffstat^{(m)},\mathcal Z_L)\to0$.

For every $u_\star\in\mathcal Z_L$,
$\hat{\mb\theta}(u_\star)\in\stationary$.
The Lipschitz property of $\hat{\mb\theta}$ gives
$$
  \dist\!\left(
    \hat{\mb\theta}(\suffstat^{(m)}),\stationary
  \right)
  \leq C_{\hat\theta}
  \dist(\suffstat^{(m)},\mathcal Z_L)
  \to0,
$$
proving \eqref{eq:result-hat}. Finally,
$$
  \dist(\mb\theta^{(m)},\stationary)
  \leq e_m^\tau
  +\dist\!\left(
    \hat{\mb\theta}(\suffstat^{(m)}),\stationary
  \right)
  \to0,
$$
by \Cref{lem:tracking}, proving \eqref{eq:result-theta}.
Since the preceding choices exist on a probability-one event,
the proposition follows.
\end{proof}

\subsection{Proof of \texorpdfstring{\Cref{prop:diagonal_convergence}}{prop3}}
\label{sec:rff_convergence}

\subsubsection{Additional assumptions}

\setcounter{assumption}{7}
\begin{assumption}[Uniform-in-$L$ random Fourier features (RFF) regularity near $K^\star$]
\label{ass:rff}
All values of $L$ use the first $L$ terms of one infinite i.i.d.
sequence $\{\xi_\ell\}_{\ell\geq1}$, $\xi_\ell=(X_\ell,\psi_\ell)$.
There are deterministic compact convex sets $K^\star$ and $K^+$ satisfying $K^\star\subseteq\operatorname{int}(K^+)$ and $K^+\subseteq\operatorname{int}(\Theta)$, where $\operatorname{int}(S)$ denotes the interior of $S$, with the following properties.
\begin{enumerate}[label=(\arabic*),leftmargin=2em]
    \item The spectral matrix is real, symmetric, and even by construction. For $X\sim g$, the map $\mb{\theta}\mapsto\mb{f}(X;\mb{\theta})$ is $C^2$ on a neighborhood of $K^+$. The penalty is the same fixed function $P$, with the same coefficient $\lambda>0$, for every $L$ and for the exact target-GP likelihood.
  \item Let $\mathcal{K}_{ij}((x,\psi);\mb{\theta})=\frac{2\cos(s_i^\top x+\psi)\cos(s_j^\top x+\psi)}{g(x)}\,\mb{f}(x;\mb{\theta}).$ There is a nonnegative random variable $M(\xi_1)$, with
  $\E M(\xi_1)<\infty$, such that for every $i,j$,
  \begin{align*}
    \sup_{\mb\theta\in K^+}
    \bigl\{
      \norm{\mathcal{K}_{ij}(\xi_1;\mb\theta)}
      +
      \norm{\nabla_{\mb\theta}
        \mathcal{K}_{ij}(\xi_1;\mb\theta)}
      +
      \norm{\nabla_{\mb\theta}^2
        \mathcal{K}_{ij}(\xi_1;\mb\theta)}
    \bigr\}
    \leq M(\xi_1),
  \end{align*}
  almost surely.

  \item The nugget map $\mb D(\mb\theta)$ is $C^1$ on a neighborhood of
    $K^\star$, and there exists $d_0>0$ such that
    $$
      \mb D(\mb\theta)\succeq d_0\mb I_{np},
      \qquad
      \mb\theta\in K^\star.
    $$
\end{enumerate}
\end{assumption}
\begin{assumption}[Uniform-in-$L$ containment of stationary sets]
\label{ass:common-compact}
For the deterministic compact set $K^\star$ in \cref{ass:rff},
$$
  \varnothing\neq\Lstar\subset K^\star,
$$
and
$$
  \Prob\!\left\{
    \exists L_0<\infty\ \text{such that}\
    \varnothing\neq\stationaryL\subset K^\star
    \ \text{for every }L\geq L_0
  \right\}=1.
$$
Thus $K^\star$ contains the exact stationary set and, almost surely,
all sufficiently large-$L$ RFF stationary sets. 
\end{assumption}

\subsubsection{Lemmas}

\begin{lemma}[Uniform convergence of RFF scores]
\label{lem:rff-uniform}
Under \cref{ass:rff}, almost surely,
\begin{equation}
  \sup_{\mb\theta\in K^\star}
  \norm{
    \nabla_{\mb\theta}\lgL(\mb\theta)
    -
    \nabla_{\mb\theta}\ell(\mb\theta)}
  \longrightarrow0.
  \label{eq:score-uniform}
\end{equation}
\end{lemma}

\begin{proof}
We first establish the uniform law used below. Let
$a(z,\mb\theta)$ be a finite-dimensional vector or matrix that is
continuously differentiable in $\mb\theta$ on a neighborhood of
$K^\star$, and suppose
$$
  \sup_{\mb\theta\in K^\star}
  \left\{
    \norm{a(\xi_1,\mb\theta)}
    +\norm{\nabla_{\mb\theta}a(\xi_1,\mb\theta)}
  \right\}
  \leq M(\xi_1),
  \qquad
  \E M(\xi_1)<\infty.
$$
Then,
\begin{equation}
  \sup_{\mb\theta\in K^\star}
  \left\lVert
    \frac1L\sum_{\ell=1}^L a(\xi_\ell,\mb\theta)
    -\E a(\xi_1,\mb\theta)
  \right\rVert
  \longrightarrow0,
  \quad\text{almost surely}.
  \label{eq:uniform-average}
\end{equation}
For a finite $\delta$-net
$\{\mb\theta_1,\ldots,\mb\theta_N\}$ of $K^\star$,
the mean-value theorem and the triangle inequality give:
$$
\begin{aligned}
  &\sup_{\mb\theta\in K^\star}
  \left\lVert
    \frac1L\sum_{\ell=1}^L a(\xi_\ell,\mb\theta)
    -\E a(\xi_1,\mb\theta)
  \right\rVert\\
  &\qquad\leq
  \max_{1\leq r\leq N}
  \left\lVert
    \frac1L\sum_{\ell=1}^L a(\xi_\ell,\mb\theta_r)
    -\E a(\xi_1,\mb\theta_r)
  \right\rVert
  +\delta\left\{
    \frac1L\sum_{\ell=1}^L M(\xi_\ell)+\E M(\xi_1)
  \right\}.
\end{aligned}
$$
The strong law of large numbers, followed by $\delta\downarrow0$,
proves \eqref{eq:uniform-average}.

Applying \eqref{eq:uniform-average} to $\mathcal{K}_{ij}$ and its first
derivatives gives uniform convergence of the covariance matrices and
their first derivatives. The nugget lower bound keeps
$(\mb{\Sigma}_L+\mb{D})(\mb\theta)$ and
$(\mb{\Sigma}+\mb{D})(\mb\theta)$ positive definite on $K^\star$.
Because the matrix inversion and the log-determinant are uniformly
continuous, the uniform convergence of the covariance matrices and
their derivatives implies uniform convergence of the Gaussian
log-likelihoods and their scores.
\end{proof}

\begin{lemma}[Outer convergence of stationary sets]
\label{lem:stationary-set}
Under \cref{ass:rff,ass:common-compact} and the fixed common-penalty
convention above,
\begin{equation}
  \Delta_L
  :=
  \begin{cases}
    \displaystyle
    \sup_{\mb\theta\in\stationaryL}
    \dist(\mb\theta,\Lstar),
    &\stationaryL\neq\varnothing,\\[4pt]
    0,&\stationaryL=\varnothing,
  \end{cases}
  \longrightarrow0
  \quad\text{almost surely}.
  \label{eq:outer-convergence}
\end{equation}
\end{lemma}

\begin{proof}
Focus on a probability-one event on which
\eqref{eq:score-uniform} and \cref{ass:common-compact} hold.
If \eqref{eq:outer-convergence} failed, there exists
$\varepsilon>0$, $L_j\to\infty$, and
$\mb\theta_j\in\Theta_{g,L_j,P}^\star$ such that
$\dist(\mb\theta_j,\Lstar)\geq\varepsilon$.
By compactness, there exists a subsequence $\mb\theta_{j_k}$
converging to $\mb\theta_\infty\in K^\star$.
We relabel $(L_{j_k},\mb{\theta}_{j_k})$ as
$(L_j,\mb{\theta}_j)$.
The stationary condition states that there are some
$z_j\in\sd P(\mb\theta_j)$, so
$z_j=\nabla\ell_{g,L_j}(\mb\theta_j)$.
Uniform score convergence and continuity of $\nabla\ell$ imply
$z_j\to\nabla\ell(\mb\theta_\infty)$.
Since $P$ is continuous and convex, passing to the limit in
$z_j\in\sd P(\mb\theta_j)$ gives
$\nabla\ell(\mb\theta_\infty)\in\sd P(\mb\theta_\infty)$.
Therefore, $\mb\theta_\infty\in\Lstar$ contradicts
$\dist(\mb\theta_j,\Lstar)\geq\varepsilon$, and proves the lemma.
\end{proof}

\subsubsection{Final proof}
\begin{proof}
For each $L$, let $G_L$ be the random-feature event on which \cref{ass:model,ass:gibbs,ass:steps,ass:stability,ass:contraction,ass:stratum,ass:curvature} hold. By hypothesis, almost surely $G_L$ holds for all sufficiently large $L$, and on $G_L$, \Cref{prop:set} ensures $\stationaryL\neq\varnothing$. Set $D_{L,m}=0$ off $G_L$, and on $G_L$ let

$$
D_{L,m}
=
\max\left\{
\dist\!\left(
  \hat{\mb\theta}_L(\suffstat_L^{(m)}),\stationaryL
\right),
\dist\!\left(
  \mb\theta_L^{(m)},\stationaryL
\right)
\right\}.
$$
By \Cref{lem:stationary-set}, $\Delta_L\to0$ almost surely and hence in
probability. For every fixed $L$, \Cref{prop:set} on $G_L$ gives $D_{L,m}\to0$ conditionally almost surely given the random features and
therefore unconditionally in probability by bounded convergence.

We construct deterministic $L_k\to\infty$ and $m_k\to\infty$ to combine
these two limits. Choose them increasingly so that
$$
  \Prob(\Delta_{L_k}>k^{-1})<2^{-k-1},
  \qquad
  \Prob(D_{L_k,m_k}>k^{-1})<2^{-k-1}.
$$
Then
$\sum_{k=1}^\infty
\Prob\!\left(\Delta_{L_k}+D_{L_k,m_k}>2k^{-1}\right)<\infty$.
The Borel--Cantelli lemma therefore gives
$\Delta_{L_k}+D_{L_k,m_k}\longrightarrow0$ almost surely. Since $G_{L_k}$ holds eventually, on it
$$
  \dist(x,\Lstar)
  \leq\dist(x,\stationaryL)+\Delta_L,
$$
applying the inequality with
$x=\hat{\mb\theta}_{L_k}(\suffstat_{L_k}^{(m_k)})$
and $x=\mb\theta_{L_k}^{(m_k)}$
proves \eqref{eq:proposition-conclusion} and
\eqref{eq:actual-proposition-conclusion}.
\end{proof}

\section{Additional Technical Details and Simulation Results}\label{app:sim}
We provide additional implementation details and numerical results in this section.

\subsection{Implementation details}
We first give additional details on the implementation of our simulation study. For algorithm initialization, one may simply set $\mb{\sigma}^{-1}$ to a diagonal matrix of inverse empirical variances or, when processes are colocated, use $[(1/n) \sum_{i=1}^n \mb{Y}(s_i)\mb{Y}(s_i)^\top]^{-1}$ as a sensible but crude initialization. The spatial parameters governing $\{f_j(x)\}_{j=1}^p$ and the nugget parameter $\mb{d}$ may be initialized using independent-process fits. 
In addition, there are a number of tuning parameters and choices in the algorithm. 

Under the SAECM framework, we found that estimates were generally robust to the number of Monte Carlo samples $T$, and increasing $L$ generally results in better accuracy up to a point, at the expense of computation. Therefore, we use different numbers of random features and Monte Carlo sample sizes for estimation ($L=20{,}000,\,T=30$), likelihood evaluation ($L_{\text{ll}}=1{,}000$), and prediction ($L_{\text{pred}}=50{,}000,\,T_{\text{pred}}=10{,}000$). 
The first $T/2$ sweeps are discarded as burn-in.
We take $\gamma_m = 200/(199+m)$ and conduct 15 replicates, each with $500$ SAECM steps or until the change in the Monte Carlo complete-data penalized log-likelihood falls below $10^{-3}$. 
For choice of the proposal distribution $g(\cdot)$, we refer the reader to \cite{arroyo_algorithm_2021}. 
Conceptually, $g(\cdot)$ should have heavier tails than each of the diagonal elements in $\mb{f}(x)$, so we choose a Mat\'ern spectral density as the proposal with $a_{\text{prop}}=3$ and $\nu_{\text{prop}}=1.25$ as the parameters. 
We consider the penalty grid $\{0.0003, 0.0006, 0.001, 0.003, 0.006, 0.012, 0.025, 0.05, 0.1\}$, together with an unpenalized dense fit.
BIC uses a sample size of $n=500$ for colocated observations and $n=300$ for misaligned observations.

\subsection{Benchmark Implementation Details}
\label{app:benchmark}
We describe the detailed steps of LMC and IOX in the following.
First, the loading matrix is estimated from the sample covariance after subtracting the pilot nugget, which is estimated from the marginal variogram. 
Then, the largest $q$ eigenvalues are retained, while the rest are set to zero. 
For each eigenvector, we project the data onto its direction and fit a univariate Gaussian process under each of the three covariance families.
The family for each process is selected using AIC. 
Finally, given the truncated loading matrix and the selected family for each latent process, maximum likelihood estimation is used to estimate the spatial parameters and a shared nugget.
In the misaligned setting, the loading matrix is formed by calculating the pairwise covariances between processes using overlapping locations and is then truncated to rank $q$. 
Because the misaligned data cannot be projected onto the eigenspace, we perform weighted least squares to regress the covariogram on the eigenvectors and estimate the correlogram of each latent process. 
Each covariance function is then fitted to the estimated correlogram, and the best is selected by AIC, subject to the constraint that each family is represented by at least one latent process.
Once the covariance functions are determined, maximum likelihood estimation can be performed. 

The IOX approach is implemented using the \texttt{spiox} package (version 0.8.1) from its GitHub repository. 
As in \citet{peruzzi2024}, we run $10{,}000$ MCMC iterations. The first $7{,}500$ are discarded as burn-in. 
Starting values are obtained from \texttt{spiox::autostart()}. The number of Vecchia neighbors is set to $15$.
We construct predictive intervals from the 2.5th and 97.5th percentiles of the posterior predictive draws. The upper bound of $\nu$ is raised to $2.9$ to cover the smoothness values in our simulation settings.

\subsection{Sensitivity analysis and convergence diagnostics}
\begin{figure}[!t]
    \centering
    \includegraphics[width=0.48\linewidth]{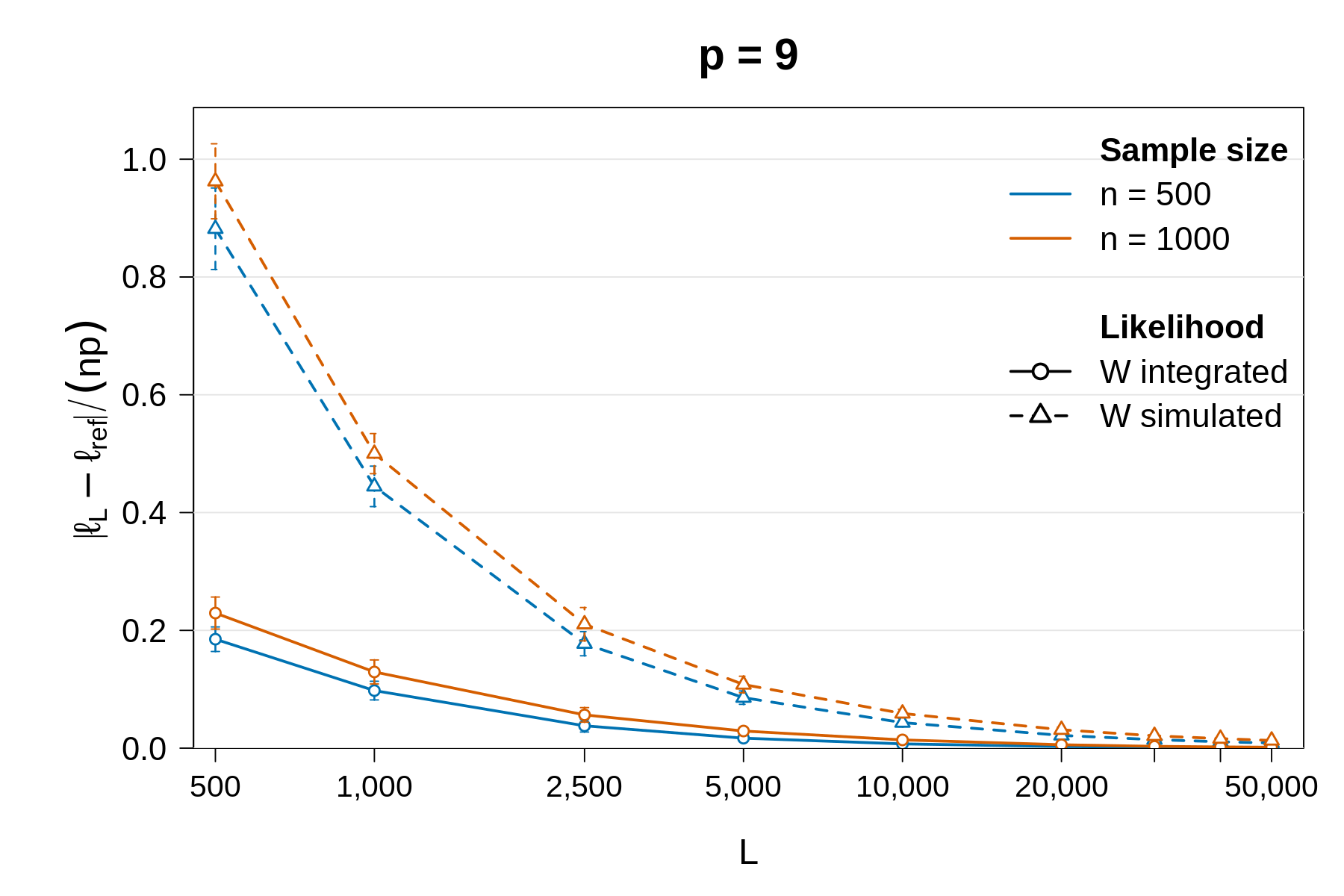}
    \includegraphics[width=0.48\linewidth]{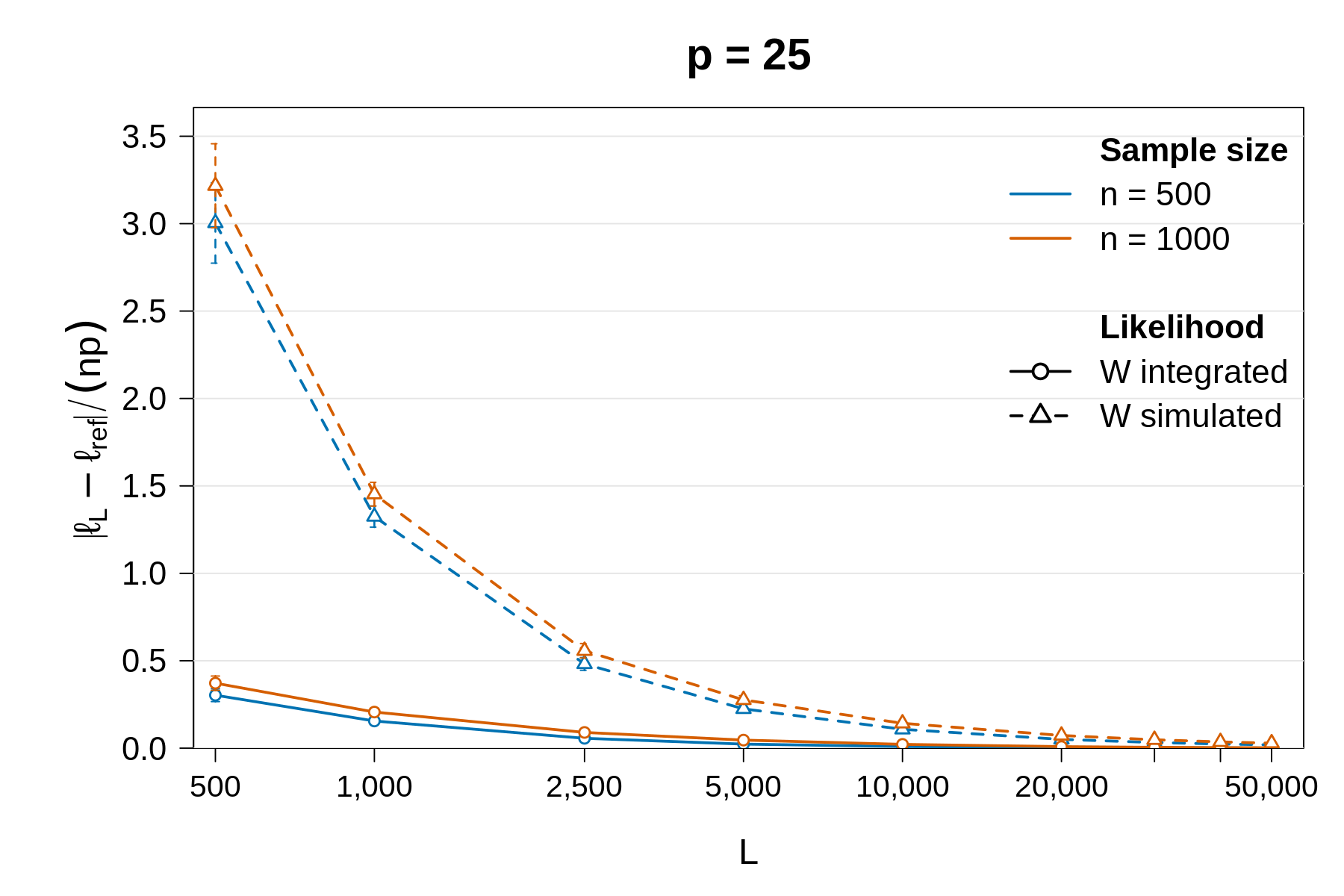}
    \includegraphics[width=0.48\linewidth]{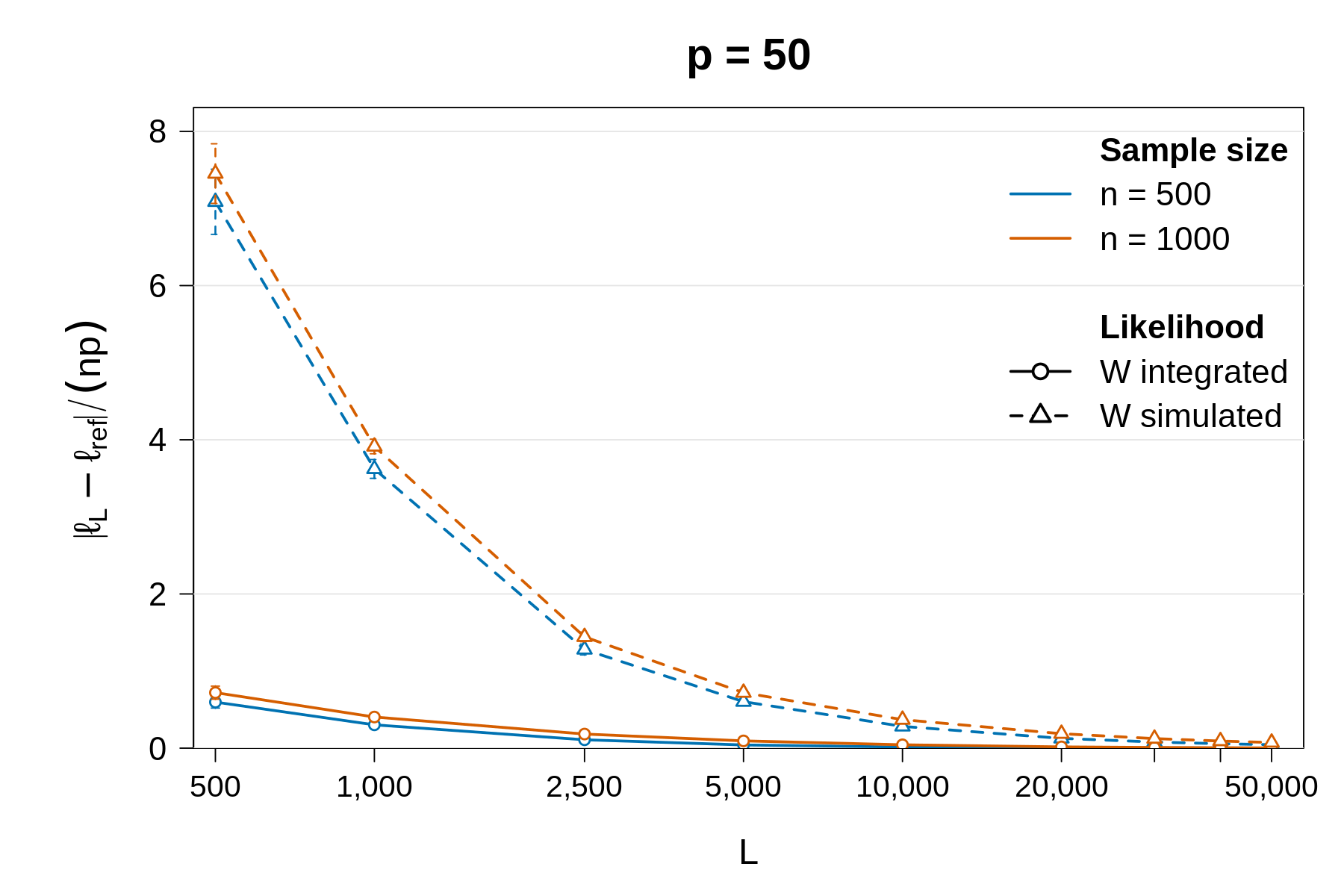}
    \caption{Mean absolute log-likelihood error divided by $np$ for mixed-family simulations with tridiagonal precision and negative off-diagonal entries, across different values of $n$, $L$, and $p$. Results are averaged across 15 replicates. Error bars indicate $\pm1$ SD.}
    \label{fig:L_sensitivity_mix}
\end{figure}

We provide a sensitivity analysis of $L$ and a convergence diagnostic of the simulation in this section. In the sensitivity analysis, we simulate data under different $n$ and $p$ and compare the random-feature approximated likelihood with a reference log-likelihood based on $100{,}000$ features for the mixed-family process. We use tridiagonal precision with negative off-diagonal entries and otherwise the same parameters as in the main-text simulations. The results are shown in Figure~\ref{fig:L_sensitivity_mix}.
The $y$-axis shows the absolute difference between each approximate likelihood and its reference likelihood, divided by $np$, and the $x$-axis shows $L$. We compare both the marginal likelihood obtained by integrating out $\mb{W}$, denoted by $\ell_{L,\textrm{int}}$, and the approximate likelihood calculated using Algorithm~\ref{sim:likeli}, denoted by $\ell_{L,\textrm{sim}}$. The plots show similar patterns across different $p$ and $n$. The absolute error of $\ell_{L,\textrm{int}}$ is smaller than that of $\ell_{L,\textrm{sim}}$ when $L$ is small. Above $L=20{,}000$, the error of $\ell_{L,\textrm{int}}$ levels off, showing limited improvement from increasing $L$, while the error of $\ell_{L,\textrm{sim}}$ continues to decrease slightly. These results support our choice of $L=20{,}000$ for estimation.

\begin{figure}[H]
\centering\tiny
\begin{tabular}{ccccc}
& \textbf{Penalized observed ll}
& \textbf{$\mb{\sigma}^{-1}$ Frobenius norm}
& \textbf{CH microergodic}
& \textbf{Nugget} \\
\simrowlabel{Positive colocated} &
\includegraphics[width=.21\linewidth]{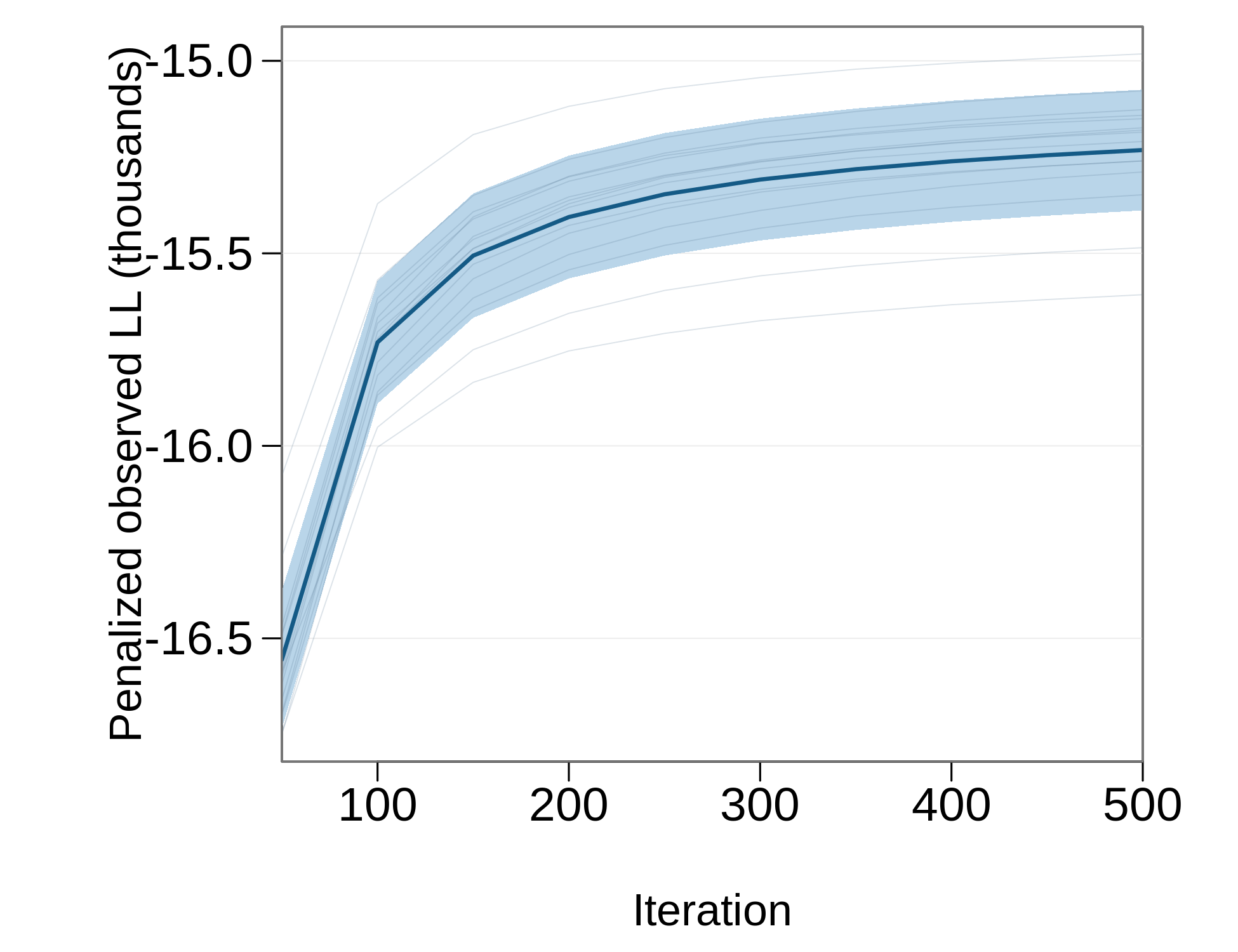} &
\includegraphics[width=.21\linewidth]{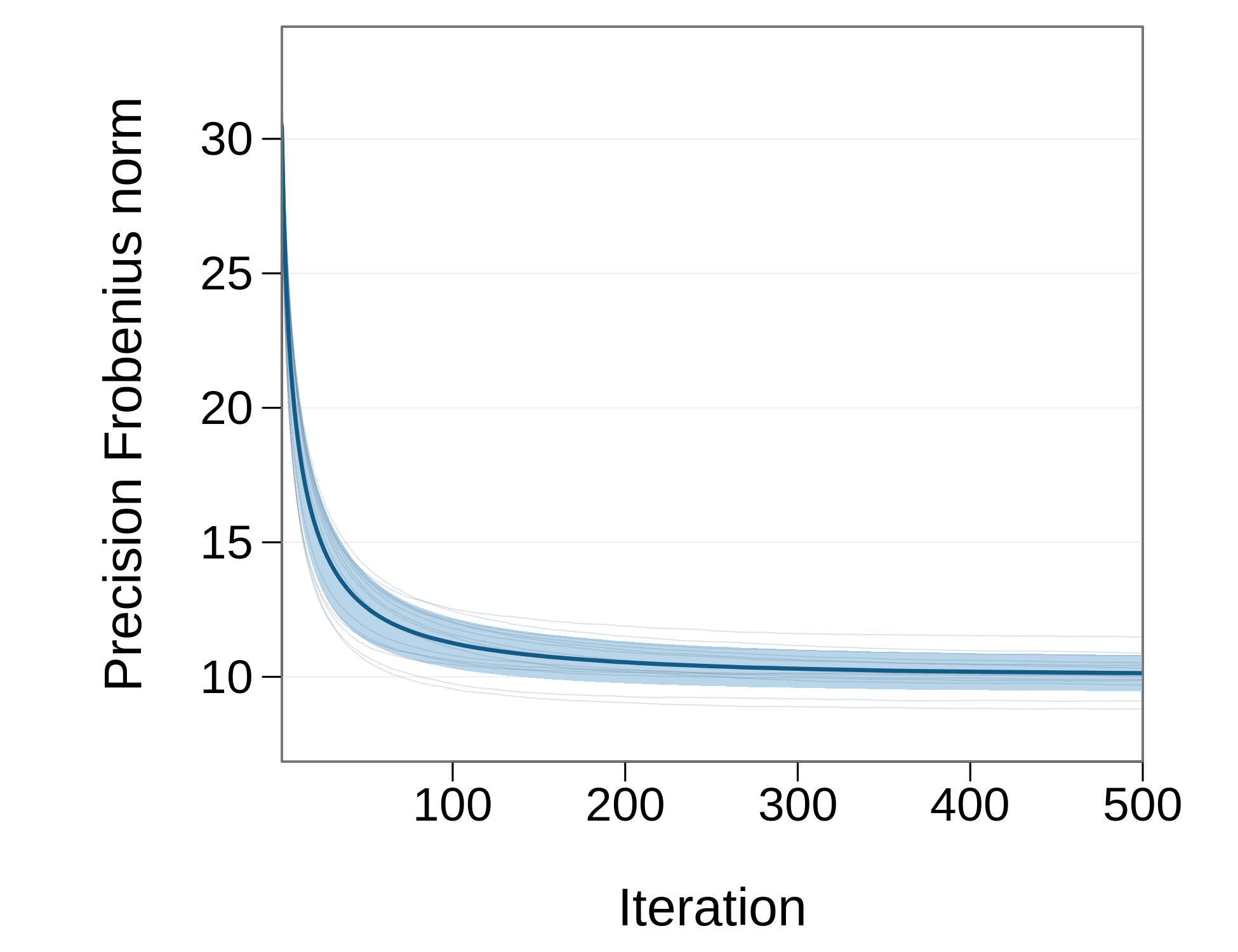} &
\includegraphics[width=.21\linewidth]{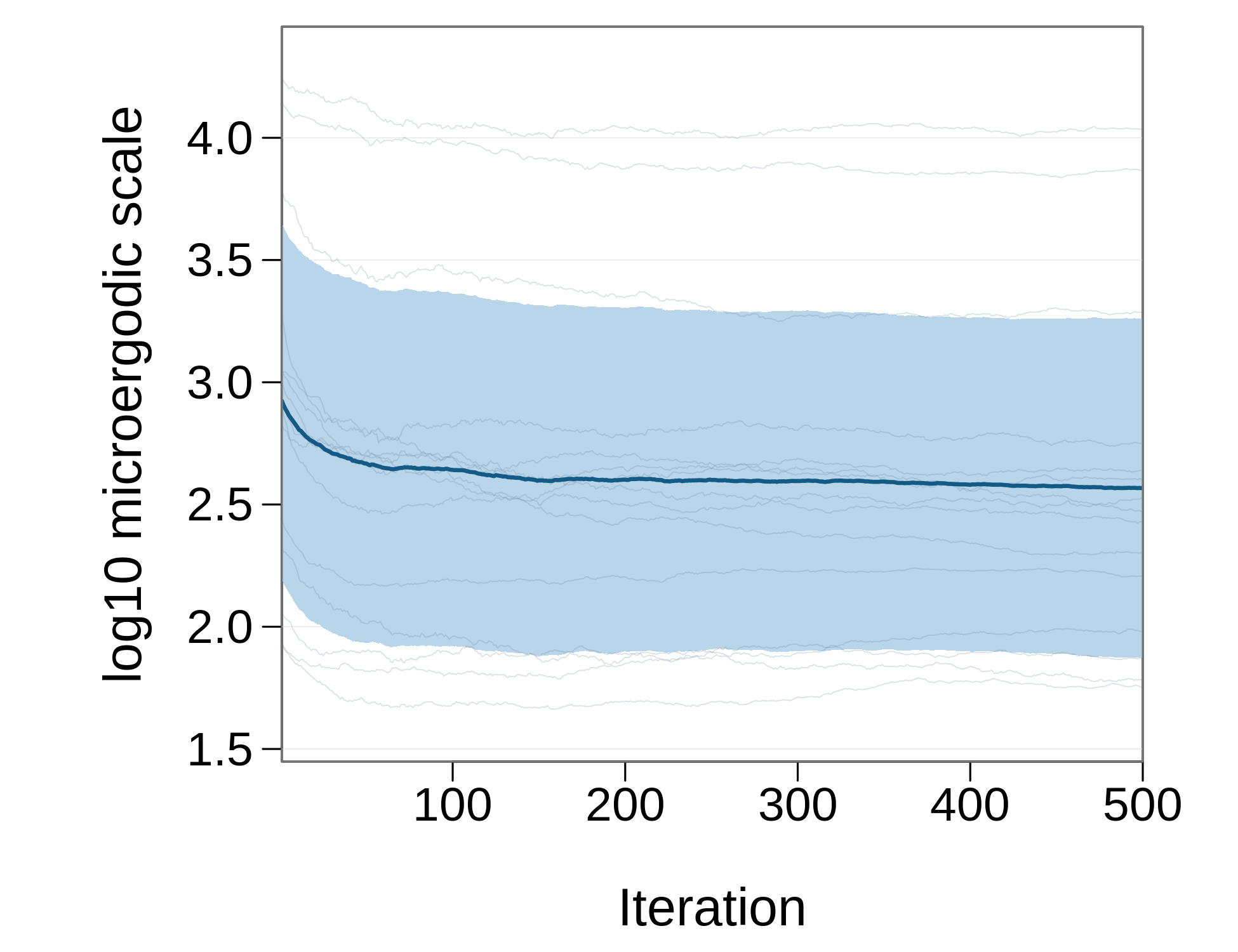} &
\includegraphics[width=.21\linewidth]{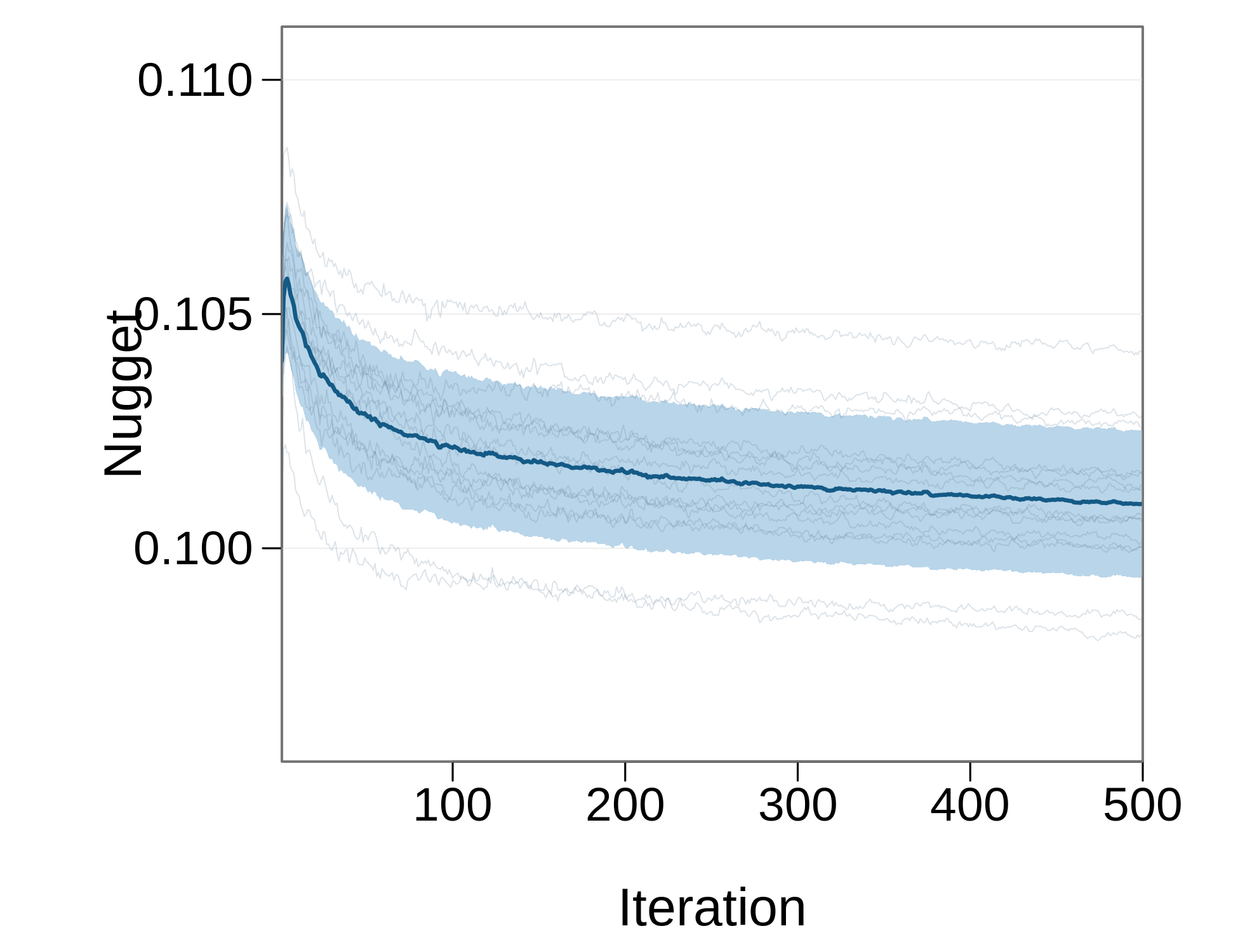} \\
\simrowlabel{Negative colocated} &
\includegraphics[width=.21\linewidth]{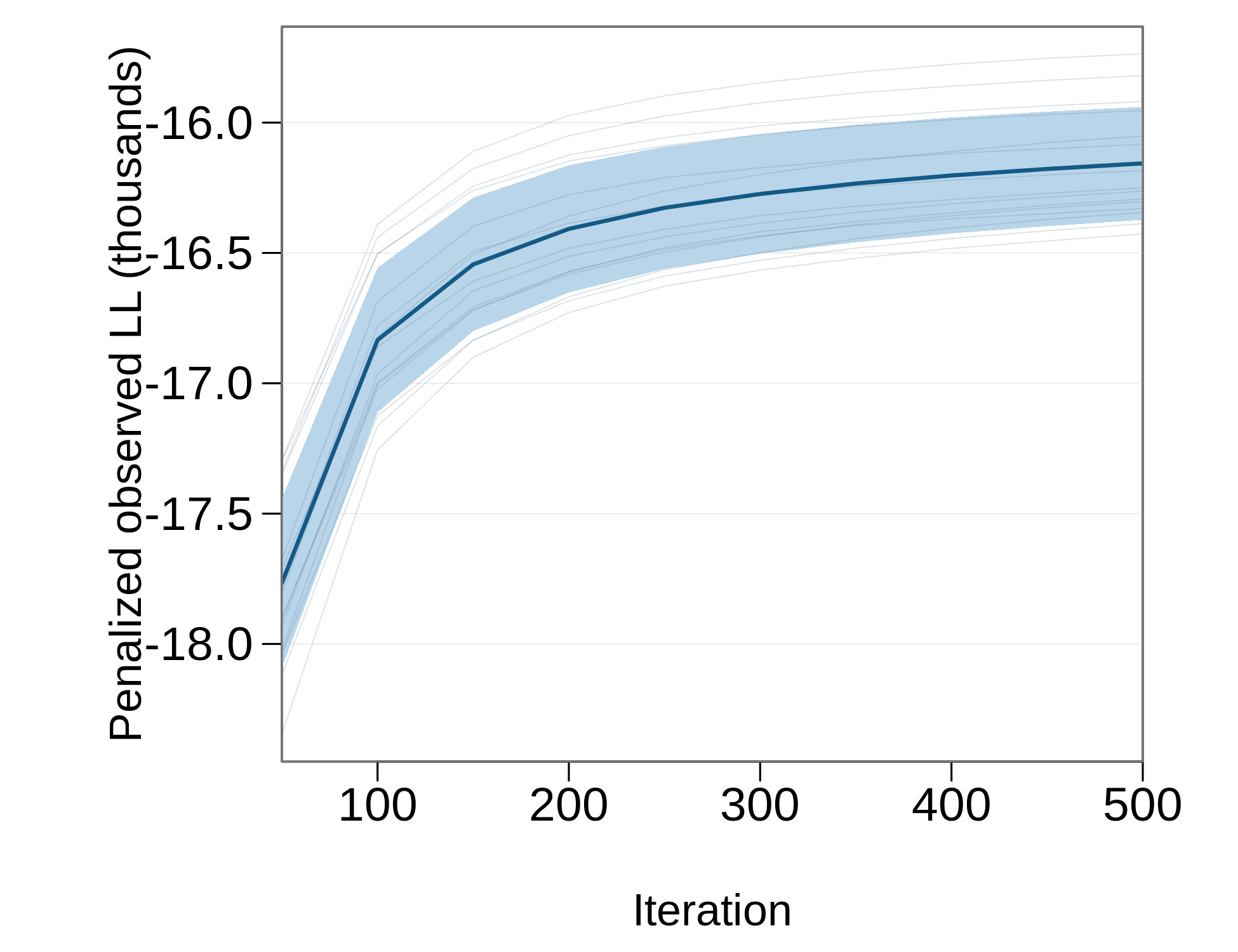} &
\includegraphics[width=.21\linewidth]{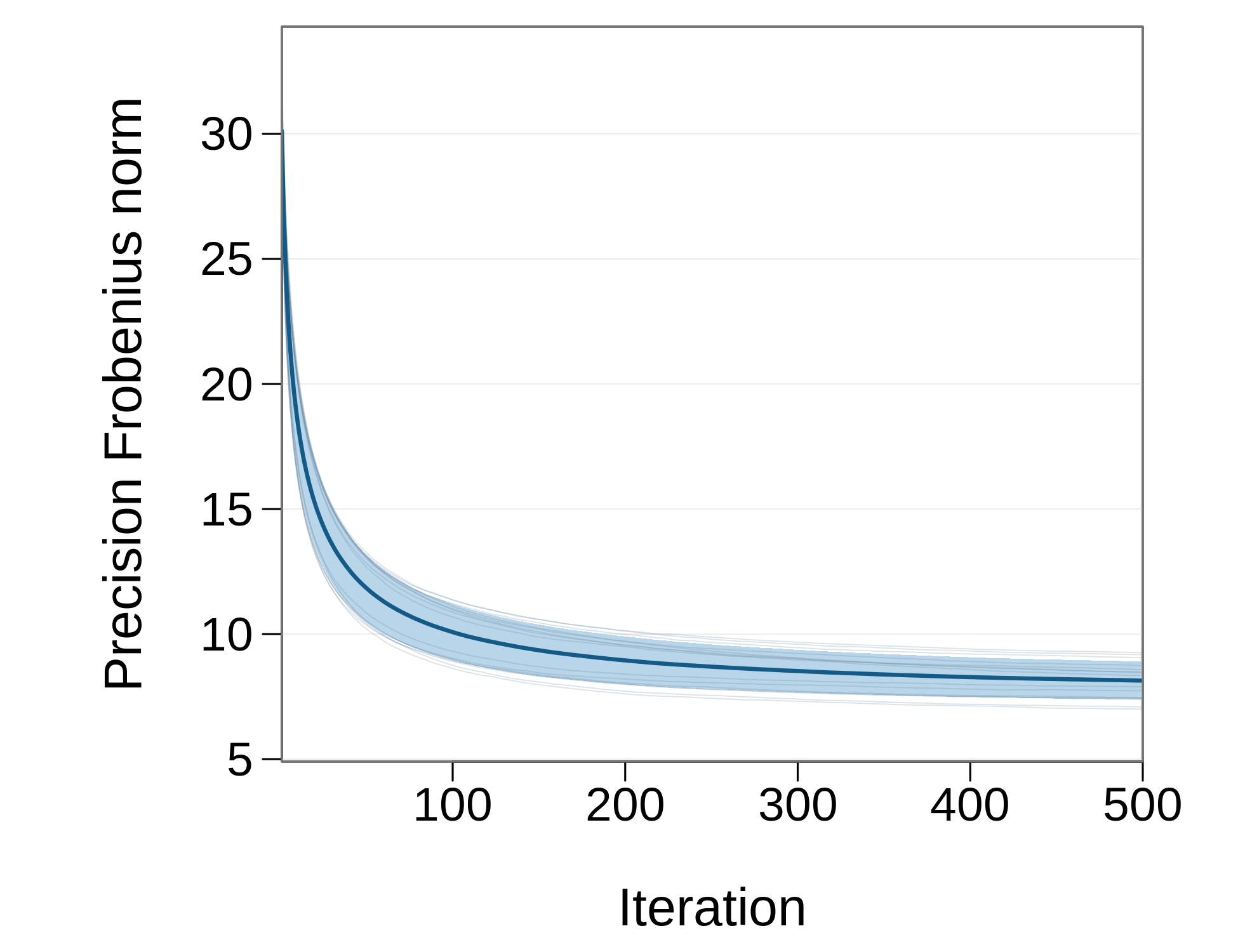} &
\includegraphics[width=.21\linewidth]{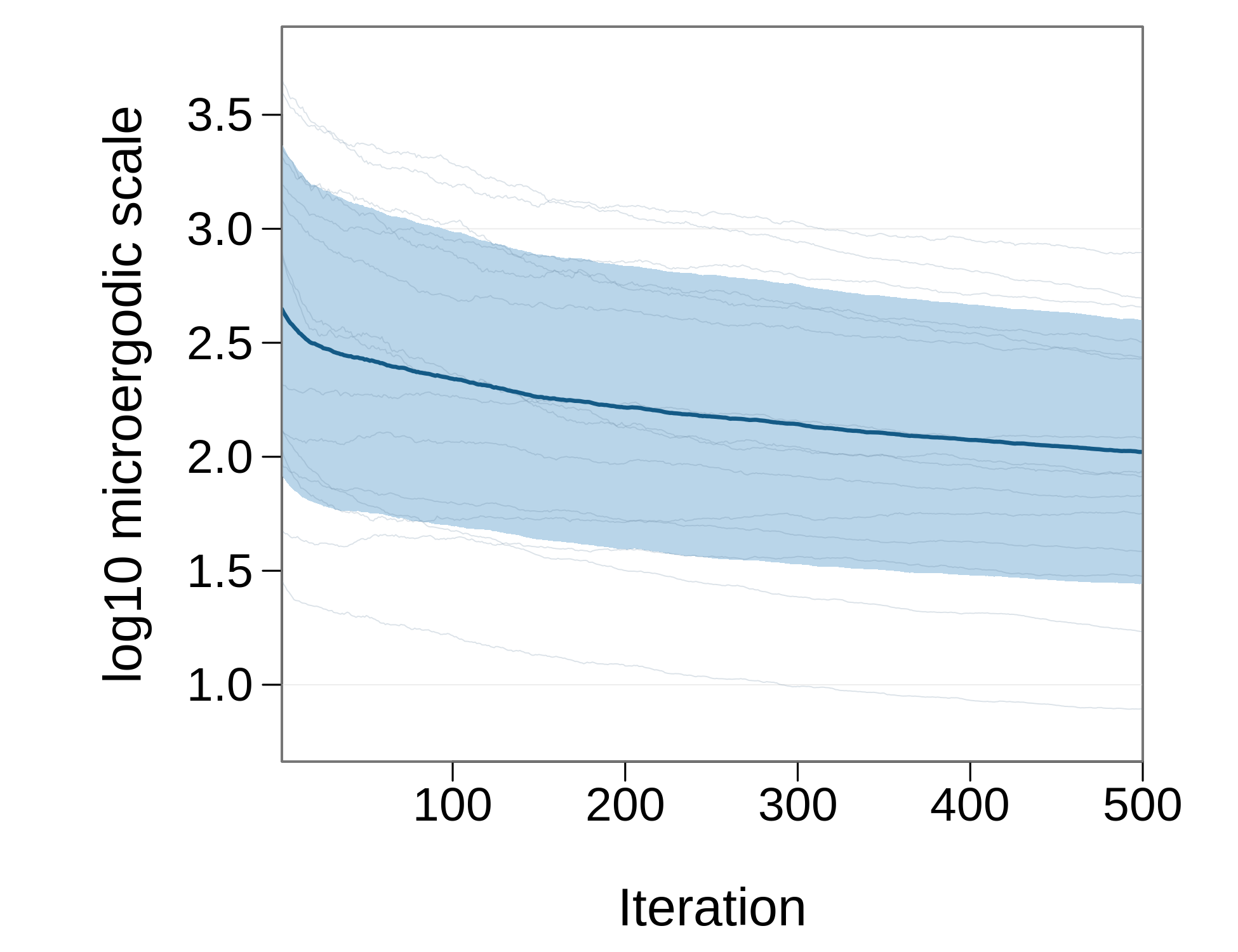} &
\includegraphics[width=.21\linewidth]{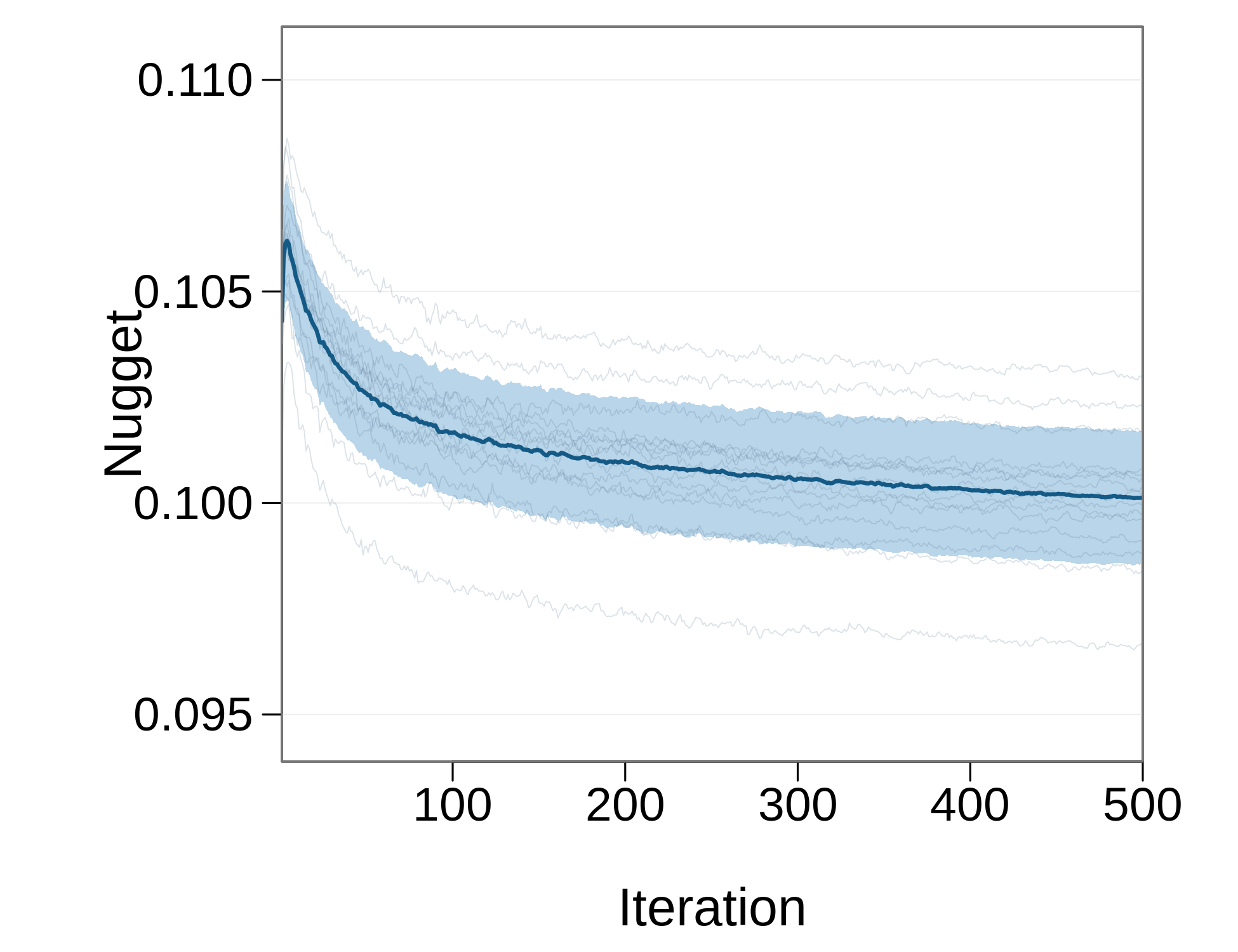}
\end{tabular}
\caption{SAECM trace diagnostics for 15 Penalized fits in the colocated mixed-family random-precision setting. Dark blue curves show means, blue bands show $\pm1$ SD, and thin curves show individual replicates. “ll” denotes log-likelihood.}
\label{fig:saecm_diagnostic_random_mix}
\end{figure}

\Cref{fig:saecm_diagnostic_random_mix} shows trace plots of the penalized observed log-likelihood based on $L=10{,}000$, the Frobenius norm of the precision matrix, the microergodic parameters, and the nugget variances for the colocated random-precision simulation in the main text. For process $j$, the CH microergodic parameter is calculated by ${\sigma_{jj}2^{\nu_j}\Gamma(\nu_j+\alpha_j)}/
({\beta_j^{2\nu_j}\Gamma(\alpha_j)})$. In our analysis, the penalized observed log-likelihood and microergodic parameters continue to drift slightly through 1,000 iterations. However, predictive performance changes little between 500 and 1,000 iterations, so we report the 500-iteration results. The nugget continues to drift, although the magnitude of the drift is small. The Frobenius norm stabilizes early. 

\subsection{Timing results}
Table~\ref{table:time_full} reports estimation times for the simulation studies. Ind is generally the fastest approach because it fits each process separately without estimating cross-covariances. LMC is relatively fast in most settings but substantially slower in the random-precision misaligned settings, where it uses more latent processes and requires a more costly likelihood calculation. The all-Mat\'ern fits also involve more covariance parameters and require more optimization steps than the mixed-family fits, which predominantly select SE latent covariance functions. The reported SAECM and IOX times correspond to $500$ and $10{,}000$ iterations. 
We fit nine positive-penalty candidates and a separate Dense benchmark, with each candidate fit run in parallel. 
Each SAECM fit used 32 threads, while each IOX and LMC fit used 4 threads and each Ind fit was allocated 8 CPU cores.
These timings reflect the computational time of the procedures used, rather than a comparison at matched estimation accuracy.

\begin{table}[H]
\centering
\small
\setlength{\tabcolsep}{3pt}
\renewcommand{\arraystretch}{0.8}
\caption{Estimation times per fit in minutes. Entries are mean (SD) across 15 replicates. Col/Mis denote colocated/misaligned observations, and Pos/Neg denote positive/negative precision off-diagonals.}
\label{table:time_full}
\begin{tabular}{lllrrrrr}
\toprule
Structure & Alignment & Sign & Penalized & Dense & Ind & LMC & IOX \\
\midrule
\multicolumn{8}{l}{\textit{Mixed-family}} \\
\addlinespace
Random & Col & Pos
& 161.73 (5.14) & 186.71 (13.65) & {25.8 (3.5)}
& 33.9 (8.2) & 184.86 (7.73) \\
& & Neg
& 167.39 (7.95) & 188.87 (7.02) & {26.7 (4.1)}
& 32.7 (6.8) & 185.78 (4.72) \\
& Mis & Pos
& 138.76 (4.18) & 166.05 (6.97) & {9.6 (1.1)}
& 97.3 (36.9) & 181.38 (6.18) \\
& & Neg
& 179.79 (2.39) & 206.44 (3.06) & {10.0 (1.9)}
& 118.5 (42.5) & 179.66 (3.60) \\
\addlinespace
Tridiagonal & Col & Pos
& 182.50 (5.76) & 185.74 (5.06) & {25.9 (5.9)}
& 32.5 (16.1) & 185.57 (5.06) \\
& & Neg
& 172.84 (3.47) & 179.20 (5.58) & {24.3 (5.2)}
& 24.5 (12.9) & 183.76 (7.05) \\
& Mis & Pos
& 184.20 (5.30) & 198.81 (4.63) & {9.8 (2.7)}
& 27.6 (13.6) & 178.94 (5.24) \\
& & Neg
& 183.67 (4.38) & 198.21 (2.53) & {9.2 (2.3)}
& 22.0 (5.9) & 179.19 (4.25) \\
\midrule
\multicolumn{8}{l}{\textit{Mat\'ern}} \\
\addlinespace
Random & Col & Pos
& 138.81 (2.66) & 166.75 (6.22) & {13.5 (0.5)}
& 13.6 (1.6) & 185.07 (5.19) \\
& & Neg
& 139.93 (4.59) & 157.87 (8.50) & 12.2 (0.5)
& {11.9 (1.4)} & 186.15 (4.08) \\
& Mis & Pos
& 148.03 (1.57) & 175.72 (3.48) & {4.7 (0.1)}
& 295.6 (35.0) & 183.85 (5.37) \\
& & Neg
& 112.21 (4.99) & 133.56 (2.16) & {4.4 (0.2)}
& 265.8 (50.5) & 181.53 (4.84) \\
\addlinespace
Tridiagonal & Col & Pos
& 159.26 (3.33) & 160.63 (3.45) & 15.4 (0.9)
& {5.7 (1.4)} & 183.61 (5.47) \\
& & Neg
& 157.24 (2.16) & 159.25 (2.01) & 14.6 (0.5)
& {6.0 (1.4)} & 185.67 (5.46) \\
& Mis & Pos
& 162.70 (10.95) & 174.21 (2.26) & {5.6 (0.2)}
& 43.9 (9.3) & 181.33 (5.15) \\
& & Neg
& 170.12 (6.05) & 176.44 (2.32) & {5.5 (0.2)}
& 35.0 (4.9) & 180.19 (4.93) \\
\bottomrule
\end{tabular}
\end{table}

\subsection{Selection results}
For completeness, we provide additional selection results below for Section~\ref{sec:sparse_rec}. Table~\ref{table:sparsity} reports the true positive rate (TPR) and false positive rate (FPR) for selection. 
\begin{table}[H]
\centering
\caption{True-positive and false-positive rates for precision-matrix edge recovery in mixed-family random-precision simulations, averaged across 15 replicates. Col/Mis denote colocated/misaligned observations, and Pos/Neg denote positive/negative precision off-diagonals.}
\label{table:sparsity}
\begin{tabular}{l cc cc}
\toprule
& \multicolumn{2}{c}{TPR} & \multicolumn{2}{c}{FPR} \\
\cmidrule(lr){2-3} \cmidrule(lr){4-5}
Setting & Penalized & Dense & Penalized & Dense \\
\midrule
Col / Pos & 0.69 & 1.00 & 0.09 & 1.00 \\
Col / Neg & 0.80 & 1.00 & 0.12 & 1.00 \\
Mis / Pos & 0.58 & 1.00 & 0.05 & 1.00 \\
Mis / Neg & 0.74 & 1.00 & 0.09 & 1.00 \\
\bottomrule
\end{tabular}
\end{table}

\subsection{Additional results for mixed-family settings}
We present additional simulation results for misaligned estimation and tridiagonal precision settings in this section. \Cref{table:estimate_random_misaligned} shows the estimation results for the mixed-family covariance model in the misaligned setting. Compared with the colocated results in \Cref{table:estimation_sim_rand}, all approaches generally have larger biases, and the Ind approach provides better estimates for some parameters. Overall, our approach still provides better estimation performance.

\begin{figure}[H]
\centering
\tiny
\begin{tabular}{@{}m{0.035\linewidth}cccc@{}}
&
\textbf{Correlation} &
\textbf{Precision} &
\textbf{Penalized} &
\textbf{Dense} \\

\rotatebox[origin=c]{90}{\textbf{Positive}} &
\includegraphics[width=.21\linewidth,align=c]{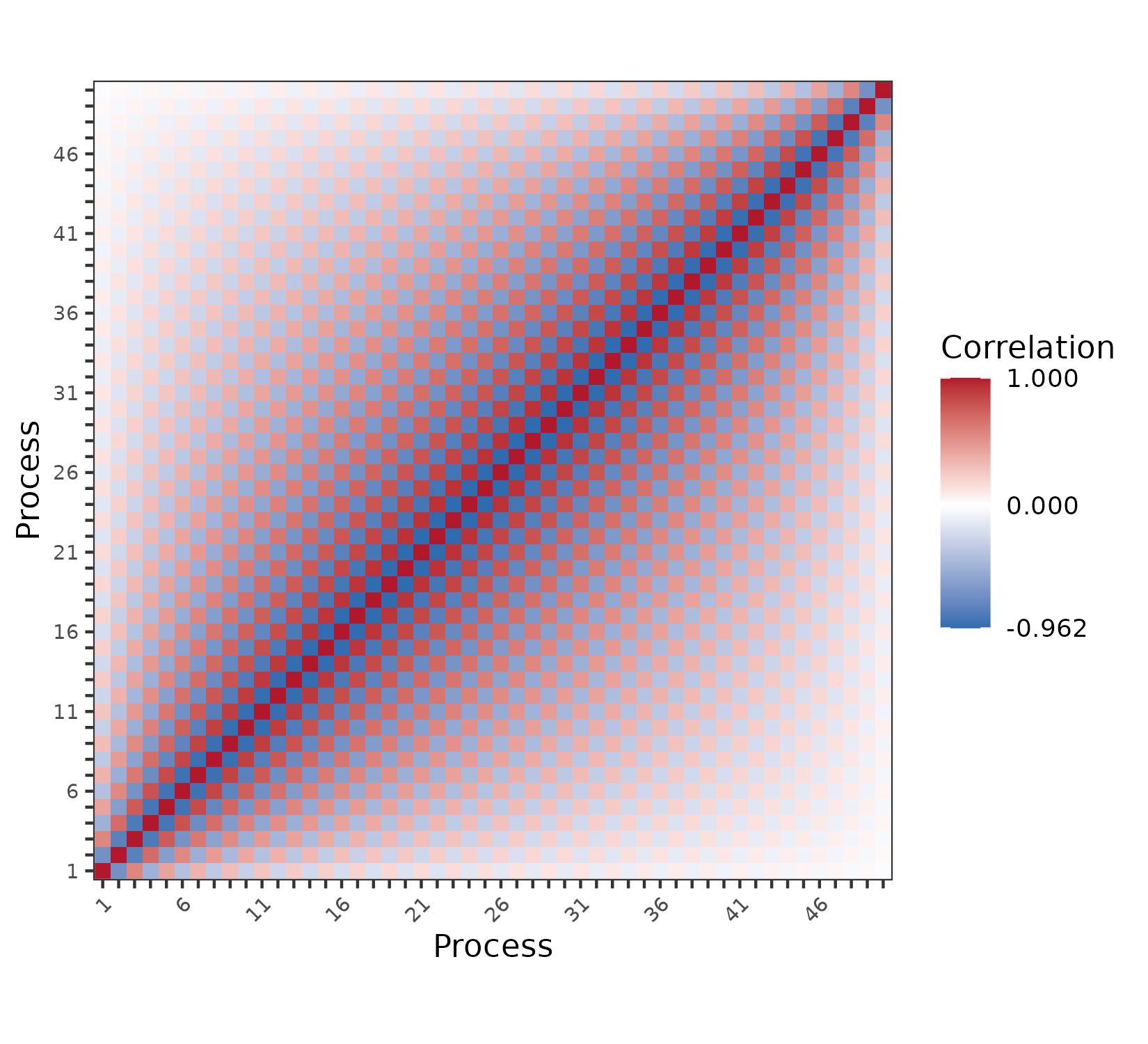} &
\includegraphics[width=.21\linewidth,align=c]{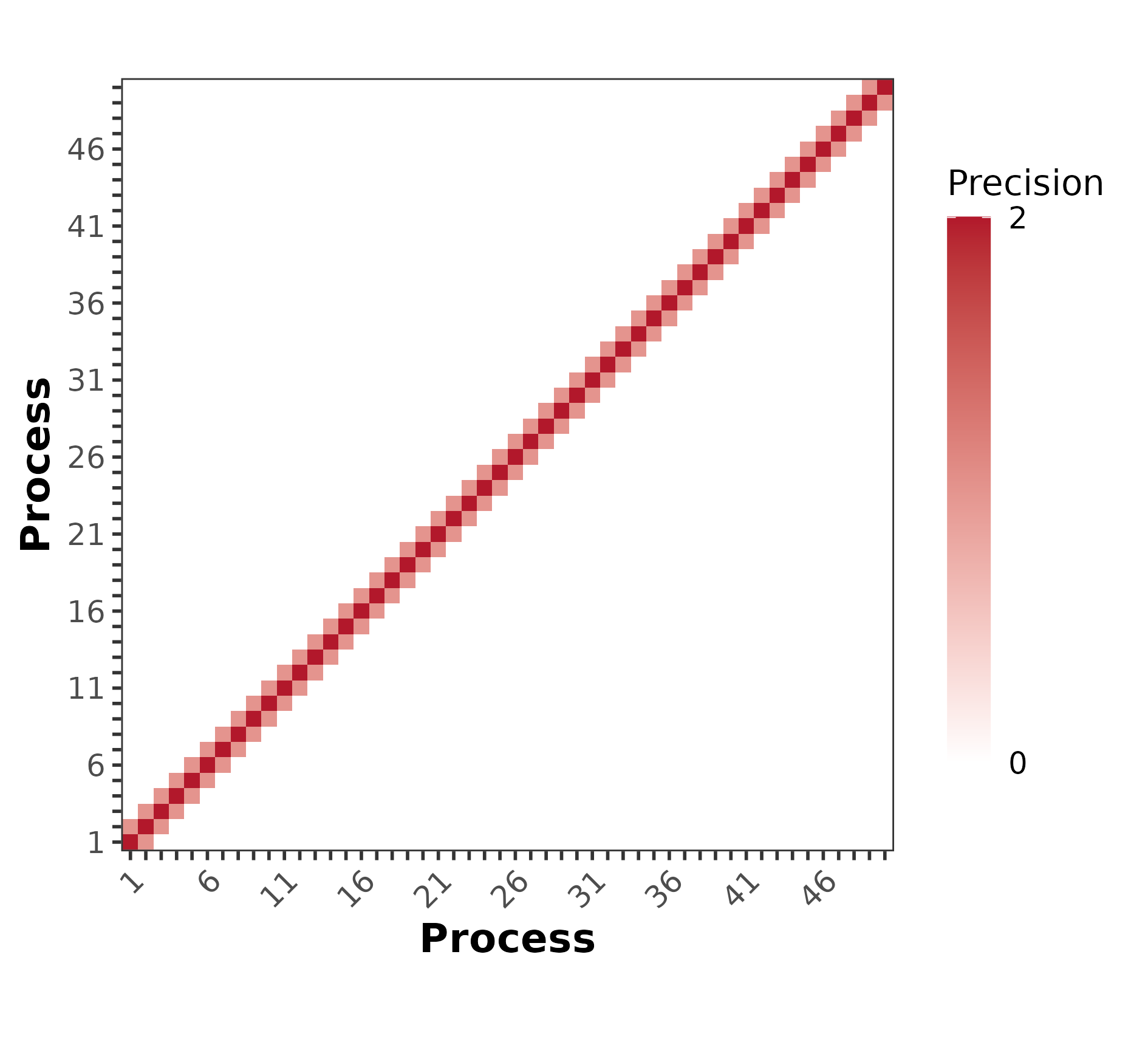} &
\includegraphics[width=.21\linewidth,align=c]{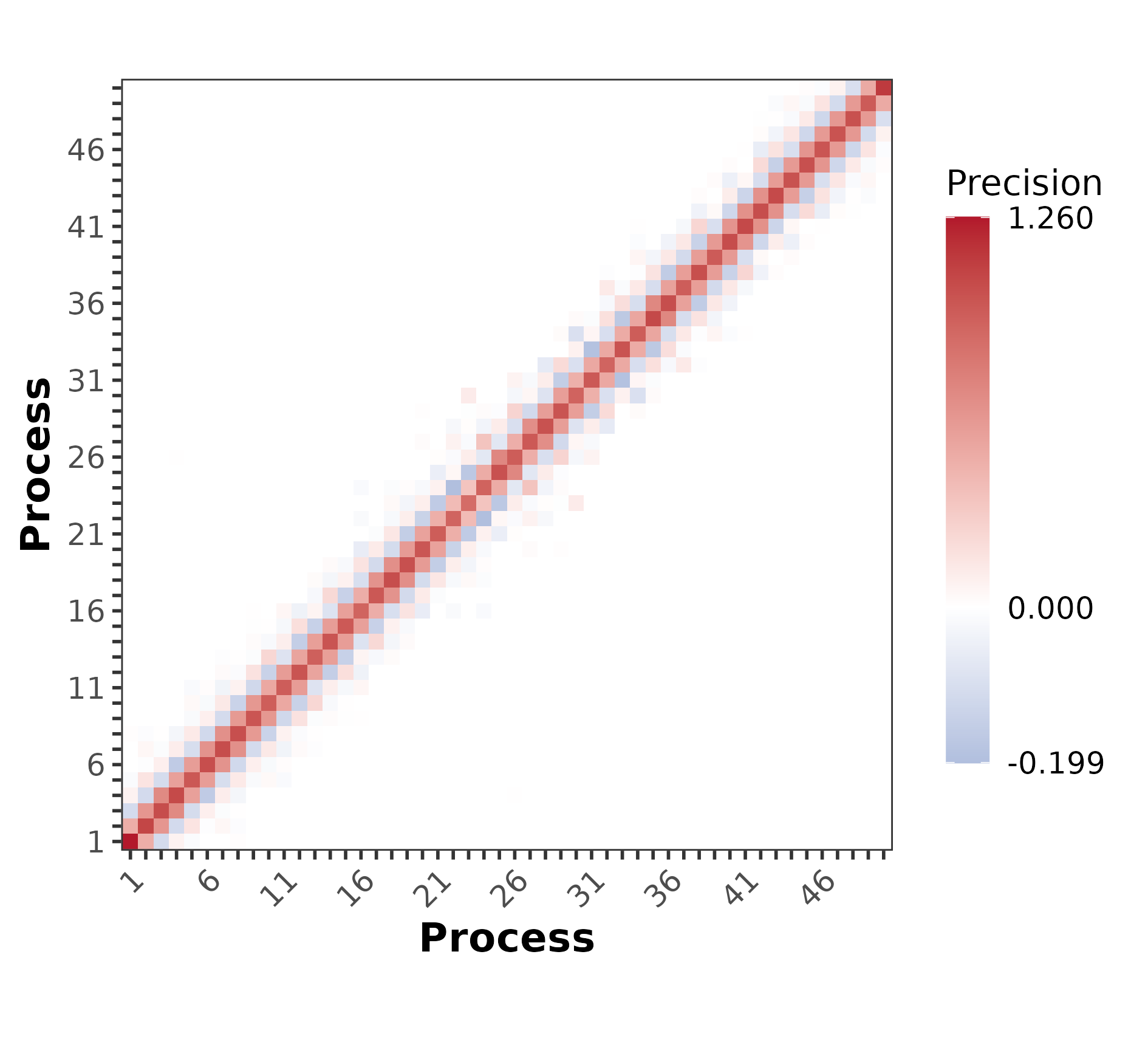} &
\includegraphics[width=.21\linewidth,align=c]{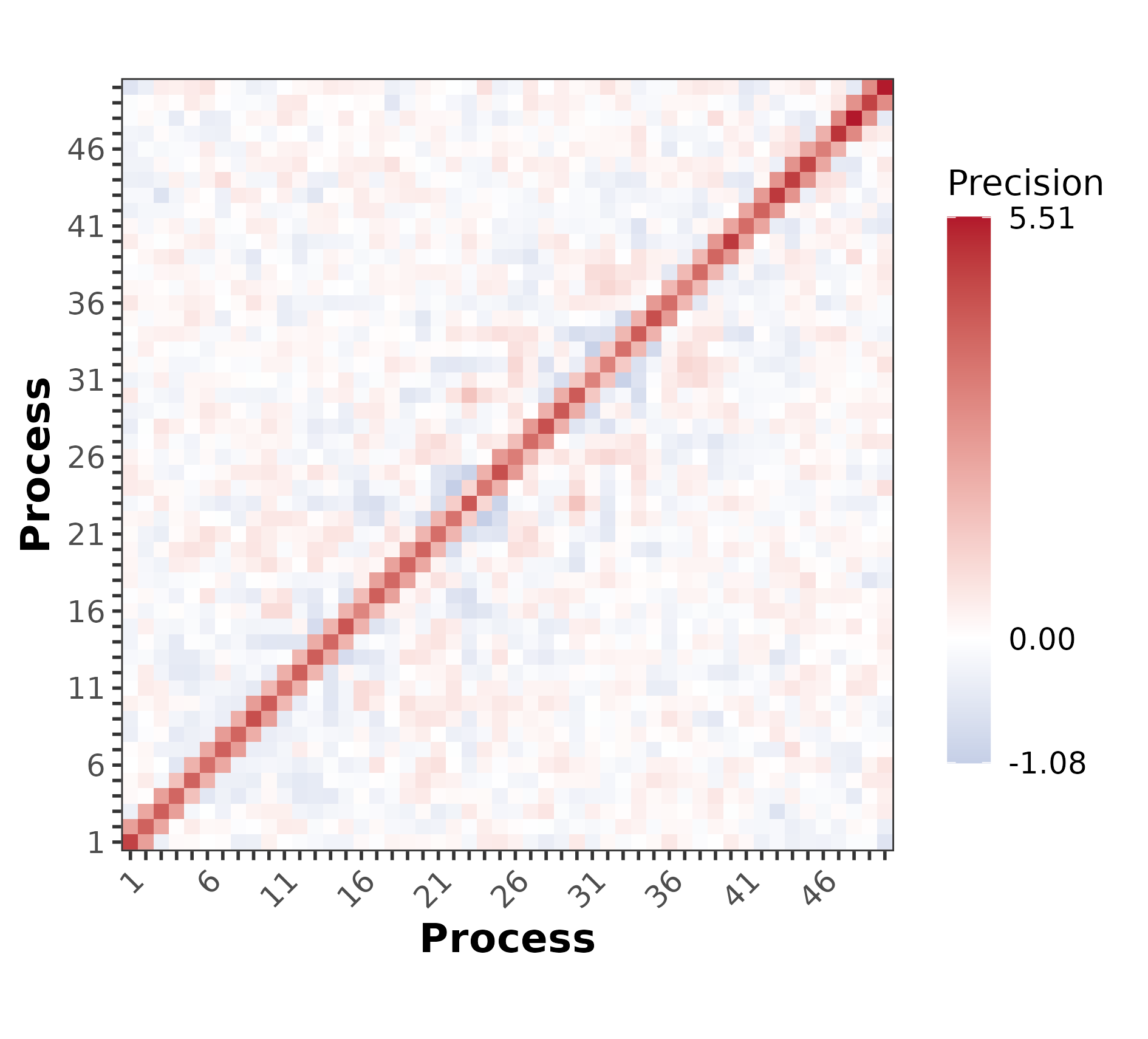} \\

\noalign{\smallskip}

\rotatebox[origin=c]{90}{\textbf{Negative}} &
\includegraphics[width=.21\linewidth,align=c]{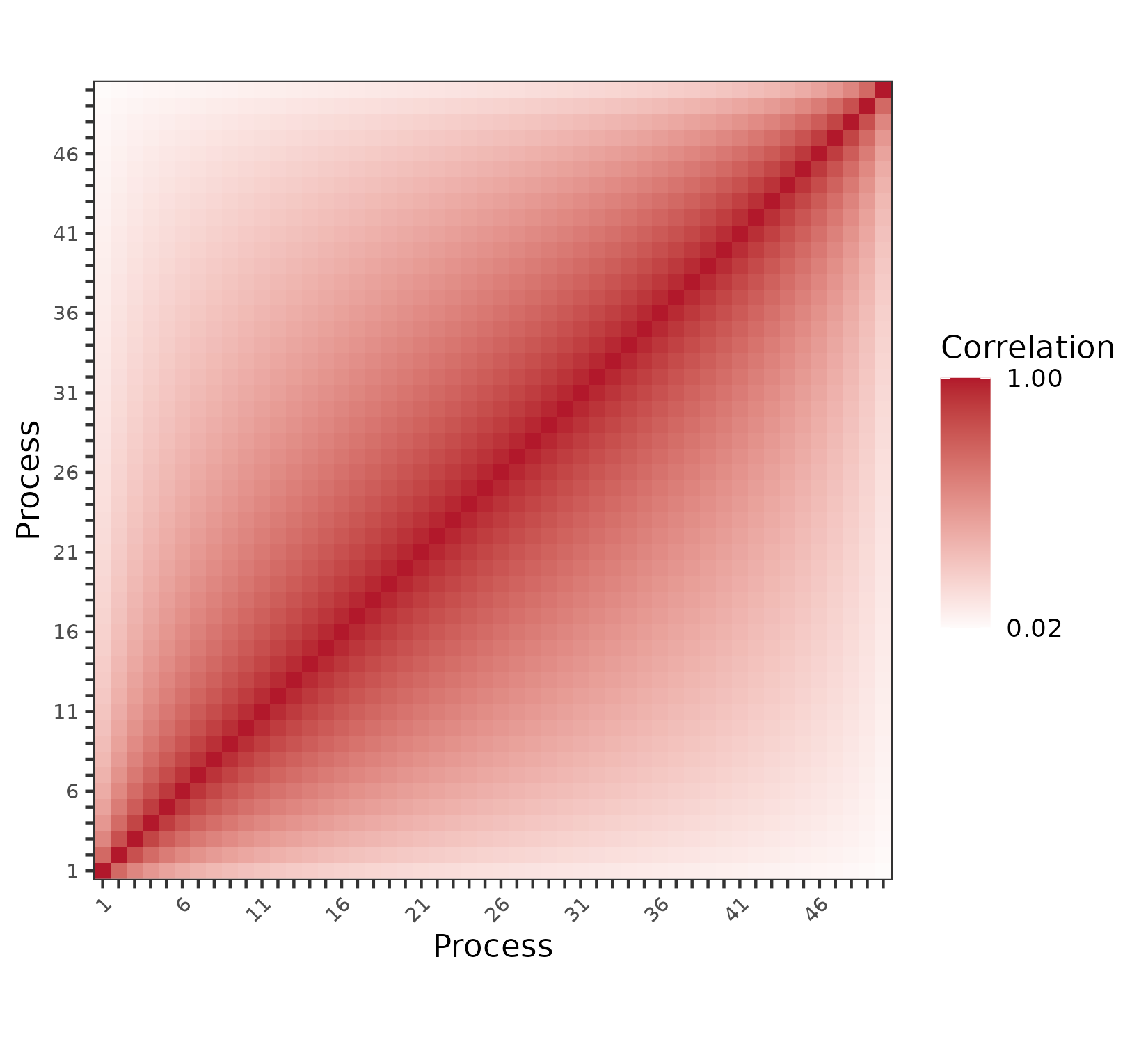} &
\includegraphics[width=.21\linewidth,align=c]{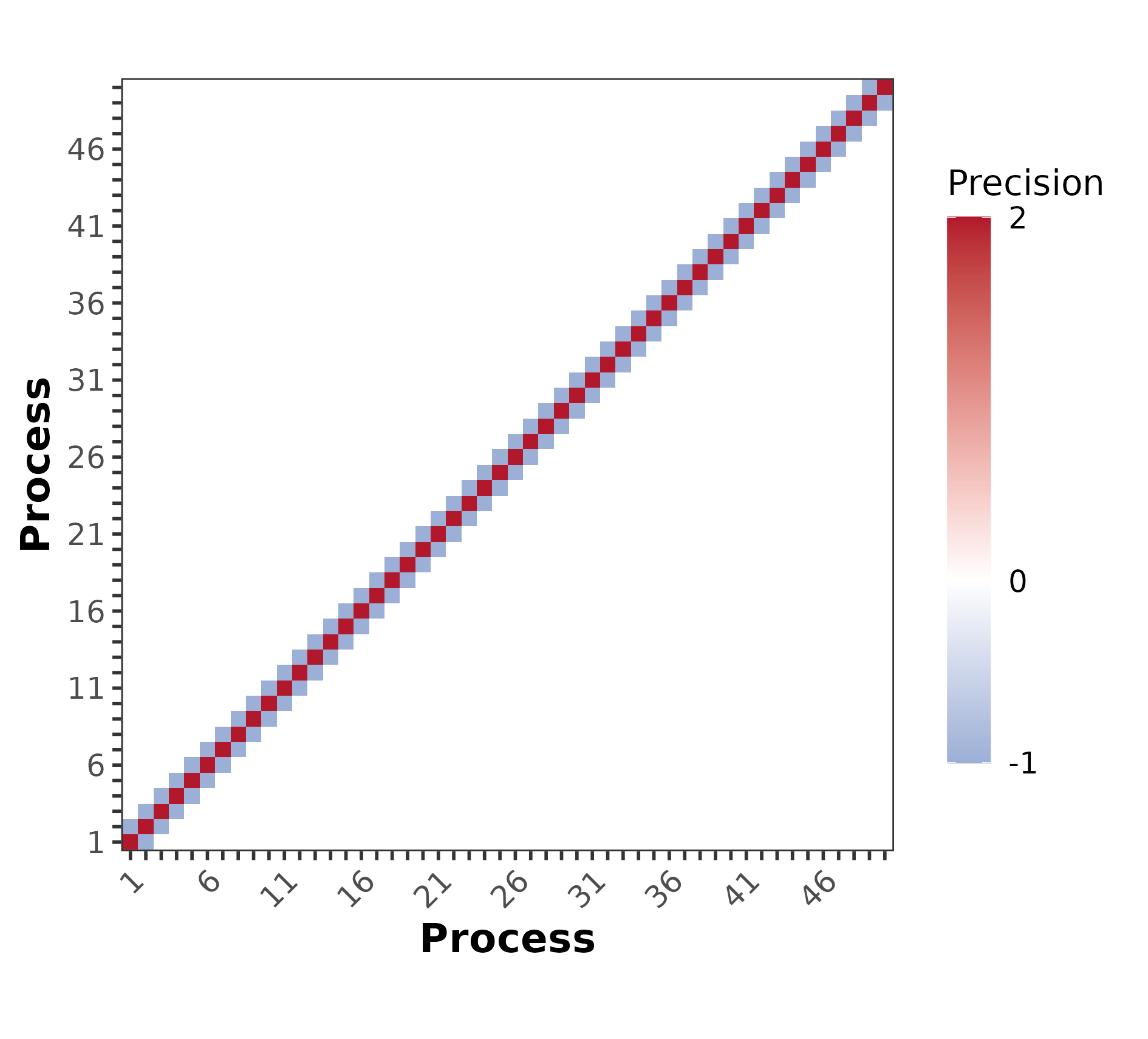} &
\includegraphics[width=.21\linewidth,align=c]{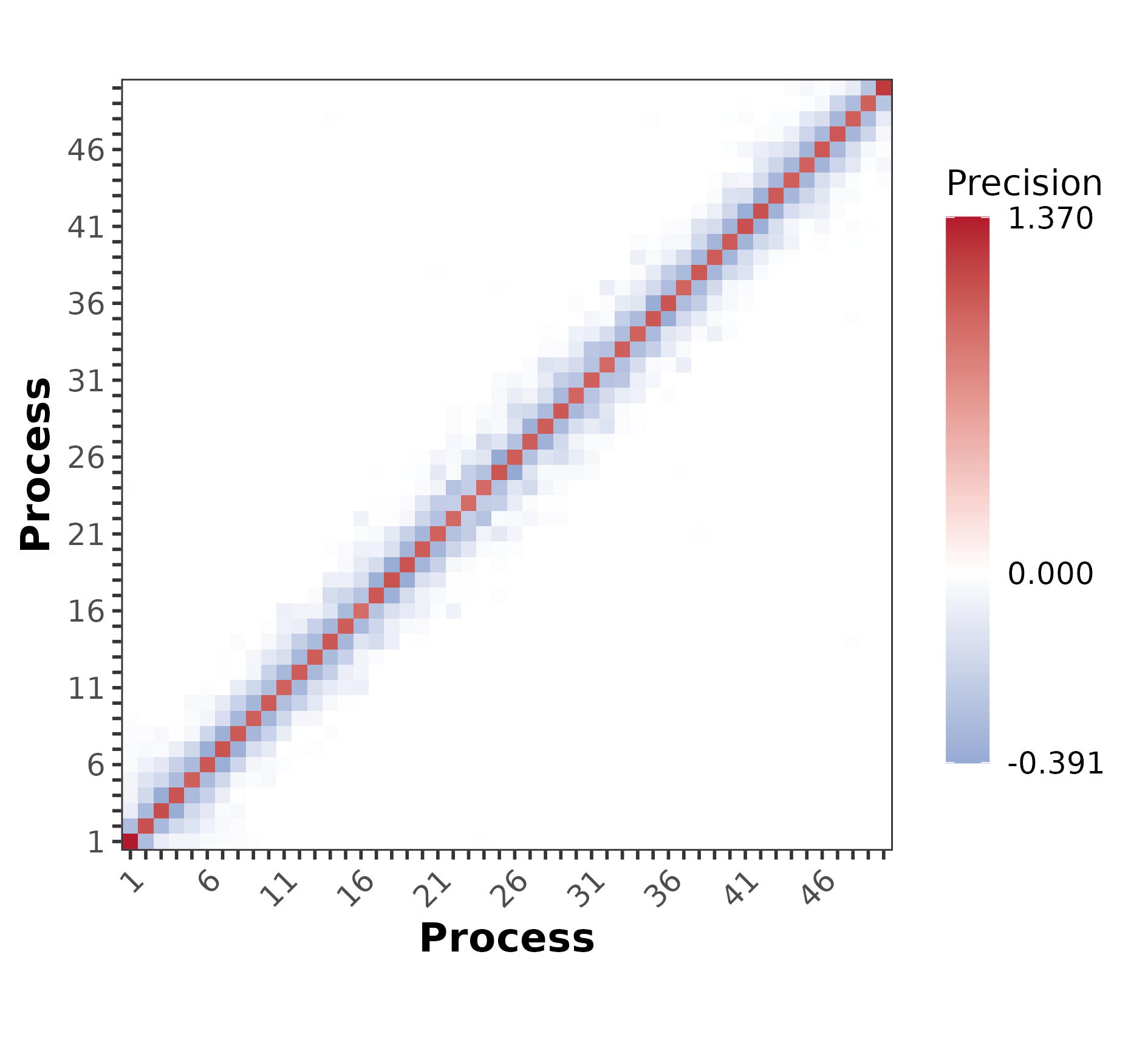} &
\includegraphics[width=.21\linewidth,align=c]{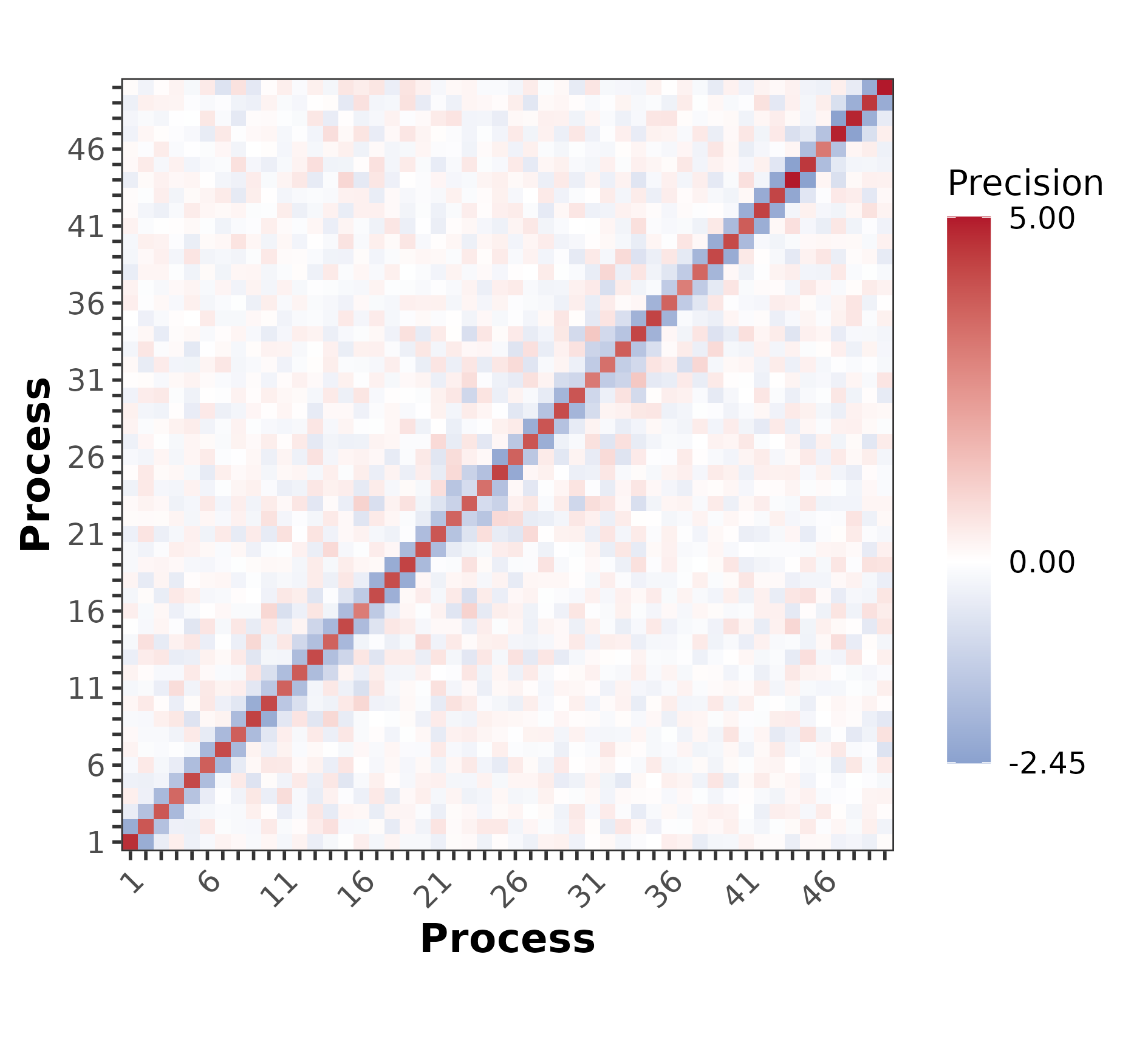} \\
\end{tabular}

\caption{True correlation and precision matrices and entrywise median precision estimates across 15 replicates for mixed-family tridiagonal-precision simulations.}
\label{fig:cov_prec_tridiagonal_mix}
\end{figure}

In the tridiagonal setting, the diagonal entries of the precision matrices are set to $2$. The off-diagonal values are set to $1$ to induce alternating signs in the correlation structure, and to $-1$ to produce uniformly positive correlations, as illustrated in Figure~\ref{fig:cov_prec_tridiagonal_mix}. The resulting variance for each process ranges from $0.98$ to $12.75$.

We use the same hyperparameters as in the random-precision setting, with the number of Monte Carlo steps $T=30$, and LMC $q=6$ to account for $90\%$ of total variance. The estimation results are shown in Figure~\ref{fig:cov_prec_tridiagonal_mix} and Table~\ref{table:estimation_tridiagona_mix}. Among all compared approaches, our method yields estimates with generally lower bias. Prediction results are provided in Table~\ref{table:prediction_tridiagonal_mix}. Both the Penalized and Dense approaches capture the true precision signals, and the Penalized fit produces sparse estimates that more closely match the true underlying structure.

The predictive performance for the tridiagonal precision structure is similar to the results obtained for the random structure. When the data are colocated, our approach provides slightly better predictive performance for individual processes than other approaches. It also provides better coverage for the sum of processes than Ind. In the misaligned setting, our method outperforms the Ind approach for both individual processes and the sum of processes.

\begin{table}[H]\centering\small
\setlength{\tabcolsep}{3.5pt}
\renewcommand{\arraystretch}{0.80}
\caption{Parameter estimation for mixed-family random-precision simulations with misaligned observations. Entries are bias (SD) across 15 replicates, averaged over the processes in each family. Bold indicates the smallest absolute bias.}
\label{table:estimate_random_misaligned}
\resizebox{\textwidth}{!}{%
\begin{tabular}{llrccc|ccc}
\toprule
&&&\multicolumn{3}{c|}{Positive off-diagonals}
&\multicolumn{3}{c}{Negative off-diagonals}\\
Family & Parameter & Truth range
& Penalized & Dense & Ind
& Penalized & Dense & Ind\\
\midrule
CH & $\alpha$ & [1.27, 1.73]
& \textbf{3.66 (5.62)} & 4.10 (6.12) & 3.73 (4.49)
& \textbf{3.55 (5.36)} & 3.87 (5.59) & 3.82 (4.41) \\

CH & $\beta$ & [0.60, 0.87]
& 0.60 (1.07) & 0.67 (1.15) & \textbf{0.60 (1.04)}
& \textbf{0.60 (0.99)} & 0.71 (1.12) & 0.71 (1.06) \\

CH & $\nu$ & [1.50, 2.50]
& 0.61 (1.32) & \textbf{0.54 (1.30)} & 0.78 (1.54)
& 0.45 (1.30) & \textbf{0.38 (1.27)} & 0.51 (1.45) \\

CH & $d$ & 0.10
& \textbf{0.00 (0.00)} & 0.00 (0.00) & 0.00 (0.02)
& \textbf{0.00 (0.00)} & 0.00 (0.00) & 0.00 (0.02) \\

SE & $a$ & [1.76, 2.16]
& 0.02 (0.13) & 0.02 (0.13) & \textbf{0.00 (0.13)}
& 0.01 (0.13) & 0.01 (0.13) & \textbf{0.00 (0.13)} \\

SE & $d$ & 0.10
& 0.00 (0.00) & 0.00 (0.00) & \textbf{0.00 (0.01)}
& \textbf{0.00 (0.00)} & 0.00 (0.00) & 0.00 (0.01) \\

Mat\'ern & $a$ & [2.52, 3.48]
& 0.51 (1.57) & \textbf{0.34 (1.49)} & 0.77 (1.77)
& 0.58 (1.44) & \textbf{0.40 (1.30)} & 0.60 (1.48) \\

Mat\'ern & $\nu$ & [1.50, 2.50]
& 0.83 (1.64) & \textbf{0.66 (1.57)} & 0.98 (1.86)
& 0.62 (1.35) & \textbf{0.47 (1.27)} & 0.61 (1.41) \\

Mat\'ern & $d$ & 0.10
& 0.00 (0.00) & 0.00 (0.00) & \textbf{0.00 (0.02)}
& 0.00 (0.00) & 0.00 (0.00) & \textbf{0.00 (0.02)} \\
\bottomrule
\end{tabular}%
}
\end{table}

\begin{table}[H]\centering\small
\setlength{\tabcolsep}{3.5pt}
\renewcommand{\arraystretch}{0.80}
\caption{Parameter estimation for mixed-family tridiagonal-precision simulations with colocated observations. Entries are bias (SD) across 15 replicates, averaged over the processes in each family. Bold indicates the smallest absolute bias.}
\label{table:estimation_tridiagona_mix}
\resizebox{\textwidth}{!}{%
\begin{tabular}{llrccc|ccc}\toprule
&&&\multicolumn{3}{c|}{Positive off-diagonals}&\multicolumn{3}{c}{Negative off-diagonals}\\
Family & Parameter & Truth range & Penalized & Dense & Ind & Penalized & Dense & Ind\\\midrule
CH & $\alpha$ & [1.27, 1.73] & \textbf{1.51 (3.28)} & 2.07 (3.04) & 2.83 (3.28) & \textbf{1.13 (3.15)} & 1.47 (2.53) & 2.11 (3.05)
 \\
CH & $\beta$ & [0.60, 0.87] & 0.36 (0.51) & \textbf{0.30 (0.49)} & 0.48 (0.66) & 0.29 (0.50) & \textbf{0.19 (0.45)} & 0.34 (0.63)
 \\
CH & $\nu$ & [1.50, 2.50] & \textbf{-0.04 (0.46)} & 0.12 (0.60) & 0.14 (0.70) & \textbf{-0.04 (0.45)} & 0.14 (0.58) & 0.15 (0.69)
 \\
CH & $d$ & 0.10 & 0.00 (0.00) & 0.00 (0.00) & \textbf{0.00 (0.02)} & 0.00 (0.00) & 0.00 (0.00) & \textbf{0.00 (0.02)}
 \\
SE & $a$ & [1.76, 2.16] & 0.00 (0.07) & \textbf{0.00 (0.07)} & 0.00 (0.07) & 0.01 (0.07) & 0.01 (0.07) & \textbf{0.00 (0.07)}
 \\
SE & $d$ & 0.10 & 0.00 (0.00) & 0.00 (0.00) & \textbf{0.00 (0.01)} & 0.00 (0.00) & 0.00 (0.00) & \textbf{0.00 (0.01)}
 \\
Mat\'ern & $a$ & [2.52, 3.48] & -0.29 (0.64) & \textbf{0.15 (0.71)} & 0.18 (0.82) & -0.24 (0.59) & \textbf{0.20 (0.66)} & 0.25 (0.81)
 \\
Mat\'ern & $\nu$ & [1.50, 2.50] & \textbf{0.02 (0.46)} & 0.15 (0.50) & 0.18 (0.57) & \textbf{0.01 (0.45)} & 0.12 (0.48) & 0.15 (0.56)
 \\
Mat\'ern & $d$ & 0.10 & 0.00 (0.00) & 0.00 (0.00) & \textbf{0.00 (0.02)} & 0.00 (0.00) & 0.00 (0.00) & \textbf{0.00 (0.02)}
 \\
\bottomrule\end{tabular}
}
\end{table}

\begin{table}[H]
\centering
\small
\setlength{\tabcolsep}{2.5pt}
\renewcommand{\arraystretch}{0.8}
\caption{Prediction summaries for mixed-family tridiagonal-precision simulations with colocated and misaligned observations. Entries are means (SDs) across 15 replicates. Col/Mis denote colocated/misaligned observations, and Pos/Neg denote positive/negative precision off-diagonals. Bold indicates coverage closest to 95\%, or the lowest mean for other metrics.}
\label{table:prediction_tridiagonal_mix}
\begin{tabular}{llccccc}
\toprule
Setting & Metric & Penalized & Dense & Ind & LMC & IOX \\
\midrule
Col/Pos & Norm
& \textbf{3.29 (0.07)} & 3.31 (0.07) & 3.51 (0.09) & 7.24 (0.45) & 3.43 (0.11) \\
& Coverage (\%)
& 94.31 (0.44) & 94.26 (0.52) & \textbf{94.81 (0.38)} & 95.21 (0.16) & 96.68 (0.37) \\
& Sum RMSE
& 2.71 (0.18) & \textbf{2.70 (0.14)} & 2.89 (0.24) & 4.84 (0.74) & 2.74 (0.16) \\
& Sum coverage (\%)
& 95.41 (2.15) & \textbf{94.93 (1.93)} & 97.65 (1.58) & 99.31 (1.13) & 96.05 (1.72) \\
& PI length
& \textbf{1.76 (0.04)} & 1.76 (0.04) & 1.92 (0.05) & 4.10 (0.26) & 2.07 (0.04) \\
& Sum PI length
& 10.75 (0.14) & \textbf{10.54 (0.16)} & 14.01 (0.35) & 27.96 (1.89) & 11.28 (0.27) \\
\addlinespace
Col/Neg & Norm
& \textbf{3.29 (0.09)} & 3.32 (0.09) & 3.52 (0.09) & 7.15 (0.28) & 3.43 (0.08) \\
& Coverage (\%)
& 94.50 (0.36) & 94.46 (0.35) & \textbf{94.84 (0.35)} & 95.18 (0.38) & 96.55 (0.22) \\
& Sum RMSE
& \textbf{8.69 (1.02)} & 9.08 (1.11) & 10.66 (1.48) & 11.33 (1.66) & 10.33 (1.21) \\
& Sum coverage (\%)
& 92.96 (3.63) & 92.96 (3.12) & 55.68 (3.75) & \textbf{96.64 (1.39)} & 98.67 (1.41) \\
& PI length
& \textbf{1.76 (0.04)} & 1.76 (0.04) & 1.92 (0.05) & 4.06 (0.17) & 2.06 (0.05) \\
& Sum PI length
& 29.43 (1.88) & 30.48 (2.12) & \textbf{14.01 (0.41)} & 41.83 (1.74) & 44.00 (1.88) \\
\addlinespace
Mis/Pos & Norm
& 3.58 (0.08) & \textbf{3.57 (0.08)} & 4.29 (0.13) & 8.92 (2.93) & 3.79 (0.11) \\
& Coverage (\%)
& 94.26 (0.48) & 94.04 (0.48) & \textbf{94.61 (0.39)} & 94.50 (0.85) & 98.07 (0.20) \\
& Sum RMSE
& 2.96 (0.22) & \textbf{2.93 (0.20)} & 3.73 (0.40) & 5.43 (0.90) & 3.02 (0.17) \\
& Sum coverage (\%)
& 96.11 (2.34) & \textbf{95.20 (2.28)} & 97.33 (1.90) & 99.09 (1.13) & 97.87 (1.23) \\
& PI length
& 1.91 (0.04) & \textbf{1.88 (0.04)} & 2.28 (0.06) & 4.95 (1.71) & 2.55 (0.05) \\
& Sum PI length
& 12.13 (0.25) & \textbf{11.53 (0.20)} & 17.10 (0.43) & 29.50 (1.87) & 13.44 (0.21) \\
\addlinespace
Mis/Neg & Norm
& 3.55 (0.09) & \textbf{3.54 (0.08)} & 4.27 (0.14) & 8.79 (2.78) & 3.76 (0.09) \\
& Coverage (\%)
& 94.49 (0.42) & 94.27 (0.46) & 94.68 (0.33) & \textbf{94.72 (0.65)} & 98.12 (0.22) \\
& Sum RMSE
& 9.19 (0.99) & \textbf{9.19 (0.98)} & 12.58 (1.51) & 20.70 (28.93) & 11.09 (1.25) \\
& Sum coverage (\%)
& 91.73 (3.35) & \textbf{92.48 (3.46)} & 54.99 (3.88) & 92.32 (3.31) & 99.36 (0.81) \\
& PI length
& 1.91 (0.04) & \textbf{1.88 (0.04)} & 2.27 (0.06) & 4.92 (1.53) & 2.53 (0.05) \\
& Sum PI length
& 30.06 (1.60) & 30.97 (1.66) & \textbf{17.06 (0.47)} & 70.10 (99.65) & 55.54 (2.31) \\
\bottomrule
\end{tabular}
\end{table}

\subsection{Multivariate Mat\'ern Process}
\label{subsec:all_matern}
We present the estimation and prediction results in Tables~\ref{table:matern_random_compact} and \ref{table:matern_tridiagonal_compact} when the data are generated from a multivariate Mat\'ern process with $a=3$ and $\nu=1.5$ to address the gap where IOX is misspecified for the marginal process, as in the main text. Under this setting, the cross-process parameters are $a_{ij}=3$ and $\nu_{ij}=1.5$ for $i\neq j$ and the covariance is separable before adding the nugget. We implemented the latent model that separately models the spatial process and the nugget described in \citet{peruzzi2024} since it aligns with our model formulation.

For random precision matrices, IOX has larger absolute bias in $a$ and smaller absolute bias in $\nu$ under colocated observations, with the pattern reversed under misalignment. For tridiagonal precision matrices, Penalized has smaller absolute biases in both parameters than IOX. For random precision matrices, Penalized and IOX have comparable prediction norms and sum RMSE. For tridiagonal precision matrices, IOX achieves the lowest prediction norm and sum RMSE under most settings, and its individual-process and sum coverages are slightly above the nominal rate. The Penalized approach has near-nominal sum coverage under positive off-diagonals but undercovers under negative off-diagonals. LMC has near-nominal sum coverage under negative off-diagonals. Overall, the correctly specified Matérn comparison shows similar performance for Penalized and IOX.

\begin{table}[H]\centering\small
\centering\small
\setlength{\tabcolsep}{3pt}
\renewcommand{\arraystretch}{0.8}
\caption{All-Mat\'ern random-precision simulations: mean (SD) across 15 replicates, with Bias rows averaged over processes. Col/Mis: colocated/misaligned; Pos/Neg: positive/negative precision off-diagonals. Bold: smallest absolute bias, coverage closest to 95\%, or lowest mean otherwise.}
\label{table:matern_random_compact}
\begin{tabular}{@{}llrrrrr@{}}
\toprule
Setting & Metric & Penalized & Dense & Ind & LMC & IOX\\
\midrule
Col/Pos & Bias($a$)
& \textbf{0.21 (1.23)} & 0.70 (1.15) & 0.45 (1.32) & -- & 0.92 (1.04)\\
& Bias($\nu$)
& 0.36 (0.83) & 0.48 (0.82) & 0.36 (0.92) & -- & \textbf{-0.03 (0.52)}\\
& Bias($d$)
& 0.00 (0.00) & 0.01 (0.00) & \textbf{0.00 (0.01)} & -- & 0.00 (0.01)\\
& Norm
& \textbf{2.95 (0.05)} & 2.96 (0.05) & 2.97 (0.05) & 4.47 (0.09) & 2.95 (0.05)\\
& Coverage (\%)
& 94.63 (0.39) & 94.46 (0.43) & 94.76 (0.33) & \textbf{94.85 (0.42)} & 95.97 (0.25)\\
& Sum RMSE
& \textbf{2.66 (0.20)} & 2.66 (0.17) & 2.72 (0.20) & 3.83 (0.23) & 2.66 (0.18)\\
& Sum coverage (\%)
& 95.84 (2.01) & \textbf{95.25 (1.70)} & 96.64 (1.69) & 97.17 (1.17) & 96.64 (1.25)\\
\addlinespace
Col/Neg & Bias($a$)
& \textbf{-0.03 (1.07)} & 0.64 (1.02) & 0.40 (1.19) & -- & 1.11 (1.00)\\
& Bias($\nu$)
& 0.22 (0.61) & 0.39 (0.61) & 0.27 (0.72) & -- & \textbf{0.06 (0.48)}\\
& Bias($d$)
& 0.00 (0.00) & 0.01 (0.00) & \textbf{0.00 (0.01)} & -- & 0.00 (0.01)\\
& Norm
& \textbf{3.02 (0.06)} & 3.03 (0.06) & 3.06 (0.06) & 4.61 (0.07) & 3.02 (0.06)\\
& Coverage (\%)
& 94.62 (0.46) & 94.67 (0.38) & \textbf{94.84 (0.43)} & 95.29 (0.25) & 96.04 (0.29)\\
& Sum RMSE
& 4.99 (0.57) & 5.03 (0.62) & 5.20 (0.48) & 5.64 (0.41) & \textbf{4.98 (0.58)}\\
& Sum coverage (\%)
& 90.45 (3.57) & 93.81 (3.03) & 78.35 (4.36) & \textbf{95.57 (2.26)} & 96.27 (2.02)\\
\addlinespace
Mis/Pos & Bias($a$)
& 0.40 (1.48) & 0.28 (1.39) & 0.71 (1.72) & -- & \textbf{-0.19 (0.79)}\\
& Bias($\nu$)
& 0.53 (1.19) & \textbf{0.43 (1.13)} & 0.65 (1.41) & -- & -0.64 (0.27)\\
& Bias($d$)
& \textbf{0.00 (0.00)} & 0.00 (0.00) & 0.00 (0.02) & -- & 0.00 (0.01)\\
& Norm
& 3.20 (0.06) & \textbf{3.19 (0.06)} & 3.30 (0.05) & 4.67 (0.09) & 3.25 (0.10)\\
& Coverage (\%)
& \textbf{94.74 (0.36)} & 94.21 (0.41) & 94.64 (0.34) & 94.65 (0.37) & 97.84 (0.26)\\
& Sum RMSE
& 2.89 (0.22) & \textbf{2.86 (0.21)} & 3.00 (0.25) & 4.04 (0.29) & 2.87 (0.20)\\
& Sum coverage (\%)
& 96.27 (2.04) & \textbf{95.52 (2.49)} & 96.85 (2.17) & 97.23 (1.08) & 98.24 (1.42)\\
\addlinespace
Mis/Neg & Bias($a$)
& 0.51 (1.53) & 0.30 (1.35) & 0.63 (1.67) & -- & \textbf{-0.03 (0.82)}\\
& Bias($\nu$)
& 0.48 (1.11) & \textbf{0.33 (0.99)} & 0.52 (1.25) & -- & -0.57 (0.28)\\
& Bias($d$)
& \textbf{0.00 (0.00)} & -0.01 (0.00) & 0.00 (0.02) & -- & 0.00 (0.01)\\
& Norm
& 3.32 (0.07) & \textbf{3.29 (0.05)} & 3.45 (0.06) & 4.85 (0.08) & 3.33 (0.06)\\
& Coverage (\%)
& 94.71 (0.34) & 94.24 (0.39) & 94.62 (0.32) & \textbf{94.91 (0.36)} & 97.78 (0.19)\\
& Sum RMSE
& 5.51 (0.56) & \textbf{5.35 (0.55)} & 6.02 (0.50) & 6.32 (0.48) & 5.47 (0.59)\\
& Sum coverage (\%)
& 84.32 (5.67) & 93.01 (2.94) & 75.95 (4.27) & \textbf{94.51 (2.53)} & 98.08 (1.04)\\
\bottomrule
\end{tabular}
\end{table}

\begin{table}[H]\centering\small
\setlength{\tabcolsep}{3pt}
\renewcommand{\arraystretch}{0.8}
\caption{All-Mat\'ern tridiagonal-precision simulations: mean (SD) across 15 replicates, with Bias rows averaged over processes. Col/Mis: colocated/misaligned; Pos/Neg: positive/negative precision off-diagonals. Bold: smallest absolute bias, coverage closest to 95\%, or lowest mean otherwise.
}
\label{table:matern_tridiagonal_compact}
\begin{tabular}{@{}llrrrrr@{}}
\toprule
Setting & Metric & Penalized & Dense & Ind & LMC & IOX\\
\midrule
Col/Pos & Bias($a$)
& \textbf{-0.09 (0.56)} & 0.11 (0.60) & 0.09 (0.73) & -- & 1.28 (0.56)\\
& Bias($\nu$)
& \textbf{-0.01 (0.20)} & 0.04 (0.22) & 0.06 (0.29) & -- & 0.15 (0.24)\\
& Bias($d$)
& 0.00 (0.00) & 0.00 (0.00) & \textbf{0.00 (0.02)} & -- & 0.01 (0.01)\\
& Norm
& 4.30 (0.18) & 4.36 (0.19) & 4.34 (0.16) & 6.99 (0.16) & \textbf{4.19 (0.19)}\\
& Coverage (\%)
& 93.88 (0.81) & 93.40 (0.92) & 94.75 (0.67) & \textbf{95.12 (0.36)} & 95.89 (0.59)\\
& Sum RMSE
& \textbf{2.59 (0.19)} & 2.59 (0.15) & 2.91 (0.26) & 3.96 (0.45) & 2.60 (0.16)\\
& Sum coverage (\%)
& 95.31 (2.53) & \textbf{94.72 (2.16)} & 98.83 (0.85) & 99.63 (0.90) & 96.16 (1.55)\\
\addlinespace
Col/Neg & Bias($a$)
& \textbf{0.03 (0.51)} & 0.24 (0.56) & 0.22 (0.72) & -- & 1.30 (0.56)\\
& Bias($\nu$)
& \textbf{0.02 (0.19)} & 0.06 (0.22) & 0.08 (0.30) & -- & 0.16 (0.24)\\
& Bias($d$)
& 0.00 (0.00) & 0.00 (0.00) & \textbf{0.00 (0.02)} & -- & 0.01 (0.01)\\
& Norm
& 4.33 (0.21) & 4.35 (0.21) & 4.36 (0.15) & 7.00 (0.15) & \textbf{4.19 (0.13)}\\
& Coverage (\%)
& 93.89 (0.77) & 93.99 (0.79) & 94.89 (0.39) & \textbf{95.07 (0.42)} & 95.87 (0.32)\\
& Sum RMSE
& 20.20 (2.88) & 20.27 (2.97) & 18.76 (2.10) & 18.83 (2.31) & \textbf{18.32 (1.81)}\\
& Sum coverage (\%)
& 92.64 (3.38) & 92.85 (3.11) & 42.29 (4.40) & \textbf{95.31 (1.24)} & 96.00 (1.71)\\
\addlinespace
Mis/Pos & Bias($a$)
& \textbf{-0.01 (1.08)} & -0.03 (0.70) & 0.18 (0.96) & -- & 0.35 (0.54)\\
& Bias($\nu$)
& \textbf{-0.04 (0.44)} & -0.15 (0.30) & 0.14 (0.50) & -- & -0.38 (0.18)\\
& Bias($d$)
& -0.01 (0.01) & -0.01 (0.00) & \textbf{0.00 (0.04)} & -- & 0.00 (0.01)\\
& Norm
& 4.67 (0.17) & 4.59 (0.17) & 5.56 (0.19) & 7.58 (0.21) & \textbf{4.54 (0.20)}\\
& Coverage (\%)
& 93.23 (0.56) & 93.35 (0.73) & 94.48 (0.56) & \textbf{94.95 (0.40)} & 97.72 (0.36)\\
& Sum RMSE
& 3.01 (0.37) & 2.86 (0.21) & 4.20 (0.48) & 4.55 (0.67) & \textbf{2.82 (0.17)}\\
& Sum coverage (\%)
& 97.01 (1.97) & \textbf{95.20 (2.38)} & 98.61 (1.19) & 99.41 (0.77) & 97.65 (1.37)\\
\addlinespace
Mis/Neg & Bias($a$)
& -0.11 (0.67) & \textbf{0.03 (0.61)} & 0.25 (0.88) & -- & 0.42 (0.55)\\
& Bias($\nu$)
& -0.14 (0.28) & -0.16 (0.25) & \textbf{0.12 (0.43)} & -- & -0.36 (0.19)\\
& Bias($d$)
& -0.02 (0.00) & -0.02 (0.00) & \textbf{0.00 (0.03)} & -- & 0.00 (0.01)\\
& Norm
& 4.58 (0.17) & 4.59 (0.17) & 5.54 (0.21) & 7.53 (0.16) & \textbf{4.51 (0.13)}\\
& Coverage (\%)
& 93.54 (0.70) & 93.54 (0.71) & 94.62 (0.57) & \textbf{94.94 (0.31)} & 97.72 (0.37)\\
& Sum RMSE
& 19.92 (2.42) & 19.99 (2.53) & 21.17 (2.37) & 19.39 (2.06) & \textbf{18.64 (1.66)}\\
& Sum coverage (\%)
& 90.72 (3.61) & 92.59 (3.00) & 44.21 (3.77) & \textbf{95.84 (1.49)} & 98.88 (0.90)\\
\bottomrule
\end{tabular}
\end{table}

\section{Additional MERRA-2 Analysis Results}\label{app:data}
We provide parameter estimation results for MERRA-2 analysis in Table~\ref{tab:merra-estimation}. 
The proposal parameters $a_{\text{prop}}$ and $\nu_{\text{prop}}$ are the median fitted $a$ and minimum fitted $\nu$ from the independent process fits. Different numbers of random features and Monte Carlo sample sizes are used for estimation ($L=20{,}000$, $T=40$), likelihood evaluation ($L_{\text{ll}}=1{,}000$), and prediction ($L_{\text{pred}}=30{,}000$, $T_{\text{pred}}=20{,}000$).
The first $T/2$ sweeps are discarded as burn-in.
In addition to an unpenalized dense fit, we consider the penalty grid $\{0.0001,\allowbreak  0.00025, \allowbreak 0.0004,\allowbreak  0.00055,\allowbreak 
  0.0007,\allowbreak  0.00085,\allowbreak  0.001, \allowbreak 0.00125,\allowbreak  0.00175,\allowbreak 
  0.0025,\allowbreak  0.003, \allowbreak 0.005, \allowbreak 0.01,\allowbreak  0.02\}$, each with $700$ iterations. 
We take $\gamma_m = 30/(29+m)$. BIC uses $n=500$ for colocated observations and $n=300$ for misaligned observations.
In the misaligned setting, $\mb{\sigma}^{-1}$ is initialized at the diagonal matrix of inverse independent-fit variances.

\begin{table}[H]
\centering\small
\setlength{\tabcolsep}{3pt}
\caption{Parameter estimation for MERRA-2. Entries are medians across three folds of the within-fold means over processes.}
\label{tab:merra-estimation}
\renewcommand{\arraystretch}{0.8}
\begin{tabular}{llrrrrr}
\toprule
Design & Metric & Penalized & Dense & Ind & LMC ($q=10$) & IOX \\
\midrule
Colocated
 & $a/\phi$ & 4.26 & 4.30 & 4.00 & 6.58 & 4.25 \\
 & $\nu$ & 1.92 & 1.90 & 1.86 & 4.92 & 2.09 \\
 & Nugget & 0.006 & 0.007 & 0.003 & 0.230 & 0.030 \\
 & Sparsity (\%) & 15.64 & 0.00 & 100.00 & -- & -- \\
\addlinespace
Misaligned
 & $a/\phi$ & 3.82 & 3.94 & 4.09 & 5.32 & 3.81 \\
 & $\nu$ & 1.96 & 1.95 & 2.00 & 3.62 & 1.78 \\
 & Nugget & 0.008 & 0.007 & 0.011 & 0.257 & 0.053 \\
 & Sparsity (\%) & 66.92 & 0.00 & 100.00 & -- & -- \\
\bottomrule
\end{tabular}
\end{table}

\end{document}